\documentclass[oneside,11pt]{article}

\usepackage[utf8]{inputenc}
\usepackage[T1]{fontenc}

\usepackage{amsthm}
\usepackage{amsmath}
\usepackage{amsfonts}
\usepackage{amssymb}

\usepackage{graphicx} 
\usepackage{float}
\usepackage{booktabs}
\usepackage{multirow}
\usepackage{rotating}
\usepackage{array}

\usepackage{breakcites}

\usepackage{enumitem}

\usepackage[top=1.5cm,bottom=2cm,right=2.5cm,left=2.5cm]{geometry}
\usepackage{titling}
\usepackage{natbib}

\usepackage{ifplatform}

\ifwindows
\fi

\ifwindows
	\usepackage{bbm}
\else
	\iflinux
		\usepackage{bbm}
	\else
		\newcommand{\mathbbm}[1]{\mathbf{#1}}
	\fi
\fi

\DeclareFontFamily{OT1}{pzc}{}
\DeclareFontShape{OT1}{pzc}{m}{it}{<-> s * [1.10] pzcmi7t}{}
\DeclareMathAlphabet{\mathpzc}{OT1}{pzc}{m}{it}

\usepackage[pdftex,final,bookmarksnumbered,bookmarksopen=false,breaklinks,colorlinks]{hyperref}
\hypersetup{
	final,
    bookmarksnumbered=true,
    bookmarksopen=true,
    bookmarksopenlevel=0,
    unicode=false,          
    pdftoolbar=true,        
    pdfmenubar=true,        
    pdffitwindow=false,     
    pdftitle={Testing parametric models in linear-directional regression},    
	pdfdisplaydoctitle=true, 
	pdftoolbar=true,
	pdfmenubar=true,
	pdflang={English},
    pdfauthor={Eduardo Garcia-Portugues, Ingrid Van Keilegom, Rosa M. Crujeiras, Wenceslao Gonzalez-Manteiga},     
    pdfsubject={arXiv paper},   
    pdfcreator={Eduardo Garcia-Portugues},   
    pdfproducer={Eduardo Garcia-Portugues}, 
    pdfkeywords={Bootstrap calibration}{Directional data}{Goodness-of-fit test}{Local linear regression}, 
    pdfnewwindow=true,      
    breaklinks=true,
    hidelinks,       
    linkcolor=black,          
    citecolor=black,        
    filecolor=magenta,      
    urlcolor=cyan           
}

\newtheorem{theo}{Theorem}

\newtheorem{rem}{Remark}

\newtheorem{lem}{Lemma}
\newtheorem{algo}{Algorithm}

\allowdisplaybreaks

\newif\ifmain
\maintrue
\ifmain

\else

\fi
\newif\ifsupplement
\supplementtrue

\begin{document}

\ifmain

\title{Testing parametric models in linear-directional regression}
\setlength{\droptitle}{-1cm}
\predate{}%
\postdate{}%

\date{}

\author{Eduardo Garc\'ia-Portugu\'es$^{1,2,3,5}$, Ingrid Van Keilegom$^4$, \\ Rosa M. Crujeiras$^3$, and Wenceslao Gonz\'alez-Manteiga$^3$}

\footnotetext[1]{
Department of Mathematical Sciences, University of Copenhagen (Denmark).}
\footnotetext[2]{
The Bioinformatics Centre, Department of Biology, University of Copenhagen (Denmark).}
\footnotetext[3]{
Department of Statistics and Operations Research, University of Santiago de Compostela (Spain).}
\footnotetext[4]{
Institute of Statistics, Biostatistics and Actuarial Sciences, Universit\'e catholique de Louvain (Belgium).}
\footnotetext[5]{Corresponding author. e-mail: \href{mailto:egarcia@math.ku.dk}{egarcia@math.ku.dk}.}
\maketitle

\begin{abstract}
This paper presents a goodness-of-fit test for parametric regression models with scalar response and directional predictor, that is, a vector on a sphere of arbitrary dimension. The testing procedure is based on the weighted squared distance between a smooth and a parametric regression estimator, where the smooth regression estimator is obtained by a projected local approach. Asymptotic behavior of the test statistic under the null hypothesis and local alternatives is provided, jointly with a consistent bootstrap algorithm for application in practice. A simulation study illustrates the performance of the test in finite samples. The procedure is applied to test a linear model in text mining.
\end{abstract}
\begin{flushleft}
\end{flushleft}

\section{Introduction}
\label{sec:intro}

Directional data (data on a general sphere of dimension $q$) appear in a variety of contexts, the simplest one being provided by observations of angles on a circle (circular data). Directional data is present in wind directions or animal orientation \citep{Mardia2000} and, recently, it has been considered in higher dimensional settings for text mining \citep{Srivastava2010}. In order to identify a statistical pattern within a certain collection of texts, these objects may be represented by a vector on a sphere where each vector component gives the relative frequency of a certain word. From this vector-space representation, text classification can be performed \citep{Banerjee2005}, but other interesting problems such as popularity prediction could be tackled. For instance, a linear-directional regression model could be used to predict the popularity of articles in news aggregators, quantified by the number of comments or views \citep{Tatar2012}, based on the news contents.\\ 

When dealing with directional and linear variables at the same time, the joint behavior could be modeled by considering a flexible density estimator \citep{Garcia-Portugues:dirlin}. Nevertheless, a regression approach may be more useful, allowing at the same time for explaining a relation between the variables and for making predictions. Nonparametric regression estimation methods for linear-directional models have been proposed by different authors. For example, \cite{Cheng2013} introduced a general local linear regression method on manifolds and, quite recently, \cite{DiMarzio2014} presented a local polynomial method when both the predictor and the response are defined on spheres. Despite the flexibility of these estimators, in terms of interpretation of the results, purely parametric models may be more convenient. In this context, goodness-of-fit tests can be designed, providing a tool for assessing a certain parametric linear-directional regression model.\\

Goodness-of-fit tests for directional data, or including a directional component in the data generating process, have not been deeply studied. For the density case, \cite{Boente2013} provide a nonparametric goodness-of-fit test for directional densities and similar ideas are used by \cite{Garcia-Portugues:clt} for directional-linear densities. Except for the exploratory tool and lack-of-fit test for linear-circular regression developed by \cite{Deschepper2008} there are no other works in the regression context. The related Euclidean literature is extensive: the reader is referred to \cite{Hart1997} for a comprehensive reference and to \cite{Hardle1993} and \cite{Alcala1999} for the most relevant works for this contribution.\\

This paper presents a goodness-of-fit test for parametric linear-directional regression models. The test is constructed from a projected local regression estimator (Section \ref{sec:reg}). The asymptotic distribution of the test statistic, based on a weighted squared distance between the nonparametric and parametric fits, is obtained under a family of local alternatives containing the null hypothesis (Section \ref{sec:gof}). A bootstrap strategy, proved to be consistent, is proposed for the calibration of the test in practice. The performance of the test is checked for finite samples in a simulation study (Section \ref{sec:simu}) and the test is applied to assess a constrained linear model for news popularity prediction in text mining (Section \ref{sec:data}). An appendix contains the proofs of the main results, whereas technical lemmas and further information on the simulation study and data application are provided as Supporting Information (SI).

\section{Nonparametric linear-directional regression}
\label{sec:reg}

Let $\Omega_q=\big\{\mathbf{x}\in\mathbb{R}^{q+1}:||\mathbf{x}||=1\big\}$ denote the $q$-sphere in $\mathbb{R}^{q+1}$ and $\omega_{q}$ denote both its associated Lebesgue measure and its surface area, $\omega_{q}=2\pi^\frac{q+1}{2}\big/\Gamma\big(\frac{q+1}{2}\big)$. A directional density $f$ satisfies $\int_{\Omega_q}f(\mathbf{x})\,\omega_q(d\mathbf{x})=1$. From a sample $\mathbf{X}_1,\ldots,\mathbf{X}_n$ of a random variable (rv) $\mathbf{X}$ with density $f$, \cite{Hall1987} and \cite{Bai1988} introduced the kernel density estimator 
\begin{equation}
\hat f_{h}(\mathbf{x})=\frac{1}{n}\sum_{i=1}^n L_h\left(\mathbf{x},\mathbf{X}_i\right),\quad L_h(\mathbf{x},\mathbf{y})=c_{h,q}(L)L\left(\frac{1-\mathbf{x}^T\mathbf{y}}{h^2}\right),\quad \mathbf{x}\in \Omega_q,
\label{KDE}
\end{equation}
where $L$ is a directional kernel, $h>0$ is the bandwidth parameter and
\begin{align}
c_{h,q}(L)^{-1}=\lambda_{h,q}(L)h^q=\lambda_{q}(L)h^{q}(1+\mathpzc{o}\left(1\right))\label{chq}
\end{align}
with $\lambda_{h,q}(L)=\omega_{q-1}\int_0^{2h^{-2}} L(r) r^{\frac{q}{2}-1}(2-rh^2)^{\frac{q}{2}-1}\,dr$ and $\lambda_q(L)=2^{\frac{q}{2}-1}\omega_{q-1}\int_0^{\infty} L(r) r^{\frac{q}{2}-1}\,dr$.\\ 

Assume that $\mathbf{X}$ is the covariate in the regression model
\begin{align}
Y=m(\mathbf{X})+\sigma(\mathbf{X}){\varepsilon}, \label{model}
\end{align}
where $Y$ is a scalar rv (response), $m$ is the regression function given by the conditional mean ($m(\mathbf{x})=\mathbb{E}\left[Y|\mathbf{X}=\mathbf{x}\right]$), and $\sigma^2$ is the conditional variance ($\sigma^2(\mathbf{x})=\mathbb{V}\mathrm{ar}\left[Y|\mathbf{X}=\mathbf{x}\right]$). Errors are collected by ${\varepsilon}$, a rv such that $\mathbb{E}\left[\varepsilon|\mathbf{X}\right]=0$, $\mathbb{E}\left[\varepsilon^2|\mathbf{X}\right]=1$ and $\mathbb{E}\left[|\varepsilon|^3|\mathbf{X}\right]$ and $\mathbb{E}\left[\varepsilon^4|\mathbf{X}\right]$ are assumed to be bounded rv's. Both $m,f:\Omega_q\longrightarrow\mathbb{R}$ can be extended from $\Omega_q$ to $\mathbb{R}^{q+1}\boldsymbol\backslash\left\{\mathbf{0}\right\}$ by considering a radial projection. This allows the consideration of easily tractable derivatives and the use of Taylor expansions. 
\begin{enumerate}[label=\textbf{A\arabic{*}}., ref=\textbf{A\arabic{*}}]
	\item $m$ and $f$ are extended from $\Omega_q$ to $\mathbb{R}^{q+1}\boldsymbol\backslash\left\{\mathbf{0}\right\}$ by $m\left(\mathbf{x}\right)\equiv m\left(\mathbf{x}/\left|\left|\mathbf{x}\right|\right|\right)$ and $f\left(\mathbf{x}\right)\equiv f\left(\mathbf{x}/\left|\left|\mathbf{x}\right|\right|\right)$. $m$ is three times and $f$ is twice continuously differentiable. $f$ is bounded away from zero. \label{assump:a1}
\end{enumerate}
Assumption \ref{assump:a1} guarantees that $f$ and $m$ are uniformly bounded in $\Omega_{q}$. More importantly, the directional derivative of $m$ (and $f$) in the direction $\mathbf{x}$ and evaluated at $\mathbf{x}$ is zero, \textit{i.e.}, $\mathbf{x}^T\boldsymbol\nabla m(\mathbf{x})=0$. This is a key fact on the construction of Taylor expansion of $m$ at $\mathbf{X}_i$:
\begin{align*}
m(\mathbf{X}_i)&=m(\mathbf{x})+\boldsymbol\nabla m(\mathbf{x})^T(\mathbf{X}_i-\mathbf{x})+\mathcal{O}\left(\left|\left|\mathbf{X}_i-\mathbf{x}\right|\right|^2\right)\\
&=m(\mathbf{x})+\boldsymbol\nabla m(\mathbf{x})^T\left(\mathbf{I}_{q+1}-\mathbf{x}\mathbf{x}^T\right)(\mathbf{X}_i-\mathbf{x})+\mathcal{O}\left(\left|\left|\mathbf{X}_i-\mathbf{x}\right|\right|^2\right)\\
&\approx \beta_0+\boldsymbol\beta_1^T \mathbf{B}_{\mathbf{x}}^T(\mathbf{X}_i-\mathbf{x}),
\end{align*}
where $\mathbf{B}_{\mathbf{x}}=(\mathbf{b}_1,\ldots,\mathbf{b}_q)_{(q+1)\times q}$ is the matrix that completes $\mathbf{x}$ to an orthonormal basis $\left\{\mathbf{x},\mathbf{b}_1,\ldots,\mathbf{b}_q\right\}$ of $\mathbb{R}^{q+1}$ and satisfies $\mathbf{B}_{\mathbf{x}}^T\mathbf{B}_{\mathbf{x}}=\mathbf{I}_{q}$ and $\mathbf{B}_{\mathbf{x}}\mathbf{B}_{\mathbf{x}}^T=\sum_{i=1}^q\mathbf{b}_i\mathbf{b}_i^T=\mathbf{I}_{q+1}-\mathbf{x}\mathbf{x}^T$, with $\mathbf{I}_q$ the identity matrix of dimension $q$.\\

With this setting, $\beta_0\in\mathbb R$ captures the constant effect in $m(\mathbf{x})$ while $\boldsymbol\beta_1\in\mathbb R^q$ contains the linear effects of the \textit{projected gradient} of $m$ given by $\mathbf{B}_{\mathbf{x}}^T\boldsymbol\nabla m(\mathbf{x})$. It should be noted that the dimension of $\boldsymbol\beta_1$ is the \textit{adequate} for the $q$-sphere $\Omega_{q}$, which would be $q+1$ if an usual Taylor expansion in $\mathbb{R}^{q+1}$ was performed. The \textit{projected local estimator} at $m(\mathbf{x})$ is obtained as the weighted average of local constant (denoted by $p=0$) or linear ($p=1$) fits given by $\beta_0$ or  $\beta_0+\boldsymbol\beta_1^T\mathbf{B}_{\mathbf{x}}^T(\mathbf{X}_i-\mathbf{x})$, respectively. Given the sample $(\mathbf{X}_1,Y_1),\ldots,(\mathbf{X}_n,Y_n)$ from (\ref{model}), comprised of independent and identically distributed (iid) rv's in $\Omega_{q} \times\mathbb{R}$, both fits can be formulated as the weighted least squares problem
\begin{align*}
\min_{\boldsymbol\beta\in\mathbb{R}^{q+1}}\sum_{i=1}^n \Big(Y_i-\beta_0-\delta_{p,1}\left(\beta_1,\ldots,\beta_q\right)^T\mathbf{B}_{\mathbf{x}}^T(\mathbf{X}_{i}-\mathbf{x})\Big)^2 L_h(\mathbf{x},\mathbf{X}_i),
\end{align*}
where $\delta_{r,s}$ is the Kronecker Delta. The solution to the minimization problem is given by
\begin{align}
\hat{\boldsymbol\beta}=\left(\boldsymbol{\mathcal{X}}_{\mathbf{x},p}^T\boldsymbol{\mathcal{W}}_{\mathbf{x}}\boldsymbol{\mathcal{X}}_{\mathbf{x},p}\right)^{-1}\boldsymbol{\mathcal{X}}_{\mathbf{x},p}^T\boldsymbol{\mathcal{W}}_{\mathbf{x}}\mathbf{Y}, \label{betaest}
\end{align}
where $\mathbf{Y}$ is the vector of observed responses, $\boldsymbol{\mathcal{W}}_{\mathbf{x}}$ is the diagonal weight matrix with $i$-th entry $L_h(\mathbf{x},\mathbf{X}_i)$, $\boldsymbol{\mathcal{X}}_{\mathbf{x},1}$ is the design matrix with $i$-th row $(1, (\mathbf{X}_i-\mathbf{x})^T\mathbf{B}_{\mathbf{x}})$ and $\boldsymbol{\mathcal{X}}_{\mathbf{x},0}=\mathbf{1}$ ($\mathbf{1}$ stands for a  vector of ones whose dimension is determined by the context). The projected local estimator at $\mathbf{x}$ is given by $\hat\beta_0=\hat m_{h,p}(\mathbf{x})$ and is a weighted linear combination of the responses ($\mathbf{e}_1$ is a null vector with one in the first component):
\begin{align}
\hat m_{h,p}(\mathbf{x})&=\mathbf{e}_1^T\left(\boldsymbol{\mathcal{X}}_{\mathbf{x},p}^T\boldsymbol{\mathcal{W}}_{\mathbf{x}}\boldsymbol{\mathcal{X}}_{\mathbf{x},p}\right)^{-1}\boldsymbol{\mathcal{X}}_{\mathbf{x},p}^T\boldsymbol{\mathcal{W}}_{\mathbf{x}}\mathbf{Y}=\sum_{i=1}^n W_n^p\left(\mathbf{x},\mathbf{X}_i\right)Y_i,\label{hatm}
\end{align}

The next assumptions ensure that $\hat m_{h,p}$ is  a consistent estimator of $m$:
\begin{enumerate}[label=\textbf{A\arabic{*}}., ref=\textbf{A\arabic{*}}]
	\setcounter{enumi}{1}
	\item The
	conditional variance $\sigma^2$ is uniformly continuous and bounded away from zero. \label{assump:a2}
	\item $L:[0,\infty)\rightarrow[0,\infty)$ is a continuous and bounded function with exponential decay.\label{assump:a3}
	\item The sequence of bandwidths $h=h_n$ is positive and satisfies $h\to 0$ and $nh^q\to \infty$. \label{assump:a4}
\end{enumerate}
Assumptions \ref{assump:a2} and \ref{assump:a4} are usual assumptions for the multivariate local linear estimator \citep{Ruppert1994}. \ref{assump:a3} allows for the use of non-compactly supported kernels, such as the popular \textit{von Mises kernel} $L(r)=e^{-r}$. 

\begin{rem}
	The proposal of \cite{DiMarzio2014} for a local linear estimator of $m$ is rooted on a Taylor expansion of the $\sin$ and $\cos$ functions of the tangent-normal decomposition. This leads to an overparametrized design matrix of $q+2$ columns which makes $\boldsymbol{\mathcal{X}}_{\mathbf{x},p}^T\boldsymbol{\mathcal{W}}_{\mathbf{x}}\boldsymbol{\mathcal{X}}_{\mathbf{x},p}$ exactly singular, a fact handled by the authors with a pseudo-inverse. It should be noted that \cite{DiMarzio2014}'s proposal and (\ref{hatm}) present some remarkable differences: for the circular case, (\ref{hatm}) corresponds to \cite{DiMarzio2009}'s proposal (with parametrization $\kappa\equiv1/h^2$), but \cite{DiMarzio2014} differs from the aforementioned reference. Although both estimators share the same asymptotics, (\ref{hatm}) somehow offers a simpler construction and a more natural connection with previous proposals.
\end{rem}

\section{Goodness-of-fit test for linear-directional regression}
\label{sec:gof}

Assuming that model (\ref{model}) holds, the goal is to test if the regression function $m$ belongs to the parametric class of functions $\mathcal{M}_\Theta=\left\{m_{\boldsymbol{\theta}}:\boldsymbol{\theta}\in\Theta\subset\mathbb{R}^{s}\right\}$. This is equivalent to testing
\[
H_0: m(\mathbf{x})=m_{\boldsymbol{\theta}_0}(\mathbf{x}),\text{ for all }\mathbf{x}\in\Omega_q,\text{ versus }H_1: m(\mathbf{x})\neq m_{\boldsymbol{\theta}_0}(\mathbf{x}),\text{ for some }\mathbf{x}\in\Omega_q,
\]
with $\boldsymbol{\theta}_0\in\Theta$ known (simple hypothesis) or unknown (composite) and where \textit{for all} holds except for a set of probability zero and \textit{for some} holds for a set of positive probability.\\

The proposed statistic to test $H_0$ compares the nonparametric estimator with a smoothed parametric estimator in $\mathcal{M}_{\Theta}$ through a squared weighted norm:
\[
T_n=\int_{\Omega_q}\left(\hat m_{h,p}(\mathbf{x})-\mathcal{L}_{h,p}m_{\hat{\boldsymbol{\theta}}}(\mathbf{x}) \right)^2\hat f_h(\mathbf{x})w(\mathbf{x})\,\omega_q(d\mathbf{x})
\]
where $\mathcal{L}_{h,p}m (\mathbf{x})=\sum_{i=1}^n W_n^p\left(\mathbf{x},\mathbf{X}_i\right)m(\mathbf{X}_i)$ represents the local smoothing of the function $m$ from measurements $\left\{\mathbf{X}_i\right\}_{i=1}^n$ and $\hat{\boldsymbol{\theta}}$ denotes either the known parameter $\boldsymbol{\theta}_0$ (simple hypothesis) or a consistent estimator (composite hypothesis; see \ref{assump:a6n} below). An equivalent expression for $T_n$, useful for computational implementation, is $T_n=\int_{\Omega_q}\big(\sum_{i=1}^nW_n^p(\mathbf{x},\mathbf{X}_i)(Y_i-m_{\hat{\boldsymbol{\theta}}}(\mathbf{X}_i))\big)^2 \allowbreak\hat f_h(\mathbf{x})w(\mathbf{x})\,\omega_q(d\mathbf{x})$. This smoothing of the (possibly estimated) parametric regression function is included to reduce the asymptotic bias \citep{Hardle1993}. Besides, in order to mitigate the effect of the difference between $\hat m_{h,p}$ and $m_{\hat{\boldsymbol{\theta}}}$ in sparse areas of the covariate, the squared difference is weighted by a kernel density estimate of $\mathbf{X}$, namely $\hat f_h$. In addition, by the inclusion of $\hat f_h$, the effects of the unknown density both on the asymptotic bias and variance are removed. Optionally, a weight function $w:\Omega_q\longrightarrow[0,\infty)$ can be considered, for example, to restrict the test to specific regions of $\Omega_{q}$ by an indicator function.\\

The limit distributions of $T_n$ are analyzed under a family of local alternatives that contains $H_0$ as a particular case and is asymptotically close to $H_0$:
\[
H_{1P}: m(\mathbf{x})=m_{\boldsymbol{\theta}_0}(\mathbf{x})+c_ng(\mathbf{x}),\text{ for all }\mathbf{x}\in\Omega_{q},
\]
where $m_{\boldsymbol{\theta}_0}\in\mathcal{M}_\Theta$, $g:\Omega_q\longrightarrow\mathbb{R}$ and $c_n$ is a positive sequence such that $c_n\to0$, for instance  $c_n=\big(nh^\frac{q}{2}\big)^{-\frac{1}{2}}$. With this framework, $H_{1P}$ becomes $H_0$ when $g$ is such that $m_{\boldsymbol{\theta}_0}+c_n^{-\frac{1}{2}}g\in\mathcal{M}_\Theta$ ($g\equiv0$, for example) and $H_1$ when the previous statement does not hold for a set of positive probability. The following regularity conditions on the parametric estimation are required: 
\begin{enumerate}[label=\textbf{A\arabic{*}}., ref=\textbf{A\arabic{*}}]
	\setcounter{enumi}{4}
	\item $m_{\boldsymbol{\theta}}$ is continuously differentiable as a function of $\boldsymbol{\theta}$, and this derivative is also continuous for $\mathbf{x}\in\Omega_q$. \label{assump:a5n}
	\item Under $H_0$, there exists an $\sqrt{n}$-consistent estimator $\hat{\boldsymbol{\theta}}$ of $\boldsymbol{\theta}_0$, \textit{i.e.} $\hat{\boldsymbol{\theta}}-\boldsymbol{\theta}_0=\mathcal{O}_\mathbb{P}\big(n^{-\frac{1}{2}}\big)$ and such that, under $H_1$, $\hat{\boldsymbol{\theta}}-\boldsymbol{\theta}_1=\mathcal{O}_\mathbb{P}\big(n^{-\frac{1}{2}}\big)$ for a certain $\boldsymbol{\theta}_1$. \label{assump:a6n}
	\item The function $g$ is continuous. \label{assump:a7n} 
	\item Under $H_{1P}$ , the $\sqrt{n}$-consistent estimator $\hat{\boldsymbol{\theta}}$ also satisfies $\hat{\boldsymbol{\theta}}-\boldsymbol{\theta}_0=\mathcal{O}_\mathbb{P}\big(n^{-\frac{1}{2}}\big)$. \label{assump:a8n}
\end{enumerate}

\begin{theo}[Limit distributions of $T_n$]
	\label{theo:limdis}
	Under $H_{1P}$, \ref{assump:a1}--\ref{assump:a6n} and  \ref{assump:a7n}--\ref{assump:a8n} if $g\not\equiv0$, 
	\begin{align*}
	&nh^\frac{q}{2}\left(T_n-\frac{\lambda_q(L^2)\lambda_q(L)^{-2}}{nh^{q}}\int_{\Omega_q}\sigma_{\boldsymbol{\theta}_0}^2(\mathbf{x})w(\mathbf{x})\,\omega_q(d\mathbf{x})\right)\\
	&\qquad\qquad\qquad\stackrel{d}{\longrightarrow} \left\{\begin{array}{ll}\infty, &c_n^2nh^{\frac{q}{2}}\to\infty,\\
	\mathcal{N}\left(\int_{\Omega_q}g(\mathbf{x})^2f(\mathbf{x})w(\mathbf{x})\,\omega_q(d\mathbf{x}),2\nu_{\boldsymbol{\theta}_0}^2\right),& c_n^2nh^{\frac{q}{2}}\to\delta,\;0<\delta<\infty,\\
	\mathcal{N}(0,2\nu_{\boldsymbol{\theta}_0}^2),&c_n^2nh^{\frac{q}{2}}\to0,\end{array}\right.
	\end{align*}
	where $\sigma_{\boldsymbol{\theta}_0}^2(\mathbf{x})=\mathbb{E}\left[(Y-m_{\boldsymbol{\theta}_0}(\mathbf{X}))^2|\mathbf{X}=\mathbf{x}\right]$ is the conditional variance under $H_0$ and
	\begin{align*}
	\nu_{\boldsymbol{\theta}_0}^2=&\,\int_{\Omega_q}\sigma_{\boldsymbol{\theta}_0}^4(\mathbf{x})w(\mathbf{x})^2\,\omega_q(d\mathbf{x})\times\gamma_q \lambda_q(L)^{-4} \int_0^{\infty} r^{\frac{q}{2}-1}\left\{\int_0^{\infty} \rho^{\frac{q}{2}-1} L(\rho) \varphi_q(r,\rho) \,d\rho\right\}^2\,dr,\\
	\varphi_q(r,\rho)=&\,\left\{ 
	\begin{array}{ll}
	L\left(r+\rho-2(r\rho)^{\frac{1}{2}}\right)+L\left(r+\rho+2(r\rho)^{\frac{1}{2}}\right), & q=1,\\
	\int_{-1}^1 \left( 1-\theta^2 \right)^{\frac{q-3}{2}} L\left(r+\rho-2\theta(r\rho)^{\frac{1}{2}}\right)\,d\theta, & q\geq2,\\
	\end{array}
	\right.\\
	\gamma_q=&\,\left\{ 
	\begin{array}{ll}
	2^{-\frac{1}{2}}, & q=1,\\
	\omega_{q-1}\omega_{q-2}^2 2^{\frac{3q}{2}-3}, & q\geq2.\\
	\end{array}
	\right.
	\end{align*}
\end{theo}

The convergence rate as well as the asymptotic bias and variance agree with the results in the Euclidean setting given by \cite{Hardle1993} and \cite{Alcala1999}, except for the cancellation of the design density in the bias and variance, achieved by the inclusion of $\hat f_h$ in the test statistic. The use of a local estimator with $p=0$ or $p=1$ does not affect the limiting distribution, given that the \textit{equivalent kernel} \citep{Fan1996} is the same (as seen in the SI). Finally, the general complex structure of the asymptotic bias and variance turns much simpler with the von Mises kernel:
\[
\nu^2=\int_{\Omega_q}\sigma^4(\mathbf{x})w(\mathbf{x})^2\,\omega_q(d\mathbf{x})\times (8\pi)^{-\frac{q}{2}},\quad \lambda_q(L^2)\lambda_q(L)^{-2}=\big(2\pi^\frac{1}{2}\big)^{-q}.
\]

\subsection{Bootstrap calibration}
\label{subsec:boot}

The distribution of $T_n$ under $H_0$ can be approximated by the one of its bootstrapped version $T_n^*$, which can be arbitrarily well approximated by Monte Carlo. Under $H_0$, the bootstrap responses are obtained from the parametric fit and bootstrap errors that imitate the conditional variance by a wild bootstrap procedure: $Y_i^*=m_{\hat{\boldsymbol{\theta}}}(\mathbf{X}_i)+\hat{\varepsilon}_iV_i^*$, where $\hat{\varepsilon}_i=Y_i-m_{\hat{\boldsymbol{\theta}}}(\mathbf{X}_i)$ and the variables $V_1^*,\ldots,V_n^*$ are independent from the observed sample and iid with $\mathbb{E}\left[V_i^*\right]=0$, $\mathbb{V}\mathrm{ar}\left[V_i^*\right]=1$ and finite third and fourth moments. A common choice is considering a binary variable with $\mathbb{P}\big\{V_i^*=(1-\sqrt{5})/2\big\}=(5+\sqrt{5})/10$ and $\mathbb{P}\big\{V_i^*=(1+\sqrt{5})/2\big\}=(5-\sqrt{5})/10$, which corresponds to the \textit{golden section} bootstrap. The test in practice for the composite hypothesis is summarized in the next algorithm (if the simple is considered, set $\boldsymbol{\theta}_0=\hat{\boldsymbol{\theta}}=\hat{\boldsymbol{\theta}}^*$).
\begin{algo}[Test in practice]
	\label{algo:boot}
	Let $\left\{\left(\mathbf{X}_i,Y_i\right)\right\}_{i=1}^n$ be a sample from (\ref{model}). To test $H_0$, set a bandwidth $h$ and (optionally) a weight function $w$ and proceed as follows:
	\begin{enumerate}[label=\textit{\roman{*}}., ref=\textit{\roman{*}}]
		\item Compute $\hat{\boldsymbol{\theta}}$, $\hat\varepsilon_i=Y_i-m_{\hat{\boldsymbol{\theta}}}(\mathbf{X}_i)$ and $T_n=\int_{\Omega_q}\big(\sum_{i=1}^nW_n^p(\mathbf{x},\mathbf{X}_i)\hat\varepsilon_i\big)^2 \hat f_h(\mathbf{x})w(\mathbf{x})\,\omega_q(d\mathbf{x})$.\label{gofreg:algo:boot:2}
		\item \textit{Bootstrap resampling}. For $b=1,\ldots,B$:\label{gofreg:algo:boot:3}
		\begin{enumerate}[label=\textit{(\alph{*})},ref=\textit{(\alph{*})}]
			\item Obtain $\left\{\left(\mathbf{X}_i,Y_i^*\right)\right\}_{i=1}^n$, where $Y_i^*=m_{\hat{\boldsymbol{\theta}}}(\mathbf{X}_i)+\hat\varepsilon_iV_i^*$ and
			compute $\hat{\boldsymbol{\theta}}^*$ as in \ref{gofreg:algo:boot:2}.\label{gofreg:algo:boot:b}
			\item Compute $T_n^{*b}=\int_{\Omega_q}\big(\sum_{i=1}^nW_n^p(\mathbf{x},\mathbf{X}_i)\hat\varepsilon^*_i\big)^2 \hat f_h(\mathbf{x})w(\mathbf{x})\,\omega_q(d\mathbf{x})$ with $\hat{\varepsilon_i}^*=Y_i^*-m_{\hat{\boldsymbol{\theta}}^*}(\mathbf{X}_i)$.\label{gofreg:algo:boot:c}
		\end{enumerate}
		\item Approximate the $p$-value by $\frac{1}{B}\sum_{b=1}^B\mathbbm{1}_{\left\{T_n\leq T_n^{*b}\right\}}$.\label{gofreg:algo:boot:4}
	\end{enumerate}
\end{algo}

In order to prove the consistency of the resampling mechanism, that is, that $T_n^*$ has the same asymptotic distribution as $T_n$, a bootstrap analogue of assumption \ref{assump:a6n} is required:
\begin{enumerate}[label=\textbf{A\arabic{*}}., ref=\textbf{A\arabic{*}}]
	\setcounter{enumi}{8}
	\item The estimator $\hat{\boldsymbol{\theta}}^*$ computed from $\left\{\left(\mathbf{X}_i,Y_i^*\right)\right\}_{i=1}^n$ is such that $\hat{\boldsymbol{\theta}}^*-\hat{\boldsymbol{\theta}}=\mathcal{O}_{\mathbb{P}^*}\big(n^{-\frac{1}{2}}\big)$, where $\mathbb{P}^*$ is the probability law conditional on $\left\{\left(\mathbf{X}_i,Y_i\right)\right\}_{i=1}^n$. \label{assump:a9n}
\end{enumerate}
From this assumption and Theorem \ref{theo:limdis} it follows that the probability distribution function (pdf) of $T_n^*$, conditionally on the sample, converges always in probability to a Gaussian pdf, which is the asymptotic pdf of $T_n$ if $H_0$ holds.
\begin{theo}[Bootstrap consistency]
	\label{theo:boot}
	Under \ref{assump:a1}--\ref{assump:a6n} and \ref{assump:a9n} and conditionally on $\left\{\left(\mathbf{X}_i,Y_i\right)\right\}_{i=1}^n$,
	\[
	nh^\frac{q}{2}\left(T_n^*-\frac{\lambda_q(L^2)\lambda_q(L)^{-2}}{nh^{q}}\int_{\Omega_q}\sigma_{\boldsymbol{\theta}_1}^2(\mathbf{x})w(\mathbf{x})\,\omega_q(d\mathbf{x})\right)\stackrel{d}{\longrightarrow} \mathcal{N}\left(0,2\nu_{\boldsymbol{\theta}_1}^2\right)
	\]
	in probability. If the null hypothesis holds, then $\boldsymbol{\theta}_1=\boldsymbol{\theta}_0$.
\end{theo}

\section{Simulation study}
\label{sec:simu}

The finite sample performance of the goodness-of-fit test is explored in four simulation scenarios, labeled S1 to S4. Their associated parametric regression models are shown in Figure \ref{fig:models} with the following codification: the radius from the origin represents the response $m(\mathbf{x})$ for an $\mathbf{x}$ direction, resulting in a distortion from a perfect circle or sphere. The design densities of the scenarios are taken from \cite{Garcia-Portugues:exact}, the noise is either heteroskedastic (S1 and S2) or homocedastic (S3 and S4) and two different deviations (for S1--S2 and for S3--S4) are considered. The tests based on the projected local constant and linear estimators are compared with  $M=1000$ Monte Carlo trials and $B=1000$ bootstrap replicates, under $H_0$ and $H_1$, for a grid of bandwidths and with $n=100$ and $q=1,2,3$. Parametric estimation is done by nonlinear least squares, which is justified by their simplicity and asymptotic normality \citep{Jennrich1969}, hence satisfying \ref{assump:a6n}. For the sake of brevity, only a coarse grained description of the scenarios and a selected output of the study is provided here. The reader is referred to the SI for the complete report.

\begin{figure}[H]
	\centering
	\includegraphics[width=0.225\textwidth]{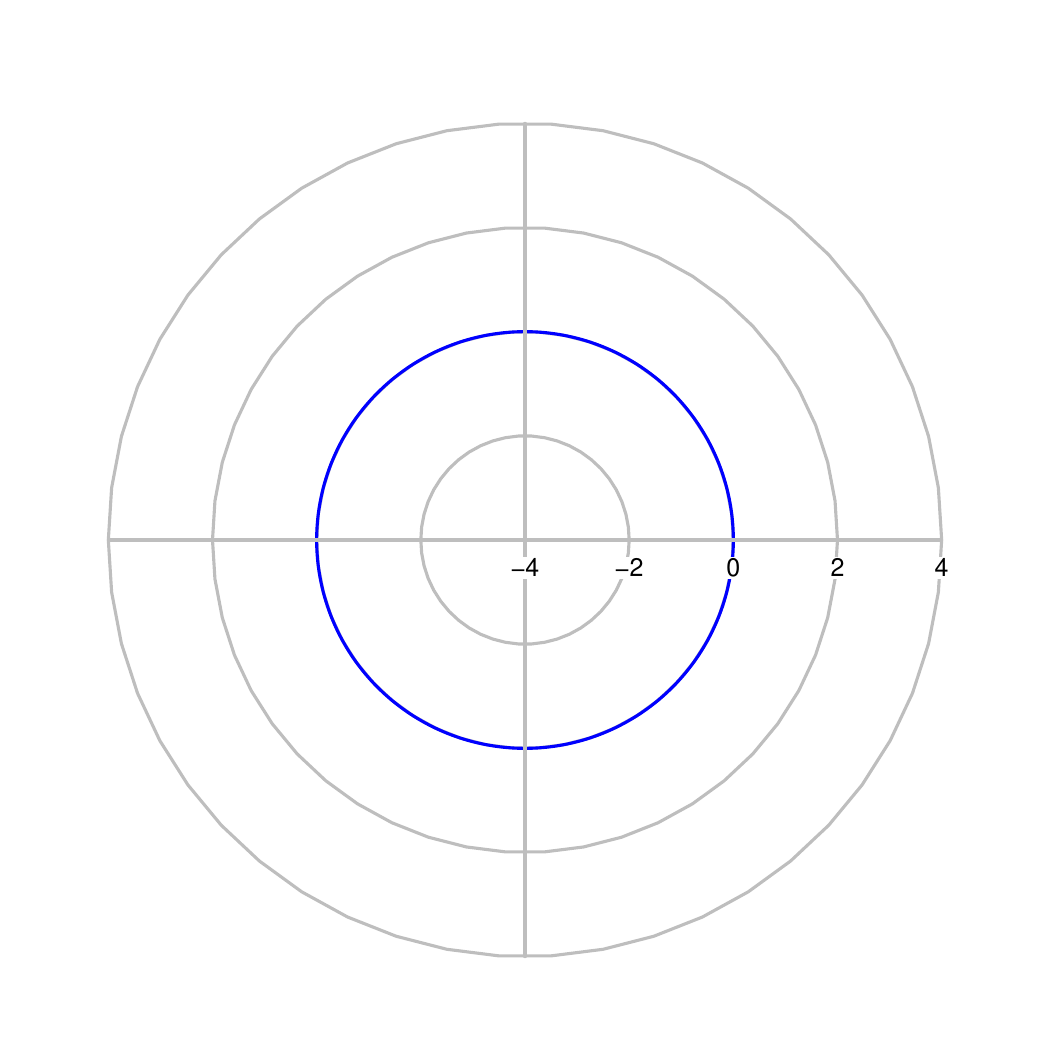}
	\includegraphics[width=0.225\textwidth]{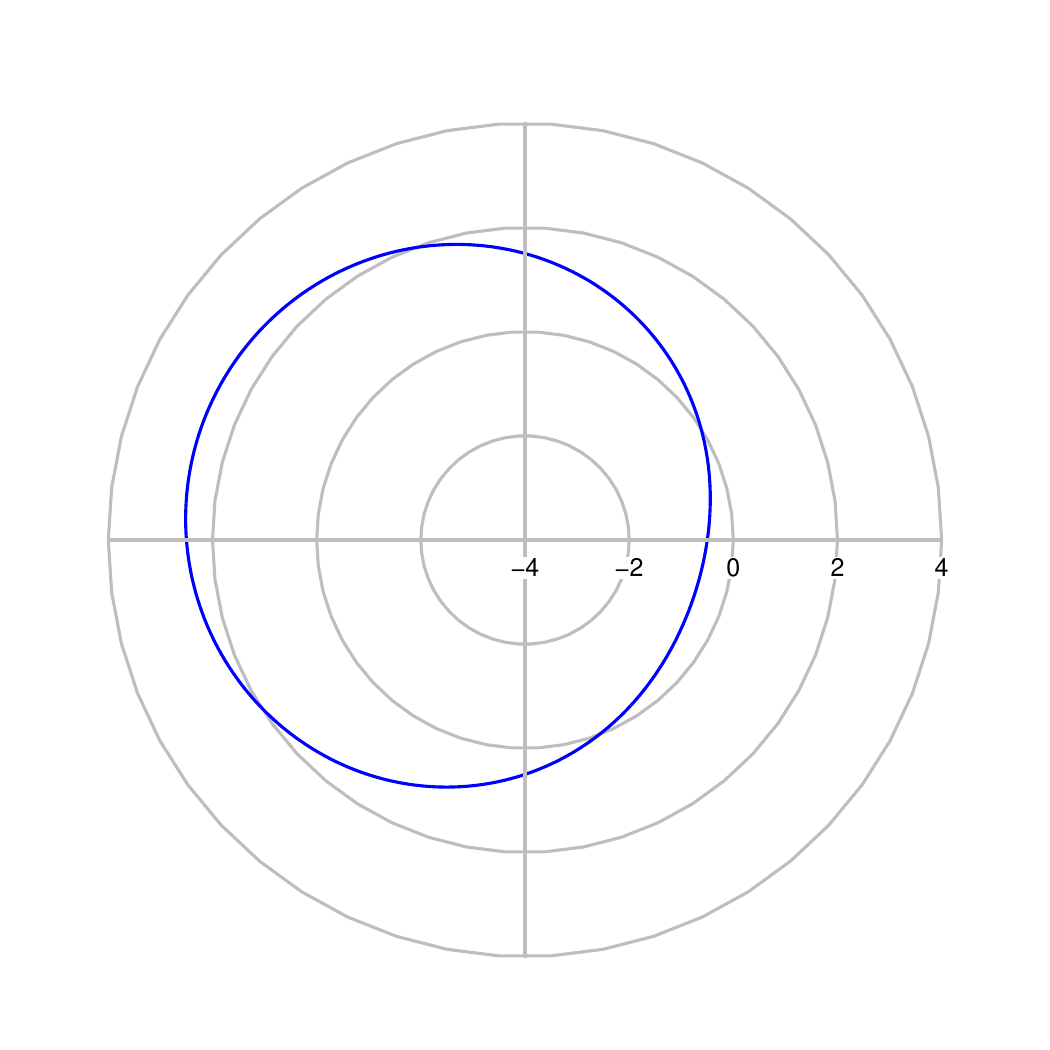}
	\includegraphics[width=0.225\textwidth]{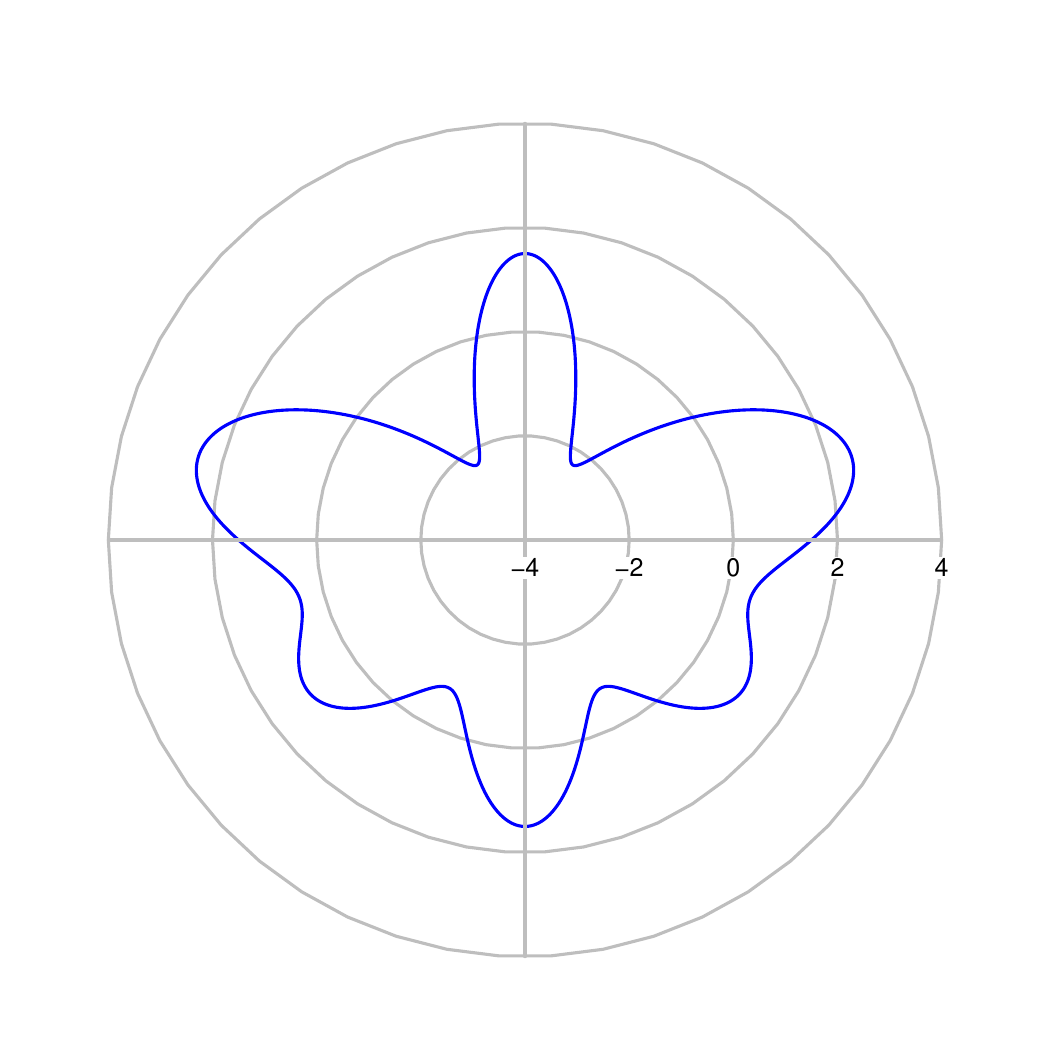}
	\includegraphics[width=0.225\textwidth]{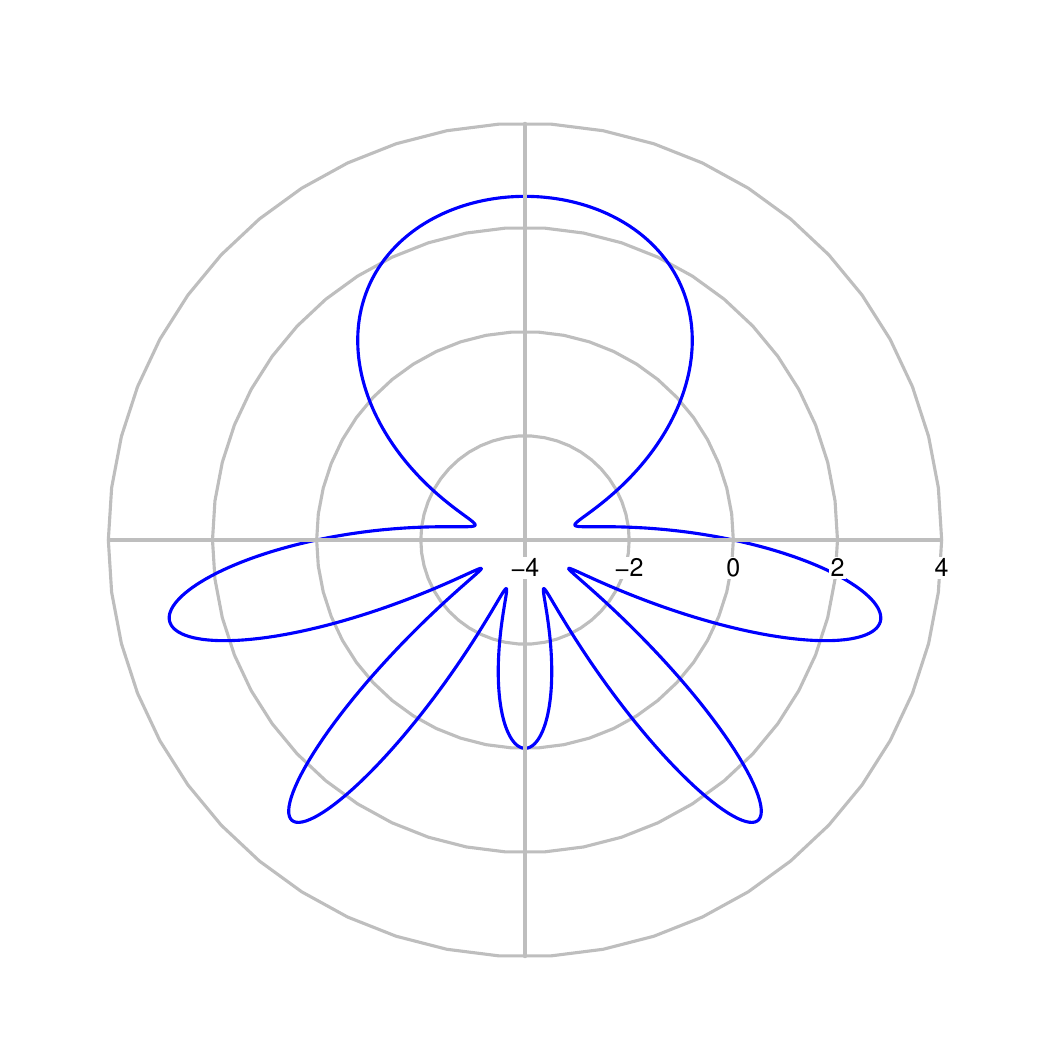}\\
	\includegraphics[width=0.225\textwidth]{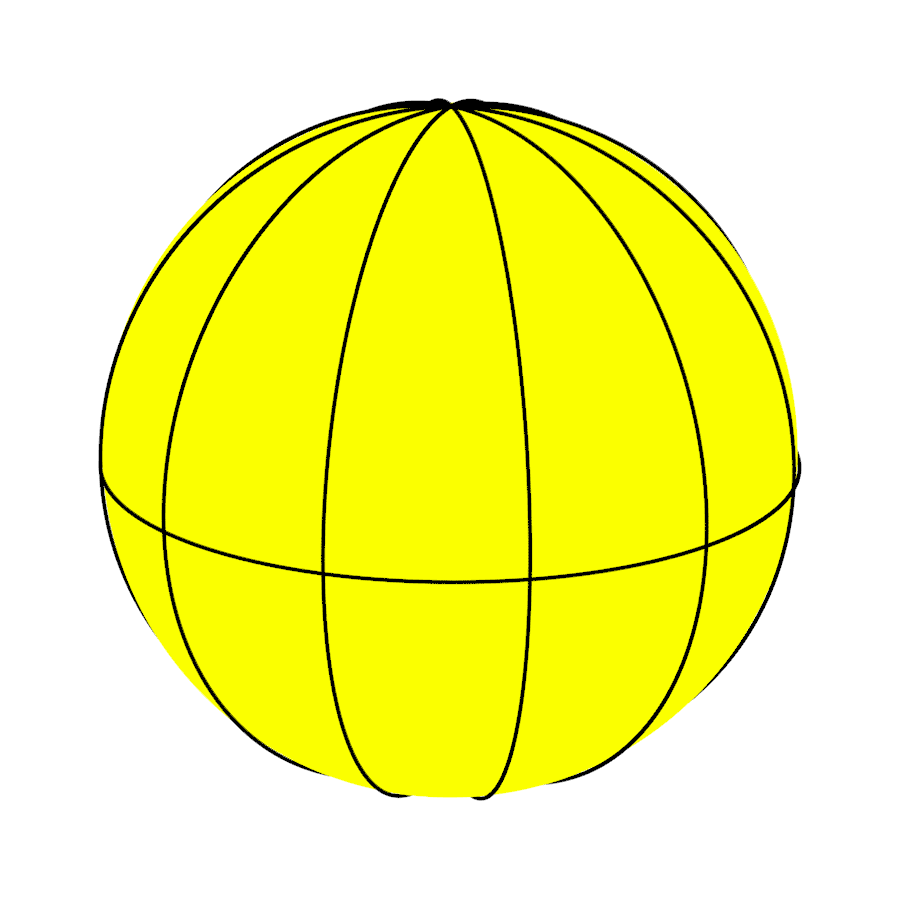}
	\includegraphics[width=0.225\textwidth]{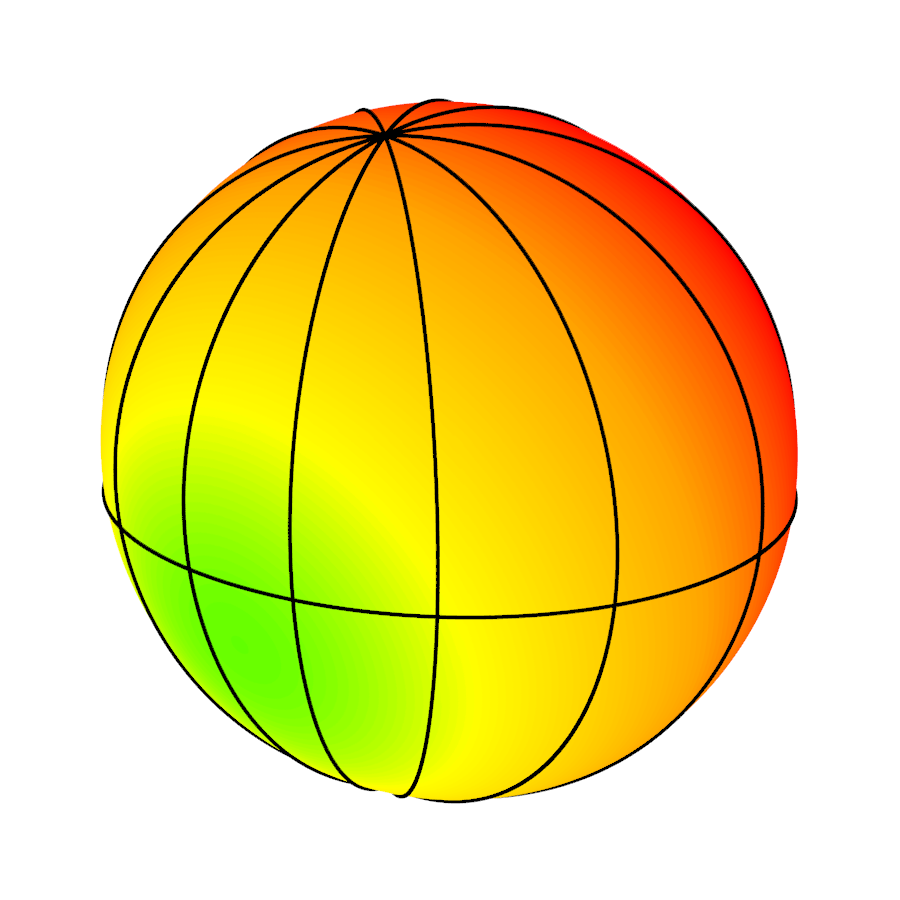}
	\includegraphics[width=0.225\textwidth]{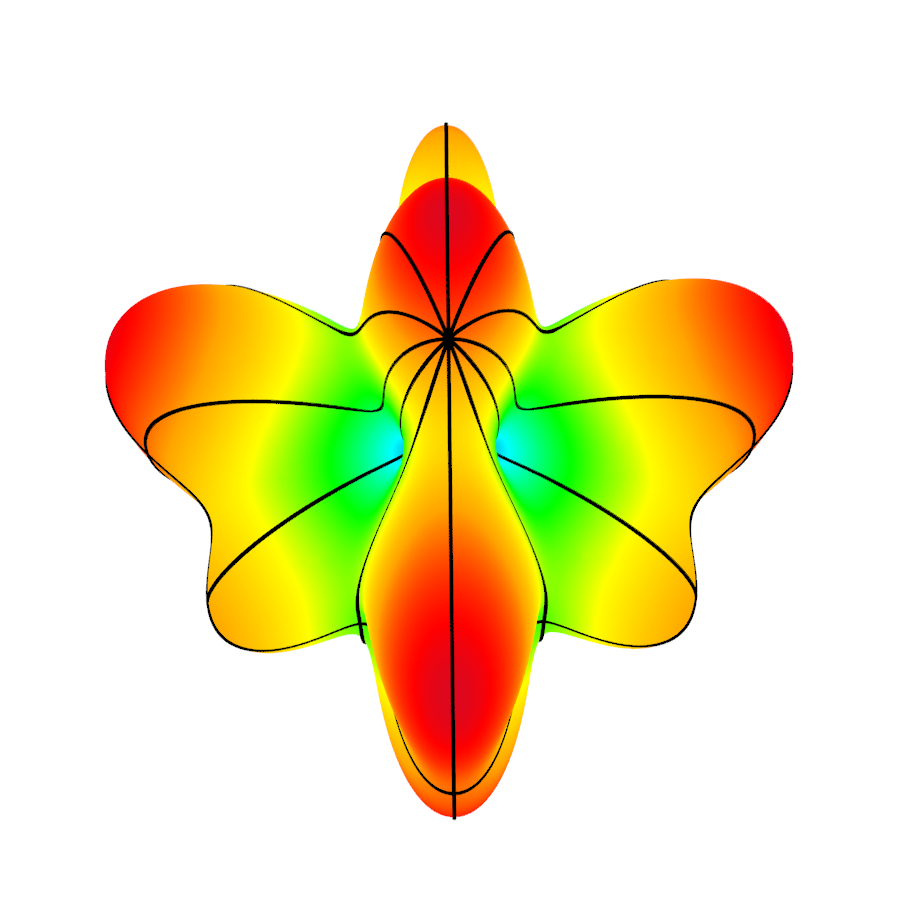}
	\includegraphics[width=0.225\textwidth]{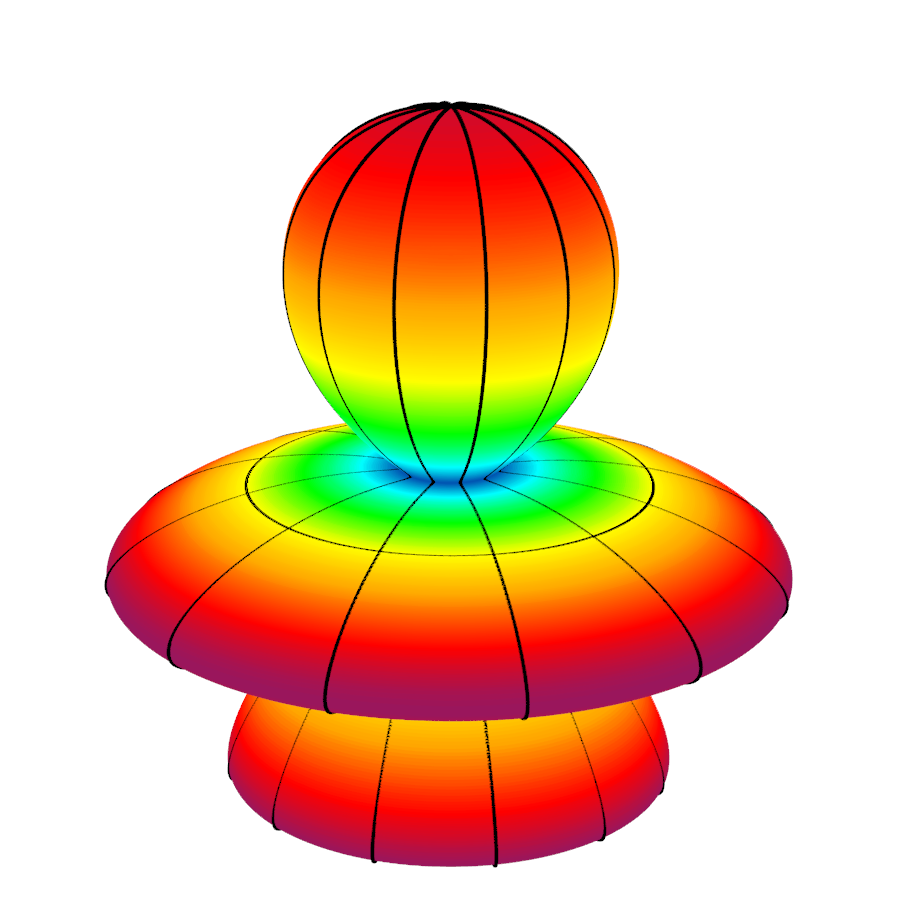}
	\caption{\small From left to right: parametric regression models for scenarios S1 to S4, for circular and spherical cases. Color shading represents the distance from the origin of the regression surface. \label{fig:models}}
\end{figure}

The empirical sizes of the goodness-of-fit tests are shown using the so called \textit{significance trace} \citep{Bowman1997}, \textit{i.e.}, the curve of percentages of empirical rejections for different bandwidths. As shown in Figure \ref{fig:pow:1}, except for very small bandwidths that result in a conservative test, the significance level is stabilized around the $95\%$ confidence band for the nominal level $\alpha=0.05$, for the different scenarios and dimensions. The power is satisfactory, given that the proposed tests succeed in detecting the mild deviations from the null hypotheses. Despite the fact that the test based on the local linear estimator ($p=1$) provides a better power for large bandwidths in certain scenarios, the overall impression is that the test with $p=0$ is hard to beat: the powers with $p=0$ and $p=1$ are almost the same for low dimensions, whereas as the dimension increases the local constant estimator performs better for a wider range of bandwidths. This effect could be explained by the spikes that local linear regression tends to show in the boundaries of the support (design densities of S3 and S4), which become more important as the dimension increases. The lower power for S1 and S4 is due to deviations happening in areas with low density or high variance.

\begin{figure}[H]
	\centering
	\includegraphics[width=0.32\textwidth]{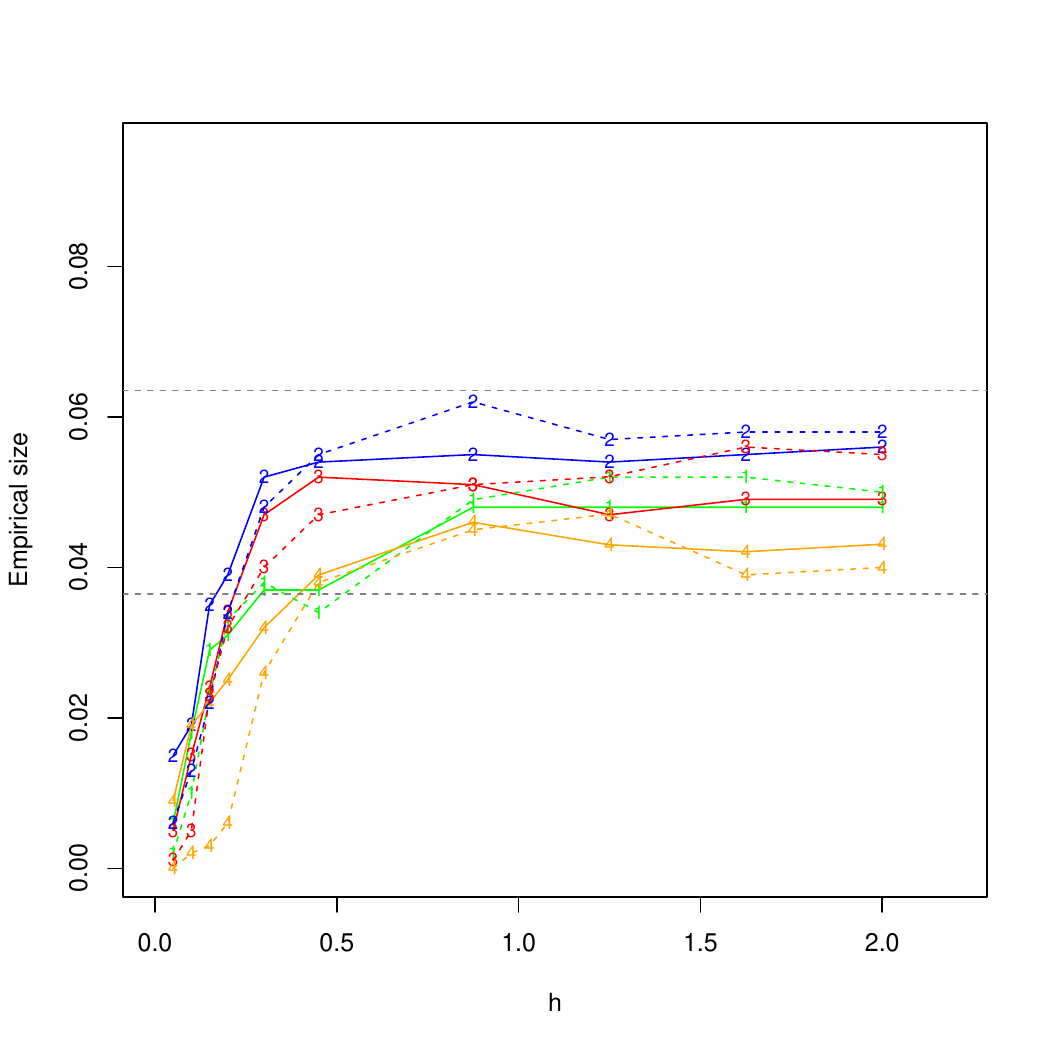}
	\includegraphics[width=0.32\textwidth]{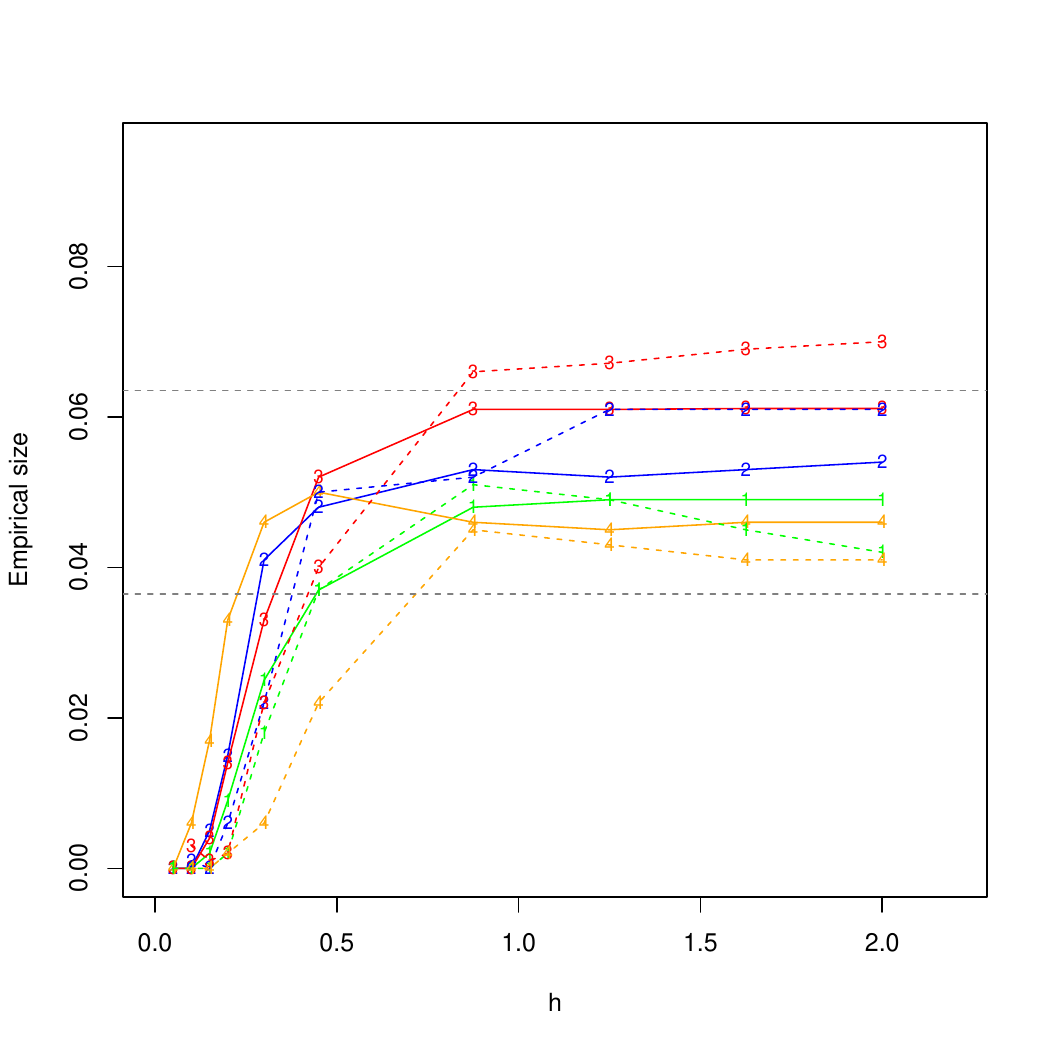}
	\includegraphics[width=0.32\textwidth]{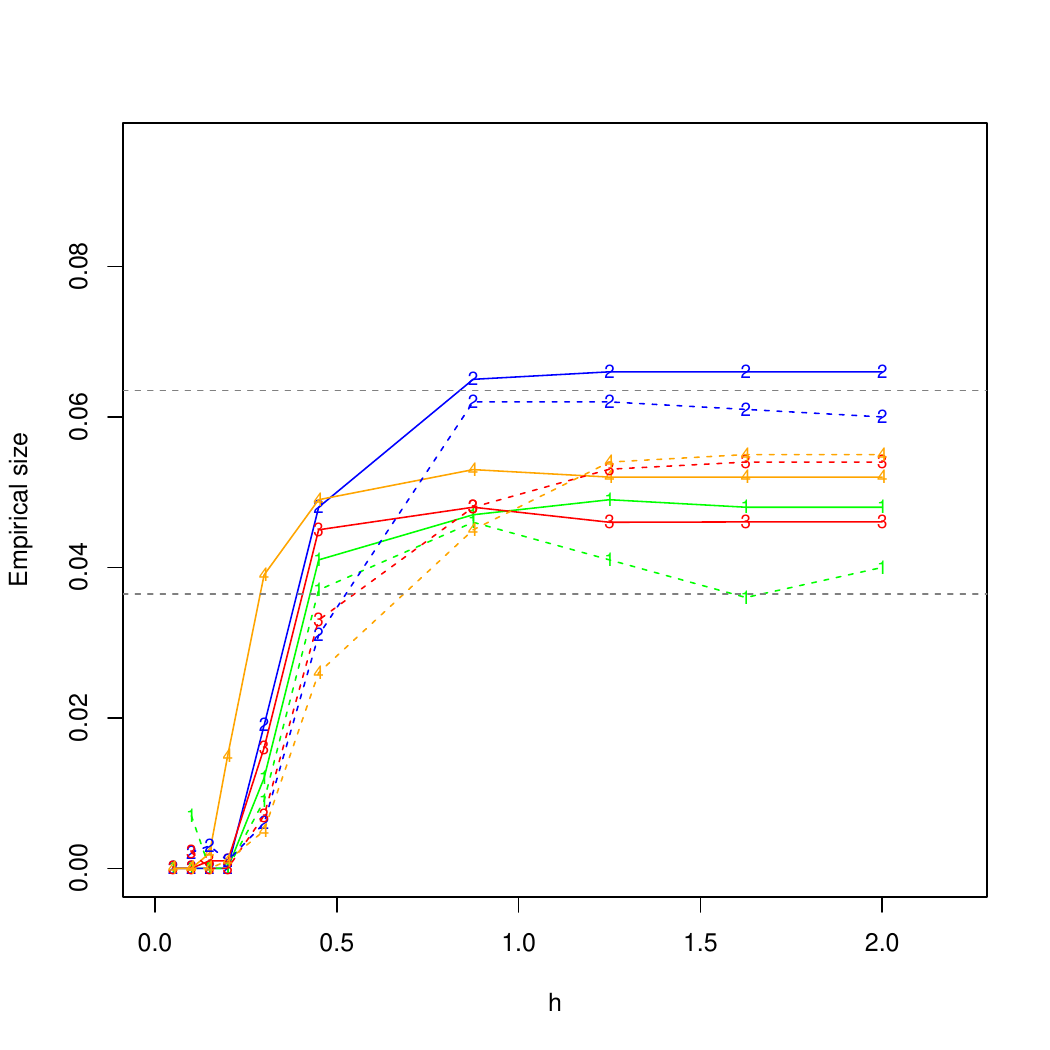}\\
	\includegraphics[width=0.32\textwidth]{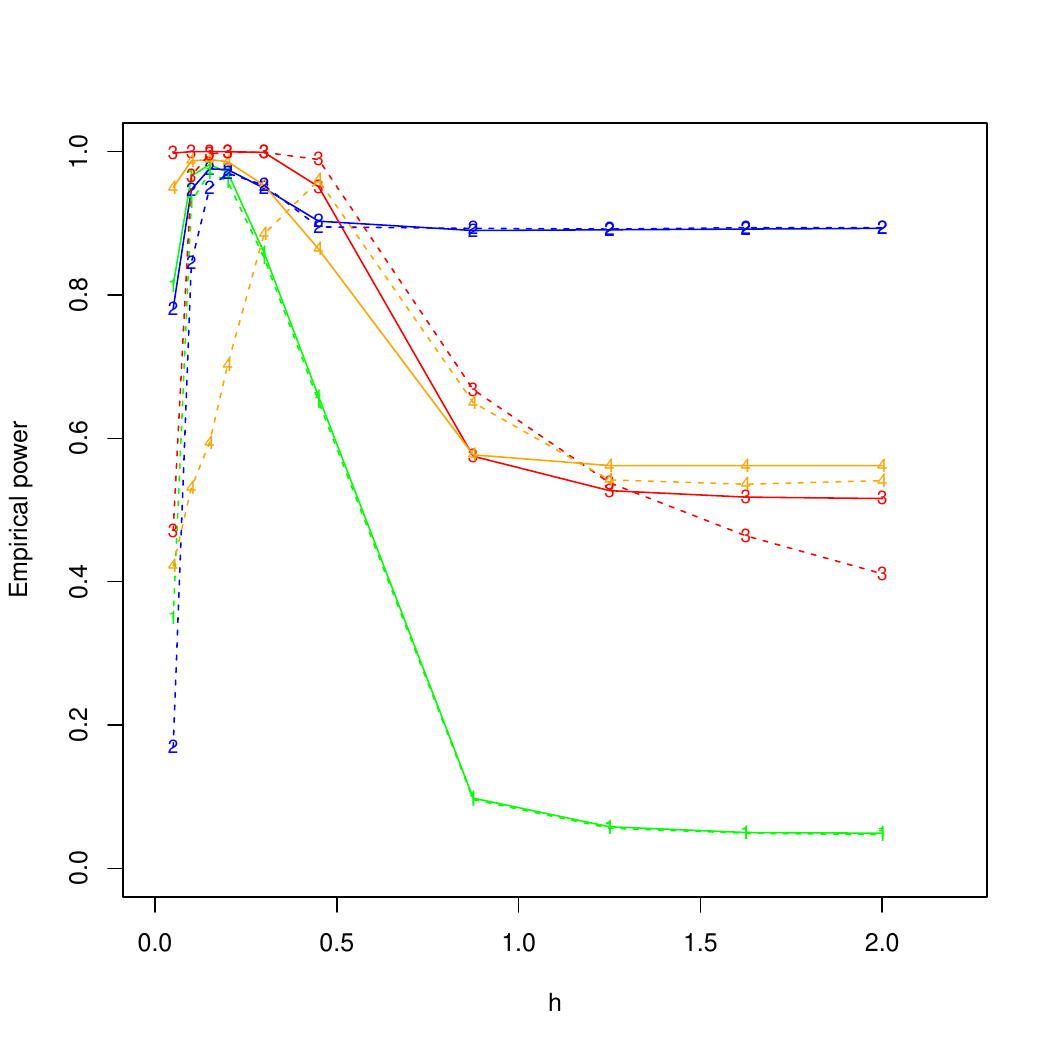}
	\includegraphics[width=0.32\textwidth]{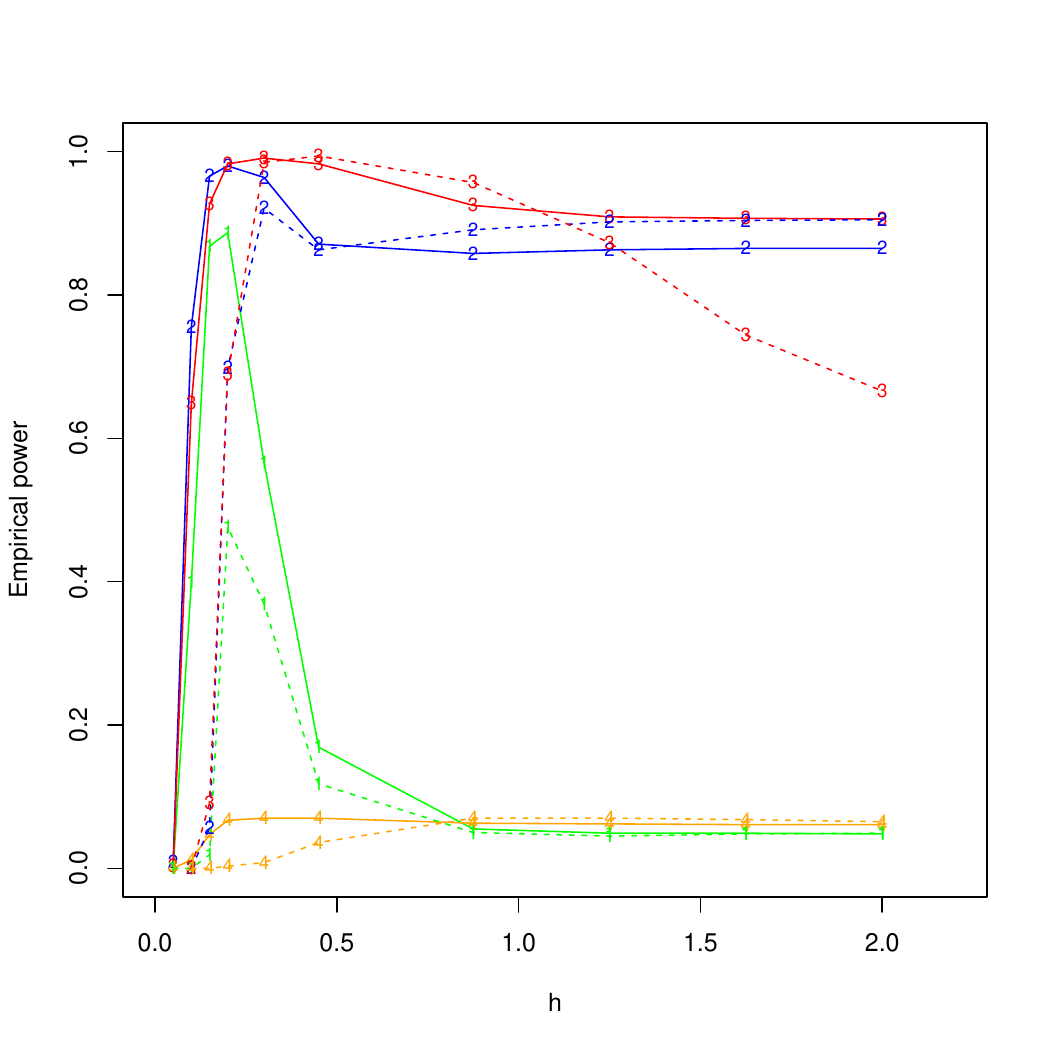}
	\includegraphics[width=0.32\textwidth]{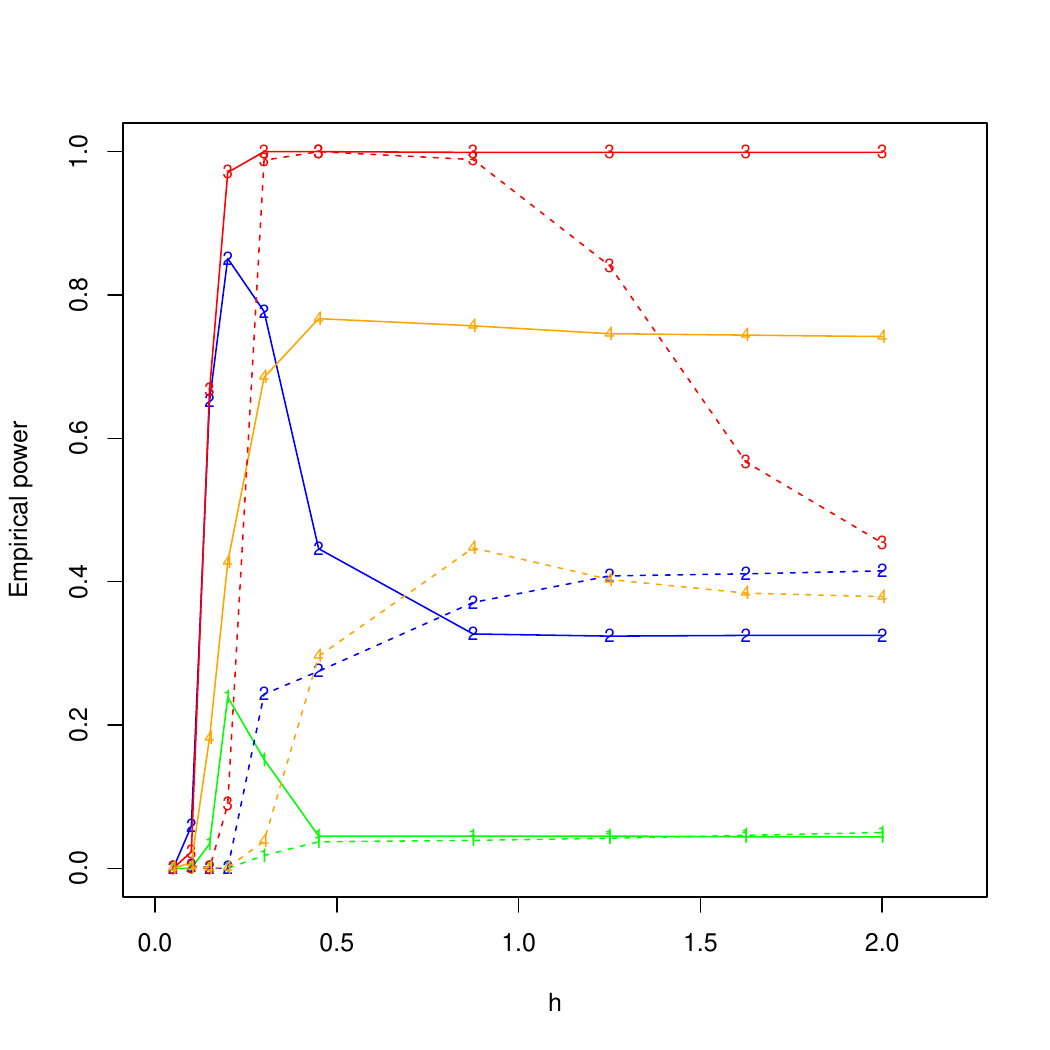}
	\caption{\small Empirical sizes (first row) and powers (second row) for significance level $\alpha=0.05$ for the different scenarios, with $p=0$ (solid line) and $p=1$ (dashed line). From left to right: columns represent dimensions $q=1,2,3$ with sample size $n=100$. Green, blue, red and orange colors correspond to scenarios S1 to S4, respectively. \label{fig:pow:1}}
\end{figure}

\section{Application to text mining}
\label{sec:data}

In different applications within text mining, it is quite common to consider a \textit{corpus} (collection of documents) and to determine the so-called vector space model: a corpus $\mathbf{d}_1,\ldots,\mathbf{d}_n$ is codified by the set of vectors $\left\{(d_{i1},\ldots,d_{iD})\right\}_{i=1}^n$ (the \textit{document-term matrix}) with respect to a dictionary (or a \textit{bag of words}) $\left\{w_1,\ldots,w_D\right\}$, such that $d_{ij}$ represents the frequency of the dictionary's $j$-th word in the document $\mathbf{d}_i$. Usually, a normalization of the document-term matrix is performed to remove length distortions and map documents with similar contents, albeit different lengths, into close vectors. If the Euclidean norm is used for this, then the documents can then be regarded as points in $\Omega_{D-1}$ providing a set of directional data.\\

\pagebreak

The corpus that is analyzed in this application was acquired from the news aggregator \textit{Slashdot} (\url{www.slashdot.org}). This website publishes summaries of news about technology and science that are submitted and evaluated by users. Each news entry includes a title, a summary with links to other related news and a discussion thread gathering users comments. The goal is to test a linear model that takes as a predictor the topic of the news (a directional variable in $\Omega_{D-1}$) and as a response the log-number of comments. This is motivated by the frequent use of simple linear models in this context (see \cite{Tatar2012} for example) and that, in text classifications, it has been checked that non-linear classifiers hardly provide any advantage with respect to linear ones \citep{Joachims2002}. After a data preprocessing process (using \cite{Meyer2008}; see SI), the $n=8121$ news collected from 2013 were represented in a document term matrix formed by $D=1508$ words. \\

In order to construct a plausible linear model, a preliminary variable selection was performed using LASSO (Least Absolute Shrinkage and Selection Operator) regression with (tuning) parameter $\lambda$ selected by an overpenalized \textit{three} standard error rule \citep{Hastie2009}. After removing some extra variables by using a backward stepwise method with BIC, a fitted vector $\hat{\boldsymbol{\eta}}\in\mathbb{R}^{D}$ with $d=77$ non-zero entries is obtained. The test is applied to check the null hypothesis of a candidate linear model with coefficient $\boldsymbol\eta$ constrained to be zero except in these previously selected $d$ words, that is $H_0: m(\mathbf{x})=c+\boldsymbol\eta^T\mathbf{x}$, with $\boldsymbol\eta$ subject to $\mathbf{A}\boldsymbol\eta=\mathbf{0}$ for an adequate choice of the matrix $\mathbf{A}_{(D-d)\times D}$. The significance trace of the test (with $p=0$; $p=1$ was not implemented due to its higher cost and to computational limitations) presents a minimum $p$-value of $0.12$, hence showing no evidence to reject the linear model for a wide grid of bandwidths. Figure \ref{fig:wordcloud} displays a graphical summary of the fitted linear model. As it can be seen, stemmed words like ``kill'', ``climat'' and ``polit'' have a strong positive impact on the number of comments, since they are prone to appear in controversial news that usually generate broad discussions. On the other hand, scientific related words like ``mission'', ``abstract'' or ``lab'' have a negative impact, as they tend to raise more objective and higher specific discussions. Experiments were conducted with a model of $d=50$ non-zero coefficients chosen with a higher overpenalization, showing a strong rejection of the null hypothesis.

\begin{figure}[H]
	\centering
	\includegraphics[scale=0.5]{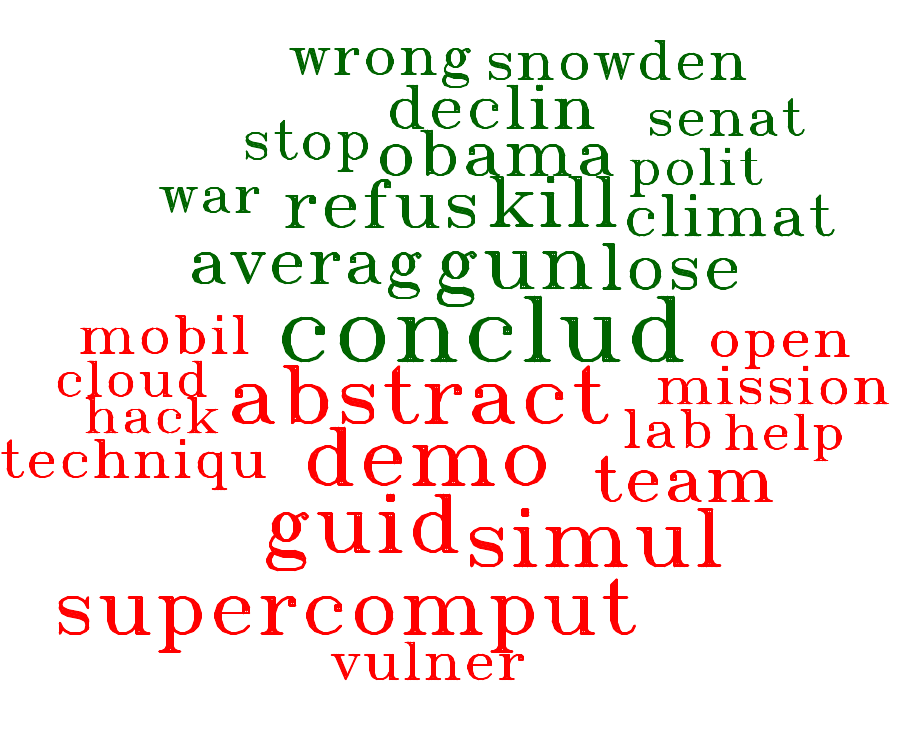}
	\caption{\small Stems of the $30$ largest coefficients (in absolute value) of the fitted constrained linear model. Green and red colors account for positive and negative impacts on news popularity, respectively, whereas the size of the stem is proportional to the magnitude of its coefficient. The linear model has an $R^2=0.25$ and the significances of each coefficient are lower than $0.002$.\label{fig:wordcloud}}
\end{figure}

\section*{Supporting information}

Additional information for this article is available online. The additional information is comprised of four extra appendices (``Technical lemmas'', ``Empirical evidence of the asymptotic distribution'', ``Further information on the simulation study'', ``Further information on the text mining application''), three tables and nine figures.

\section*{Acknowledgements}

We thank professors David E. Losada for his guidance in the data application and Ir\`{e}ne Gijbels for her useful theoretical comments. This research was supported by Project MTM2008-03010, Spanish Ministry of Science and Innovation; Project 10MDS207015PR, Direcci\'on Xeral de I+D of the Xunta de Galicia; the IAP research network grant nr. P7/06, Belgian government (Belgian Science Policy); the European Research Council under the European Community's Seventh Framework Programme (FP7/2007-2013) / ERC Grant agreement No. 203650; contract ``Projet d'Actions de Recherche Concert\'ees'' (ARC) 11/16-039 of the ``Communaut\'e fran\c{c}aise de Belgique'' (granted by the ``Acad\'emie universitaire Louvain''); the Dynamical Systems Interdisciplinary Network, University of Copenhagen. Work of the first author has been supported by a grant from Fundaci\'on Barri\'e and FPU grant AP2010--0957 from the Spanish Ministry of Education. The Authors gratefully acknowledge the computational resources used at the CESGA Supercomputing Center and valuable suggestions by three anonymous referees. 

\appendix

\section{Proofs of the main results}
\label{appendix:main}

\begin{proof}[Proof of Theorem \ref{theo:limdis}]
	The proof follows the steps of \cite{Hardle1993} and \cite{Alcala1999}. $T_n$ can be separated into three addends by adding and subtracting the true smoothed regression function $T_n=(T_{n,1}+T_{n,2}-2T_{n,3})(1+\mathpzc{o}_\mathbb{P}\left(1\right))$, where
	\begin{align*}
	T_{n,1}=&\,\int_{\Omega_q}\bigg(\sum_{i=1}^n W_n^p\left(\mathbf{x},\mathbf{X}_i\right)\left(Y_i-m_{\boldsymbol{\theta}_0}(\mathbf{X}_i)\right) \bigg)^2f(\mathbf{x})w(\mathbf{x})\,\omega_q(d\mathbf{x}),\\
	T_{n,2}=&\,\int_{\Omega_q}\left(\mathcal{L}_{h,p}\left(m_{\boldsymbol{\theta}_0}-m_{\hat{\boldsymbol{\theta}}}\right)(\mathbf{x})\right)^2f(\mathbf{x})w(\mathbf{x})\,\omega_q(d\mathbf{x}),\\
	T_{n,3}=&\,\int_{\Omega_q}\left(\hat m_{h,p}(\mathbf{x})-\mathcal{L}_{h,p}m_{\boldsymbol{\theta}_0}(\mathbf{x})\right)\mathcal{L}_{h,p}\left(m_{\boldsymbol{\theta}_0}-m_{\hat{\boldsymbol{\theta}}}\right)(\mathbf{x})f(\mathbf{x})w(\mathbf{x})\,\omega_q(d\mathbf{x}),
	\end{align*}
	because of 
	\ifsupplement
	\ref{lem:1:d0} of Lemma \ref{lem:1}. 
	\else
	\textit{i} from Lemma S4. 
	\fi
	The proof is divided into the analysis of each addend.\\
	
	\textit{Terms $T_{n,2}$ and $T_{n,3}$.} By a Taylor expansion on $m_{\boldsymbol{\theta}}(\mathbf{x})$ as a function of $\boldsymbol{\theta}$ (see \ref{assump:a5n}),
	\begin{align*}
	T_{n,2}=&\,\int_{\Omega_q}\left(\big(\hat{\boldsymbol{\theta}}-\boldsymbol{\theta}_0\big)^T\mathcal{L}_{h,p}\left(\mathcal{O}_\mathbb{P}\left(\mathbf{1}\right)\right)(\mathbf{x})\right)^2f(\mathbf{x})w(\mathbf{x})\,\omega_q(d\mathbf{x})=\mathcal{O}_\mathbb{P}\left(n^{-1}\right),
	\end{align*}
	because of the boundedness of $\frac{\partial m_{\boldsymbol{\theta}}(\mathbf{x})}{\partial \boldsymbol{\theta}}$ for $\mathbf{x}\in\Omega_q$, \ref{assump:a6n} and \ref{assump:a8n}. On the other hand,
	\begin{align*}
	T_{n,3}=&\,\mathcal{O}_\mathbb{P}\big(n^{-\frac{1}{2}}\big)\int_{\Omega_q}\left(\hat m_{h,p}(\mathbf{x})-\mathcal{L}_{h,p}m_{\boldsymbol{\theta}_0}(\mathbf{x})\right)f(\mathbf{x})w(\mathbf{x})\,\omega_q(d\mathbf{x})=\mathcal{O}_\mathbb{P}\big(n^{-1}\big),
	\end{align*}
	because of the previous considerations and 
	\ifsupplement
	\ref{lem:5:4} from Lemma \ref{lem:5}. 
	\else
	\textit{i} from Lemma S6. 
	\fi
	As a consequence, by \ref{assump:a3} it happens that $
	nh^\frac{q}{2}T_{n,3}\stackrel{p}{\longrightarrow}0$ and $nh^\frac{q}{2}T_{n,2}\stackrel{p}{\longrightarrow}0$.\\
	
	\textit{Term $T_{n,1}$.} $T_{n,1}$ is dealt with $\tilde{L}_h\left(\mathbf{x},\mathbf{X}_i\right)=\frac{1}{nh^q\lambda_q(L)f(\mathbf{x})}L\Big(\frac{1-\mathbf{x}^T\mathbf{X}_i}{h^2}\Big)$ from Lemma 
	\ifsupplement
	\ref{coro:equiv}:
	\else
	S5:
	\fi
	\begin{align*}
	T_{n,1}=\int_{\Omega_q}\bigg(\sum_{i=1}^n \tilde{L}_h\left(\mathbf{x},\mathbf{X}_i\right)\left(1+\mathpzc{o}_\mathbb{P}\left(1\right)\right)\left(Y_i-m_{\boldsymbol{\theta}_0}(\mathbf{X}_i)\right) \bigg)^2f(\mathbf{x})w(\mathbf{x})\,\omega_q(d\mathbf{x})
	=\widetilde{T}_{n,1}\left(1+\mathpzc{o}_\mathbb{P}\left(1\right)\right).
	\end{align*}
	Now it is possible to split
	$
	\widetilde{T}_{n,1}=\widetilde{T}^{(1)}_{n,1}+\widetilde{T}^{(2)}_{n,1}+2\widetilde{T}^{(3)}_{n,1}
	$
	by recalling that $Y_i-m_{\boldsymbol{\theta}_0}(\mathbf{X}_i)=\sigma(\mathbf{X}_i)\varepsilon_i+c_ng(\mathbf{X}_i)$ by (\ref{model}) and $H_{1P}$. Specifically, under $H_{1P}$ the conditional variance can be expressed as $\sigma^2(\mathbf{x})=\mathbb{E}\big[(Y-m_{\boldsymbol{\theta}_0}(\mathbf{X})-c_ng(\mathbf{X})\big)^2|\mathbf{X}=\mathbf{x}\big]=\sigma_{\boldsymbol{\theta}_0}^2(\mathbf{x})(1+\mathpzc{o}\left(1\right))$, uniformly in $\mathbf{x}\in\Omega_{q}$ since $g$ and $\sigma_{\boldsymbol{\theta}_0}$ are continuous and bounded by \ref{assump:a2} and \ref{assump:a7n}. Therefore:
	\begin{align*}
	\widetilde{T}_{n,1}^{(1)}=&\,\int_{\Omega_q}\bigg(\sum_{i=1}^n \tilde{L}_h\left(\mathbf{x},\mathbf{X}_i\right)\sigma(\mathbf{X}_i)\varepsilon_i \bigg)^2f(\mathbf{x})w(\mathbf{x})\,\omega_q(d\mathbf{x}),\\
	\widetilde{T}_{n,1}^{(2)}=&\,c_n^2\int_{\Omega_q}\bigg(\sum_{i=1}^n \tilde{L}_h\left(\mathbf{x},\mathbf{X}_i\right)g(\mathbf{X}_i) \bigg)^2f(\mathbf{x})w(\mathbf{x})\,\omega_q(d\mathbf{x}),\\
	\widetilde{T}_{n,1}^{(3)}=&\,c_n\int_{\Omega_q}\sum_{i=1}^n\sum_{j=1}^n \tilde{L}_h\left(\mathbf{x},\mathbf{X}_i\right)\tilde{L}_h\left(\mathbf{x},\mathbf{X}_j\right)\sigma(\mathbf{X}_i)\varepsilon_ig(\mathbf{X}_j) f(\mathbf{x})w(\mathbf{x})\,\omega_q(d\mathbf{x}).
	\end{align*}
	By results 
	\ifsupplement
	\ref{lem:5:1} and \ref{lem:5:2} of Lemma \ref{lem:5}, 
	\else
	\textit{ii} and \textit{iii} of Lemma S6, 
	\fi
	the behavior of the two last terms is
	\begin{align}
	nh^\frac{q}{2}\widetilde{T}_{n,1}^{(2)}=nh^\frac{q}{2}c_n^2\int_{\Omega_q}g(\mathbf{x})^2 f(\mathbf{x})w(\mathbf{x})\,\omega_q(d\mathbf{x})(1+\mathpzc{o}_\mathbb{P}\left(1\right))\text{ and }nh^\frac{q}{2}\widetilde{T}_{n,1}^{(3)}=\mathpzc{o}_\mathbb{P}\left(1\right).\label{theo:limdis:5}
	\end{align}
	If $c_n^2nh^\frac{q}{2}\to\infty$, then $nh^\frac{q}{2}\widetilde{T}_{n,1}^{(2)}\to\infty$, yielding a degenerate asymptotic distribution. If $c_n^2nh^\frac{q}{2}\to0$, then $nh^\frac{q}{2}\widetilde{T}_{n,1}^{(2)}=\mathpzc{o}_\mathbb{P}\left(1\right)$. For these reasons, $c_n=\big(nh^\frac{q}{2}\big)^{-\frac{1}{2}}$ is assumed from now on. For the first addend, let consider 
	\begin{align*}
	\widetilde{T}_{n,1}^{(1)}=&\,\int_{\Omega_q}\sum_{i=1}^n \left(\tilde{L}_h\left(\mathbf{x},\mathbf{X}_i\right)\sigma(\mathbf{X}_i)\varepsilon_i\right)^2f(\mathbf{x})w(\mathbf{x})\,\omega_q(d\mathbf{x})\\
	&+\int_{\Omega_q}\sum_{i\neq j} \tilde{L}_h\left(\mathbf{x},\mathbf{X}_i\right)\tilde{L}_h\left(\mathbf{x},\mathbf{X}_j\right)\sigma(\mathbf{X}_i)\sigma(\mathbf{X}_j)\varepsilon_i\varepsilon_j f(\mathbf{x})w(\mathbf{x})\,\omega_q(d\mathbf{x})=\widetilde{T}_{n,1}^{(1a)}+\widetilde{T}_{n,1}^{(1b)}.
	\end{align*}
	From result 
	\ifsupplement
	\ref{lem:5:3} of Lemma \ref{lem:5} 
	\else
	\textit{iv} of Lemma S6 
	\fi
	and because $\sigma^2(\mathbf{x})=\sigma_{\boldsymbol{\theta}_0}^2(\mathbf{x})(1+\mathpzc{o}\left(1\right))$ uniformly,
	\begin{align*}
	nh^\frac{q}{2}\widetilde{T}_{n,1}^{(1a)}=\frac{\lambda_q(L^2)\lambda_q(L)^{-2}}{h^\frac{q}{2}}\int_{\Omega_q}\sigma_{\boldsymbol{\theta}_0}^2(\mathbf{x})w(\mathbf{x})\,\omega_q(d\mathbf{x})(1+\mathpzc{o}\left(1\right))+\mathpzc{o}_\mathbb{P}\left(1\right).
	\end{align*}
	
	The asymptotics of $\widetilde{T}_{n,1}^{(1b)}$ are obtained checking the conditions of Theorem 2.1 in \cite{Jong1987}: \textit{a)} $\mathbb{E}\left[W_{ijn}+W_{jin}|X_i\right]=0$, $1\leq i<j\leq n$; \textit{b)} $\mathbb{V}\mathrm{ar}\left[W_n\right]\to v^2$; \textit{c)} $\big(\max_{1\leq i\leq n}\sum_{j=1}^n\allowbreak \mathbb{V}\mathrm{ar}\left[W_{ijn}\right]\big)v^{-2}\to 0$; \textit{d)} $\mathbb{E}\left[W_n^4\right]v^{-4}\to3$. To that end, let denote
	\begin{align*}
	W_{ijn}=\delta_{i,j}
	nh^\frac{q}{2}\int_{\Omega_q} \tilde{L}_h\left(\mathbf{x},\mathbf{X}_i\right)\tilde{L}_h\left(\mathbf{x},\mathbf{X}_j\right)\sigma(\mathbf{X}_i)\sigma(\mathbf{X}_j)\varepsilon_i\varepsilon_j f(\mathbf{x})w(\mathbf{x})\,\omega_q(d\mathbf{x}).
	\end{align*}
	Then, $nh^\frac{q}{2}\widetilde{T}_{n,1}^{(1b)}=W_n=\sum_{i\neq j}W_{ijn}$ and the rv's on which $W_{ijn}$ depends are $(\mathbf{X}_i,\varepsilon_i)$ and $(\mathbf{X}_j,\varepsilon_j)$. \textit{a)} is easily seen to hold by $\mathbb{E}\left[\varepsilon|\mathbf{X}\right]=0$ and the tower property, which implies that $\mathbb{E}\left[W_{ijn}\right]=0$. Because of this, the fact that $W_{ijn}=W_{jin}$ and Lemma 2.1 in \cite{Jong1987},
	\begin{align}
	\mathbb{V}\mathrm{ar}\left[W_n\right]=\mathbb{E}\bigg[\Big(\sum_{i\neq j} W_{ijn}\Big)^2\bigg]=2\mathbb{E}\bigg[\sum_{i\neq j} W_{ijn}^2\bigg]=2n(n-1)\mathbb{E}\left[W_{ijn}^2\right].\label{VWn}
	\end{align}
	Then, by 
	\ifsupplement
	\ref{lem:5:5} in Lemma \ref{lem:5} 
	\else
	\textit{v} in Lemma S6 
	\fi
	and the fact that $\sigma^2(\mathbf{x})=\sigma_{\boldsymbol{\theta}_0}^2(\mathbf{x})\allowbreak(1+\mathpzc{o}\left(1\right))$, $\mathbb{E}\big[W_{ijn}^2\big]=n^{-2}\nu_{\boldsymbol{\theta}_0}^2\left(1+\mathpzc{o}\left(1\right)\right)$ and as a consequence $\mathbb{V}\mathrm{ar}\left[W_n\right]\to 2\nu_{\boldsymbol{\theta}_0}^2$. Condition \textit{c)} follows easily:
	\[
	\bigg(\max_{1\leq i\leq n}\sum_{j=1}^n \mathbb{V}\mathrm{ar}\left[W_{ijn}\right]\bigg)v^{-2}\leq\left(\max_{1\leq i\leq n}n^{-1}\nu_{\boldsymbol{\theta}_0}^2\left(1+\mathpzc{o}\left(1\right)\right)\right)(2\nu_{\boldsymbol{\theta}_0}^{2})^{-1}=(2n)^{-1}(1+\mathpzc{o}\left(1\right))\to0.
	\]
	To check \textit{d)}, note that $\mathbb{E}\left[W_n^4\right]$ can be split in the following form in virtue of Lemma 2.1 in \cite{Jong1987}, as \cite{Hardle1993} stated:
	\begin{align}
	\mathbb{E}\left[W_n^4\right] =&\,8\sum_{i,j}\!^{\neq}\,\mathbb{E}\left[W_{ijn}^4\right]+12\sum_{i, j, k, l}\!^{\neq}\,\mathbb{E}\left[W_{ijn}^2W_{kln}^2\right]+48\sum_{i,j,k}\!^{\neq}\,\mathbb{E}\left[W_{ijn}W_{ikn}^2W_{jkn}\right]\nonumber\\
	&+192\sum_{i,j,k,l}\!^{\neq}\,\mathbb{E}\left[W_{ijn}W_{jkn}W_{kln}W_{lin}\right],\label{EWn4}
	\end{align}
	where $\sum\!^{\neq}$ stands for the summation over all \textit{pairwise different} indexes (\textit{i.e.}, such that $i\neq j$ for their associated $W_{ijn}$). By 
	\ifsupplement
	\ref{lem:5:5} of Lemma \ref{lem:5},
	\else
	\textit{v} of Lemma S6,
	\fi
	$\mathbb{E}\big[W_{ijn}^4\big]=\mathcal{O}\left((n^4h^{q})^{-1}\right)$, $\mathbb{E}\left[W_{ijn}W_{jkn}W_{kln}W_{lin}\right]=\mathcal{O}\left(n^{-4}h^{2q}\right)$ and $\mathbb{E}\left[W_{ijn}W_{ikn}^2W_{jkn}\right]=\mathcal{O}\left(n^{-4}\right)$. Therefore, by (\ref{VWn}) and (\ref{EWn4}),
	\[
	\mathbb{E}\left[W_n^4\right]=12\sum_{i\neq j}\sum_{k\neq l}\mathbb{E}\left[W_{ijn}^2W_{kln}^2\right]+\mathpzc{o}\left(1\right)=3\Big(2\sum_{i\neq j}\mathbb{E}\left[W_{ijn}^2\right]\Big)^2+\mathpzc{o}\left(1\right)=3\mathbb{V}\mathrm{ar}\left[W_n\right]^2+\mathpzc{o}\left(1\right)
	\]
	and by \ref{assump:a4}, $\mathbb{E}\left[W_n^4\right]=3\mathbb{V}\mathrm{ar}\left[W_n\right]^2+\mathpzc{o}\left(1\right)$, so \textit{d)} is satisfied, having that
	\begin{align}
	nh^\frac{q}{2}\widetilde{T}_{n,1}^{(1b)}\stackrel{d}{\longrightarrow}\mathcal{N}\left(0,2\nu_{\boldsymbol{\theta}_0}^2\right). \label{theo:limdis:7}
	\end{align}
	Using the decomposition for $T_n$ with the dominant terms $\widetilde{T}_{n,1}^{(1a)}$, $\widetilde{T}_{n,1}^{(1b)}$ and $\widetilde{T}_{n,1}^{(2)}$, it holds
	\begin{align*}
	nh^\frac{q}{2}T_n	=&\,\bigg(\frac{\lambda_q(L^2)\lambda_q(L)^{-2}}{h^\frac{q}{2}}\int_{\Omega_q}\sigma_{\boldsymbol{\theta}_0}^2(\mathbf{x})w(\mathbf{x})\,\omega_q(d\mathbf{x})+nh^\frac{q}{2}\widetilde{T}_{n,1}^{(1b)}\\
	&\qquad+\int_{\Omega_q}g(\mathbf{x})^2 f(\mathbf{x})w(\mathbf{x})\,\omega_q(d\mathbf{x})\bigg)(1+\mathpzc{o}_\mathbb{P}\left(1\right))
	\end{align*}
	and the limit distribution follows by Slutsky's theorem and (\ref{theo:limdis:7}).
\end{proof}

\begin{proof}[Proof of Theorem \ref{theo:boot}]
	Analogously as in Theorem \ref{theo:limdis}, $T_n^*=T^*_{n,1}+T^*_{n,2}-2T^*_{n,3}$.\\
	
	\textit{Terms $T_{n,2}^*$ and $T_{n,3}^*$}.
	By \ref{assump:a9n} it is seen that $nh^\frac{q}{2}T_{n,2}^*\stackrel{p^*}{\longrightarrow}0$ and $nh^\frac{q}{2}T_{n,3}^*\stackrel{p^*}{\longrightarrow}0$, where the convergence is stated in the probability law $\mathbb{P}^*$ that is conditional on the sample. \\
	
	\textit{Term $T_{n,1}^*$}. By $\hat\varepsilon_iV_i^*=(Y_i-m_{\hat {\boldsymbol{\theta}}}(\mathbf{X}_i))V_i^*$ the dominant term can be split into
	\begin{align*}
	T_{n,1}^{*}=&\,\int_{\Omega_q}\sum_{i=1}^n \left(W_n^p\left(\mathbf{x},\mathbf{X}_i\right)\hat\varepsilon_iV_i^*\right)^2\hat f_h(\mathbf{x})w(\mathbf{x})\,\omega_q(d\mathbf{x})\\
	&+\int_{\Omega_q}\sum_{i\neq j} W_n^p\left(\mathbf{x},\mathbf{X}_i\right)W_n^p\left(\mathbf{x},\mathbf{X}_j\right)\hat\varepsilon_iV_i^*\hat\varepsilon_jV_j^* \hat f_h(\mathbf{x})w(\mathbf{x})\,\omega_q(d\mathbf{x})=T_{n,1}^{*(1)}+T_{n,1}^{*(2)}.
	\end{align*}
	From result 
	\ifsupplement
	\ref{lem:6:1} of Lemma \ref{lem:6}, 
	\else
	\textit{i} of Lemma S7, 
	\fi
	the first term is 
	\begin{align}
	nh^\frac{q}{2}T_{n,1}^{*(1)}=\frac{\lambda_q(L^2)\lambda_q(L)^{-2}}{h^\frac{q}{2}}\int_{\Omega_q}\sigma_{\boldsymbol{\theta}_1}^2(\mathbf{x})w(\mathbf{x})\,\omega_q(d\mathbf{x})(1+\mathpzc{o}_\mathbb{P}\left(1\right))+\mathpzc{o}_{\mathbb{P}^*}(1), \label{theo:boot:2}
	\end{align}
	so the dominant term is $T_{n,1}^{*(2)}$, whose asymptotic behavior is obtained using Theorem 2.1 in \cite{Jong1987} conditionally on the sample. For that aim, let denote
	\begin{align*}
	W_{ijn}^*=\delta_{i,j}nh^\frac{q}{2}\int_{\Omega_q} W_n^p\left(\mathbf{x},\mathbf{X}_i\right)W_n^p\left(\mathbf{x},\mathbf{X}_j\right)\hat\varepsilon_iV_i^*\hat\varepsilon_jV_j^* \hat f_h(\mathbf{x})w(\mathbf{x})\,\omega_q(d\mathbf{x}).
	\end{align*}
	Then, $nh^\frac{q}{2}T_{n,1}^{*(2)}=W_n^*=\sum_{i\neq j}W_{ijn}^*$ and the rv's on which $W_{ijn}^*$ depends are now $V_i^*$ and $V_j^*$. Condition \textit{a)} follows immediately by the properties of the $V_i^*$'s: $\mathbb{E^*}\big[W_{ijn}^*+W_{jin}^*|V_i^*\big]=0$. On the other hand, analogously to (\ref{VWn}),
	\[
	\mathbb{V}\mathrm{ar}^*\left[W_n^*\right]=\!2\sum_{i\neq j}\mathbb{E}^*\left[W_{ijn}^{*2}\right]\!=2n^2h^q\sum_{i\neq j} \!\left[\int_{\Omega_q}\!\! W_n^p\left(\mathbf{x},\mathbf{X}_i\right)W_n^p\left(\mathbf{x},\mathbf{X}_j\right)\hat\varepsilon_i\hat\varepsilon_j \hat f_h(\mathbf{x})w(\mathbf{x})\,\omega_q(d\mathbf{x})\right]^2
	\]
	and by result 
	\ifsupplement
	\ref{lem:6:2} of Lemma \ref{lem:6},
	\else
	\textit{ii} of Lemma S7,
	\fi
	$\mathbb{V}\mathrm{ar}^*\left[W_n^*\right]\stackrel{p}{\longrightarrow} 2\nu_{\boldsymbol{\theta}_1}^2$, resulting in the verification of \textit{c)} in probability. Condition \textit{d)} is checked using the same decomposition for $\mathbb{E}^*\big[W_n^{*4}\big]$ and the results collected in 
	\ifsupplement
	\ref{lem:6:2} of Lemma \ref{lem:6}. 
	\else
	\textit{ii} of Lemma S7. 
	\fi
	Hence $\mathbb{E^*}\big[W_n^{*4}\big]=3\mathbb{V}\mathrm{ar}^*\left[W_n^*\right]^2+\mathpzc{o}_\mathbb{P}\left(1\right)$ and \textit{d)} is satisfied in probability, from which it follows that, conditionally on $\left\{(\mathbf{X}_i,Y_i)\right\}_{i=1}^n$ the pdf of $nh^\frac{q}{2}T_{n,1}^{*(2)}$ converges in probability to the pdf of $\mathcal{N}(0,2\nu_{\boldsymbol{\theta}_1}^2)$, that is:
	\begin{align}
	nh^\frac{q}{2}T_{n,1}^{*(2)}\stackrel{d}{\longrightarrow}\mathcal{N}\left(0,2\nu_{\boldsymbol{\theta}_1}^2\right)\text{ in probability}. \label{theo:boot:3}
	\end{align}
	Using the decomposition of $T_n^*$, (\ref{theo:boot:3}) and applying Slutsky's theorem:
	\begin{align*}
	nh^\frac{q}{2}T_n^* 
	=&\,\bigg(\frac{\lambda_q(L^2)\lambda_q(L)^{-2}}{h^\frac{q}{2}}\int_{\Omega_q}\sigma_{\boldsymbol{\theta}_1}^2(\mathbf{x})w(\mathbf{x})\,\omega_q(d\mathbf{x})+nh^\frac{q}{2}T_{n,1}^{*(2)}\bigg)(1+\mathpzc{o}_\mathbb{P}\left(1\right))+\mathpzc{o}_{\mathbb{P}^*}(1).
	\end{align*}\end{proof}

\fi

\ifsupplement


\newpage
\title{Supporting information for ``Testing parametric models in linear-directional regression''}
\setlength{\droptitle}{-1cm}
\predate{}%
\postdate{}%
\date{}

\author{Eduardo Garc\'ia-Portugu\'es$^{1,2,3,5}$, Ingrid Van Keilegom$^4$, \\ Rosa M. Crujeiras$^3$, and Wenceslao Gonz\'alez-Manteiga$^3$}

\footnotetext[1]{
	Department of Mathematical Sciences, University of Copenhagen (Denmark).}
\footnotetext[2]{
	The Bioinformatics Centre, Department of Biology, University of Copenhagen (Denmark).}
\footnotetext[3]{
	Department of Statistics and Operations Research, University of Santiago de Compostela (Spain).}
\footnotetext[4]{
	Institute of Statistics, Biostatistics and Actuarial Sciences, Universit\'e catholique de Louvain (Belgium).}
\footnotetext[5]{Corresponding author. e-mail: \href{mailto:egarcia@math.ku.dk}{egarcia@math.ku.dk}.}

\maketitle

\begin{abstract}
	This supplement is organized as follows. Section \ref{appendix:prooftechlemmas} gives the technical lemmas used to prove the main results in the paper. Section \ref{appendix:empev} shows an empirical evidence of the asymptotic distribution of the test statistic. Section \ref{appendix:simus} provides a complete description on the simulation study. Finally, Section \ref{appendix:data} describes data preprocessing and results omitted from data application.     
\end{abstract}
\begin{flushleft}
	\small\textbf{Keywords:} Bootstrap calibration; Directional data; Goodness-of-fit test; Local linear regression.
\end{flushleft}

\section{Technical lemmas}
\label{appendix:prooftechlemmas}

\begin{lem}[Tangent-normal change of variables]
	\label{lem:2}
	Let $f$ be a function defined in $\Omega_q$ and $\mathbf{x}\in\Omega_q$. Then
	$\int_{\Omega_q}f(\mathbf{z})\,\omega_q(d\mathbf{z})=\int_{-1}^{1}\int_{\Omega_{q-1}} f\big(t\mathbf{x}+(1-t^2)^\frac{1}{2}\mathbf{B}_{\mathbf{x}}\boldsymbol\xi\big) (1-t^2)^{\frac{q}{2}-1}\,\omega_{q-1}(d\boldsymbol\xi)\,dt,$
	where $\mathbf{B}_{\mathbf{x}}=\left(\mathbf{b}_1,\ldots,\mathbf{b}_q\right)_{(q+1)\times q}$ is the projection matrix given in Section \ref{sec:reg}.
\end{lem}

\begin{proof}[Proof of Lemma \ref{lem:2}]
	See Lemma 2 of \cite{Garcia-Portugues:dirlin}.
\end{proof}

\begin{lem}
	\label{lem:3}
	Set $\mathbf{x}=(x_1,\ldots,x_{q+1})\in\Omega_q$. For all $i,j,k=1,\ldots,q+1$, $\int_{\Omega_q}x_i\,\omega_q(d\mathbf{x})=0$, $\int_{\Omega_q}x_ix_j\,\omega_q(d\mathbf{x})=\delta_{ij}\frac{\omega_{q}}{q+1}$ and $\int_{\Omega_q}x_ix_jx_k\,\omega_q(d\mathbf{x})=0$.
\end{lem}

\begin{proof}[Proof of Lemma \ref{lem:3}]
	Apply Lemma \ref{lem:2} considering $\mathbf{x}=\mathbf{e}_i\in\Omega_q$. Then
	$\int_{\Omega_q}x_i\,\omega_q(d\mathbf{x})=\omega_{q-1}\int_{-1}^1 t(1-t^2)^{\frac{q}{2}-1}\,dt=0$
	as the integrand is an odd function. As a consequence, and applying the same change of variables, for $i\neq j$:
	\[
	\int_{\Omega_q}x_ix_j\,\omega_q(d\mathbf{x})=\int_{-1}^1(1-t^2)^{\frac{q-1}{2}}\,dt\int_{\Omega_{q-1}} \mathbf{e}_j^T\mathbf{B}_{\mathbf{x}}\boldsymbol\xi\,\omega_{q-1}(d\boldsymbol\xi)=0.
	\]
	For $i=j$, $\int_{\Omega_q}x_i^2\,\omega_q(d\mathbf{x})=\frac{1}{q+1}\int_{\Omega_q}\sum_{j=1}^q x_j^2\,\omega_q(d\mathbf{x})=\frac{\omega_{q}}{q+1}$. For the trivariate case,
	\begin{align*}
	\int_{\Omega_q}x_i^3\,\omega_q(d\mathbf{x})&=\omega_{q-1}\int_{-1}^1 t^3(1-t^2)^{\frac{q}{2}-1}\,dt=0,\\
	\int_{\Omega_q}x_i^2x_j\,\omega_q(d\mathbf{x})&=\int_{-1}^1t^2(1-t^2)^{\frac{q-1}{2}}\,dt\int_{\Omega_{q-1}} \mathbf{e}_j^T\mathbf{B}_{\mathbf{x}}\boldsymbol\xi\,\omega_{q-1}(d\boldsymbol\xi)=0,\,i\neq j,\\
	\int_{\Omega_q}x_ix_jx_k\,\omega_q(d\mathbf{x})&=\int_{-1}^1t(1-t^2)^{\frac{q}{2}}\,dt\int_{\Omega_{q-1}} \mathbf{e}_j^T\mathbf{B}_{\mathbf{x}}\boldsymbol\xi\mathbf{e}_k^T\mathbf{B}_{\mathbf{x}}\boldsymbol\xi\,\omega_{q-1}(d\boldsymbol\xi)=0,\,i\neq j\neq k,
	\end{align*}
	using that the integrand is odd and the first statement.
\end{proof}

\begin{lem}[\cite{Bai1988}]
	\label{lem:4}
	Let $\varphi:\Omega_q\longrightarrow\mathbb{R}$ be a continuous function and denote $L_h \varphi(\mathbf{x})=c_{h,q}(L)\int_{\Omega_q}L\left(\frac{1-\mathbf{x}^T\mathbf{y}}{h^2}\right)\varphi(\mathbf{y})\,\omega_q(d\mathbf{y})$. Under \ref{assump:a3}--\ref{assump:a4}, $L_h \varphi(\mathbf{x})=\varphi(\mathbf{x})+\mathpzc{o}\left(1\right)$, where the remaining order is uniform for any $\mathbf{x}\in\Omega_q$.
\end{lem}

\begin{proof}[Proof of Lemma \ref{lem:4}]
	This corresponds to Lemma 5 in \cite{Bai1988}, but with slightly different conditions and notation. \ref{assump:a1} and \ref{assump:a3} imply conditions (a), (b), (c$_1$) and (d) stated in Theorem 1 of the aforementioned paper. 
\end{proof}

\begin{lem}
	\label{lem:1}
	Under \ref{assump:a1}--\ref{assump:a4}, for a random sample $\left\{(\mathbf{X}_i,Y_i)\right\}_{i=1}^n$ the following statements hold with uniform orders for any point $\mathbf{x}\in\Omega_q$:
	\begin{enumerate}[label=\roman{*}., ref=\textit{\roman{*}}]
		\item $\hat f_h(\mathbf{x})=f(\mathbf{x})+\mathpzc{o}_\mathbb{P}\left(1\right)$.\label{lem:1:d0}
		
		\item $\frac{1}{n} \sum_{i=1}^n L_h(\mathbf{x},\mathbf{X}_i)\mathbf{B}_{\mathbf{x}}^T(\mathbf{X}_i-\mathbf{x})=\frac{2b_q(L)}{q}\mathbf{B}_{\mathbf{x}}^T\boldsymbol\nabla f(\mathbf{x})h^2+\mathpzc{o}\left(h^2\mathbf{1}\right)+\mathcal{O}_\mathbb{P}\left(\frac{h}{\sqrt{nh^q}}\mathbf{1}\right)$.\label{lem:1:d1b} 
		
		\item $\frac{1}{n} \sum_{i=1}^n L_h(\mathbf{x},\mathbf{X}_i)\mathbf{B}_{\mathbf{x}}^T(\mathbf{X}_i-\mathbf{x})Y_i=\mathcal{O}\left(h^2\mathbf{1}\right)+\mathcal{O}_\mathbb{P}\left(\frac{h}{\sqrt{nh^q}}\mathbf{1}\right)$.\label{lem:1:d1c} 
		
		\item $\frac{1}{n} \sum_{i=1}^n L_h(\mathbf{x},\mathbf{X}_i)\mathbf{B}_{\mathbf{x}}^T(\mathbf{X}_i-\mathbf{x})(\mathbf{X}_i-\mathbf{x})^T\mathbf{B}_{\mathbf{x}}=\frac{2b_q(L)}{q}\mathbf{I}_q f(\mathbf{x})h^2+\mathpzc{o}_\mathbb{P}\big(h^2\mathbf{1}\mathbf{1}^T\big)$.\label{lem:1:d2b}
	\end{enumerate}
\end{lem}

\begin{proof}[Proof of Lemma \ref{lem:1}] 
	
	\mbox{}\\\textit{Proof of \ref{lem:1:d0}}. 
	By Chebychev's inequality, $\hat f_h(\mathbf{x})=\mathbb{E}\big[\hat f_h(\mathbf{x})\big]+\mathcal{O}_\mathbb{P}\Big(\sqrt{\mathbb{V}\mathrm{ar}\big[\hat f_h(\mathbf{x})\big]}\Big)$. It follows by Lemma \ref{lem:4} that $\mathbb{E}\big[\hat f_h(\mathbf{x})\big]=f(\mathbf{x})+\mathpzc{o}\left(1\right)$ and that $\mathbb{V}\mathrm{ar}\big[\hat f_h(\mathbf{x})\big]=\frac{1}{nh^q\lambda_q(L)}(f(\mathbf{x})+\mathpzc{o}\left(1\right))$, with the remaining orders being uniform in $\mathbf{x}\in\Omega_q$. Then, as $f$ is continuous in $\Omega_q$ by assumption \ref{assump:a1} it is also bounded, so by \ref{assump:a4} $\mathbb{V}\mathrm{ar}\big[\hat f_h(\mathbf{x})\big]=\mathpzc{o}\left(1\right)$ uniformly, which results in $\hat f_h(\mathbf{x})=f(\mathbf{x})+\mathpzc{o}_\mathbb{P}\left(1\right)$ uniformly in $\mathbf{x}\in\Omega_q$.\\
	
	\textit{Proof of \ref{lem:1:d1b}}. Applying Lemma \ref{lem:2} and the change of variables $r=\frac{1-t}{h^2}$,
	\begin{align}
	\mathbb{E}\Bigg[\frac{1}{n}\sum_{i=1}^n &L_h(\mathbf{x},\mathbf{X}_i)\mathbf{B}_{\mathbf{x}}^T(\mathbf{X}_i-\mathbf{x})\Bigg]\nonumber\\
	=\,&c_{h,q}(L)\int_{\Omega_q}L\left(\frac{1-\mathbf{x}^T\mathbf{y}}{h^2}\right)\mathbf{B}_{\mathbf{x}}^T(\mathbf{y}-\mathbf{x})f(\mathbf{y})\,\omega_q(d\mathbf{y})\nonumber\\
	=\,&c_{h,q}(L)h^{q+1}\int_{0}^{2h^{-2}} L\left(r\right)r^{\frac{q-1}{2}}(2-rh^2)^{\frac{q-1}{2}} \int_{\Omega_{q-1}} f\left(\mathbf{x}+\boldsymbol\alpha_{\mathbf{x},\boldsymbol\xi}\right) \boldsymbol\xi\,\omega_{q-1}(d\boldsymbol\xi)\,dr,\label{lem:1:1}
	\end{align}
	where $\boldsymbol\alpha_{\mathbf{x},\boldsymbol\xi}=-rh^2\mathbf{x}+\left[rh^2(2-rh^2)\right]^\frac{1}{2}\mathbf{B}_{\mathbf{x}}\boldsymbol\xi$. The inner integral in (\ref{lem:1:1}) is computed by a Taylor expansion
	\begin{align}
	f(\mathbf{x}+\boldsymbol\alpha_{\mathbf{x},\boldsymbol\xi})=f(\mathbf{x})+\boldsymbol\alpha_{\mathbf{x},\boldsymbol\xi}^T\boldsymbol\nabla f(\mathbf{x})+\mathcal{O}\left(\boldsymbol\alpha_{\mathbf{x},\boldsymbol\xi}^T\boldsymbol\alpha_{\mathbf{x},\boldsymbol\xi}\right),\label{lem:1:2}
	\end{align}
	where the remaining order involves the second derivative of $f$, which is bounded, thus being the order uniform in $\mathbf{x}$. Using Lemma \ref{lem:3}, the first and second addends are:
	\begin{align*}
	\int_{\Omega_{q-1}} f\left(\mathbf{x}\right)\boldsymbol\xi \,\omega_{q-1}(d\boldsymbol\xi)=&\,0,\\
	\int_{\Omega_{q-1}} \boldsymbol\alpha_{\mathbf{x},\boldsymbol\xi}^T\boldsymbol\nabla f\left(\mathbf{x}\right)\boldsymbol\xi \,\omega_{q-1}(d\boldsymbol\xi)=&\,\left[rh^2(2-rh^2)\right]^\frac{1}{2}\int_{\Omega_{q-1}}(\mathbf{B}_{\mathbf{x}}\boldsymbol\xi)^T\boldsymbol\nabla f\left(\mathbf{x}\right)\boldsymbol\xi \,\omega_{q-1}(d\boldsymbol\xi)\\
	=&\,\left[rh^2(2-rh^2)\right]^\frac{1}{2}\int_{\Omega_{q-1}}\sum_{i,j=1}^q \xi_i\mathbf{B}_{\mathbf{x}}^T\boldsymbol\nabla f\left(\mathbf{x}\right)\xi_j \,\omega_{q-1}(d\boldsymbol\xi)\\
	=&\,\frac{\omega_{q-1}}{q}\left[rh^2(2-rh^2)\right]^\frac{1}{2} \mathbf{B}_{\mathbf{x}}^T\boldsymbol\nabla f(\mathbf{x}).
	\end{align*}
	The third addend is $\mathcal{O}\big(\boldsymbol\alpha_{\mathbf{x},\boldsymbol\xi}^T\boldsymbol\alpha_{\mathbf{x},\boldsymbol\xi}\big)=\mathcal{O}\left(h^2\mathbf{1}\right)$, because $\mathbf{B}_{\mathbf{x}}^T\mathbf{x}=\mathbf{0}$ and $(\mathbf{B}_{\mathbf{x}}\boldsymbol\xi)^T\mathbf{B}_{\mathbf{x}}\boldsymbol\xi=\boldsymbol\xi^T\mathbf{I}_q\boldsymbol\xi=1$. Therefore, (\ref{lem:1:1}) becomes
	\begin{align}
	(\mathrm{\ref{lem:1:1}})=\,&c_{h,q}(L)h^{q+2}\frac{\omega_{q-1}}{q}\int_{0}^{2h^{-2}} L\left(r\right)r^{\frac{q
		}{2}}(2-rh^2)^{\frac{q}{2}} \,dr\, \mathbf{B}_{\mathbf{x}}^T\boldsymbol\nabla f(\mathbf{x})\nonumber\\
	&+c_{h,q}(L)h^{q+1}\int_{0}^{2h^{-2}} L\left(r\right)r^{\frac{q}{2}}(2-rh^2)^{\frac{q}{2}}\,dr\,\mathcal{O}\left(h^2\mathbf{1} \right)
	\nonumber\\
	=\,&\left(b_q(L)+\mathpzc{o}\left(1\right)\right)\frac{2b_q(L)}{q}\mathbf{B}_{\mathbf{x}}^T\boldsymbol\nabla f(\mathbf{x})h^2+\mathcal{O}\left(h^3\mathbf{1}\right)\nonumber\\
	=\,&\frac{2b_q(L)}{q}\mathbf{B}_{\mathbf{x}}^T\boldsymbol\nabla f(\mathbf{x})h^2+\mathpzc{o}\left(h^2\mathbf{1}\right),\label{lem:1:3}
	\end{align}
	where the second last equality follows from applying the Dominated Convergence Theorem (DCT), (\ref{chq}) and the definition of $b_q(L)$. See the proof of Theorem 1 in \cite{Garcia-Portugues:dirlin} for the technical details involved in a similar situation.\\
	
	As the Chebychev inequality is going to be applied componentwise, the interest is now in the order of the variance vector. To that end, the square of a vector will denote the vector with correspondent squared components. By analogous computations,
	\begin{align}
	\mathbb{V}\mathrm{ar}\Bigg[\frac{1}{n}\sum_{i=1}^n &L_h(\mathbf{x},\mathbf{X}_i)\mathbf{B}_{\mathbf{x}}^T(\mathbf{X}_i-\mathbf{x})\Bigg]\nonumber\\
	\leq\,&\frac{1}{n}\mathbb{E}\left[L_h(\mathbf{x},\mathbf{X})^2(\mathbf{B}_{\mathbf{x}}^T(\mathbf{X}-\mathbf{x}))^2\right]\nonumber\\
	=\,&\frac{c_{h,q}(L)^2h^{q+2}}{n}\int_{0}^{2h^{-2}} L^2\left(r\right)r^{\frac{q}{2}}(2-rh^2)^{\frac{q}{2}} \int_{\Omega_{q-1}} f\left(\mathbf{x}+\boldsymbol\alpha_{\mathbf{x},\boldsymbol\xi}\right)\boldsymbol\xi^2 \,\omega_{q-1}(d\boldsymbol\xi)\,dr\nonumber\\
	=\,&\frac{c_{h,q}(L)^2h^{q+2}}{n}\int_{0}^{2h^{-2}} L^2\left(r\right)r^{\frac{q}{2}}(2-rh^2)^{\frac{q}{2}} \mathcal{O}\left(\mathbf{1}\right)\,dr\nonumber\\
	=\,&\mathcal{O}\left(\frac{h^2}{nh^q}\mathbf{1}\right).\label{lem:1:4}
	\end{align}
	The result follows from Chebychev's inequality, (\ref{lem:1:3}) and (\ref{lem:1:4}).\\
	
	\textit{Proof of \ref{lem:1:d1c}}. The result is proved form the previous proof and the tower property of the conditional expectation. The expectation can be expressed as
	\[
	\mathbb{E}\Bigg[\frac{1}{n}\sum_{i=1}^nL_h(\mathbf{x},\mathbf{X}_i)\mathbf{B}_{\mathbf{x}}^T(\mathbf{X}_i-\mathbf{x})Y_i\Bigg]
	=c_{h,q}(L)\int_{\Omega_q}L\left(\frac{1-\mathbf{x}^T\mathbf{y}}{h^2}\right)\mathbf{B}_{\mathbf{x}}^T(\mathbf{y}-\mathbf{x})m(\mathbf{y})f(\mathbf{y})\,\omega_q(d\mathbf{y}).
	\]
	Then, replicating the proof of \ref{lem:1:d1b}, it is easily seen that the order is $\mathcal{O}\left(h^2\mathbf{1}\right)$. The order of the variance is obtained in the same way:
	\begin{align*}
	\mathbb{V}\mathrm{ar}\left[\frac{1}{n}\sum_{i=1}^nL_h(\mathbf{x},\mathbf{X}_i)\mathbf{B}_{\mathbf{x}}^T(\mathbf{X}_i-\mathbf{x})Y_i\right] 
	\leq&\,\frac{1}{n}\mathbb{E}\left[L_h(\mathbf{x},\mathbf{X})(\mathbf{B}_{\mathbf{x}}^T(\mathbf{X}-\mathbf{x}))^2(\sigma^2(\mathbf{X})+m(\mathbf{X})^2)\right]\\
	=&\,\mathcal{O}\left(\frac{h^2}{nh^q}\mathbf{1}\right).
	\end{align*}
	As a consequence, $\frac{1}{n} \sum_{i=1}^n L_h(\mathbf{x},\mathbf{X}_i)\mathbf{B}_{\mathbf{x}}^T(\mathbf{X}_i-\mathbf{x})Y_i=\mathcal{O}\left(h^2\mathbf{1}\right)+\mathcal{O}_\mathbb{P}\left(\frac{h}{\sqrt{nh^q}}\mathbf{1}\right)$.\\
	
	\textit{Proof of \ref{lem:1:d2b}}. The steps of the proof of \ref{lem:1:d1b} are replicated:
	\begin{align}
	\mathbb{E}\Bigg[\frac{1}{n}\sum_{i=1}^n& L_h(\mathbf{x},\mathbf{X}_i)\mathbf{B}_{\mathbf{x}}^T(\mathbf{X}_i-\mathbf{x})(\mathbf{X}_i-\mathbf{x})^T\mathbf{B}_{\mathbf{x}}\Bigg]\nonumber\\
	=\,&c_{h,q}(L)h^{q+2}\int_{0}^{2h^{-2}} L\left(r\right)r^{\frac{q}{2}}(2-rh^2)^{\frac{q}{2}} \int_{\Omega_{q-1}} f\left(\mathbf{x}+\boldsymbol\alpha_{\mathbf{x},\boldsymbol\xi}\right)\boldsymbol\xi\boldsymbol\xi^T \,\omega_{q-1}(d\boldsymbol\xi)\,dr.\label{lem:1:5}
	\end{align}
	The second integral of (\ref{lem:1:5}) is obtained by expansion (\ref{lem:1:2}) and Lemma \ref{lem:3}:
	\begin{align*}
	\int_{\Omega_{q-1}} f\left(\mathbf{x}\right)\boldsymbol\xi\boldsymbol\xi^T\mathbf{B}_{\mathbf{x}} \,\omega_{q-1}(d\boldsymbol\xi)=&\,\frac{\omega_{q-1}}{q}\mathbf{I}_q f(\mathbf{x}),\\
	\int_{\Omega_{q-1}} \boldsymbol\alpha_{\mathbf{x},\boldsymbol\xi}^T\boldsymbol\nabla f\left(\mathbf{x}\right)\boldsymbol\xi\boldsymbol\xi^T \,\omega_{q-1}(d\boldsymbol\xi)=&\,\int_{\Omega_{q-1}} -rh^2\mathbf{x}^T\boldsymbol\xi\boldsymbol\xi^T\,\omega_{q-1}(d\boldsymbol\xi)=\mathcal{O}\left(h^2\mathbf{1}\mathbf{1}^T\right).
	\end{align*}
	As the third addend given by expansion (\ref{lem:1:2}) has order $\mathcal{O}\left(h^2\mathbf{1}\mathbf{1}^T\right)$, it results that:
	\begin{align}
	(\mathrm{\ref{lem:1:5}})=\,&c_{h,q}(L)h^{q+2}\int_{0}^{2h^{-2}} L\left(r\right)r^{\frac{q}{2}}(2-rh^2)^{\frac{q}{2}} \left\{\frac{\omega_{q-1}}{q}\mathbf{I}_qf\left(\mathbf{x}\right)+\mathcal{O}\left(h^2\mathbf{1}\mathbf{1}^T\right)\right\}\,dr\nonumber\\
	=\,&\frac{2b_q(L)}{q}\mathbf{I}_qf\left(\mathbf{x}\right)h^2+\mathpzc{o}\left(h^2\mathbf{1}\mathbf{1}^T\right),\label{lem:1:6}
	\end{align}
	using the same arguments as in \ref{lem:1:d1b}. The order of the variance is
	\begin{align}
	\mathbb{V}\mathrm{ar}\Bigg[\frac{1}{n}\sum_{i=1}^n& L_h(\mathbf{x},\mathbf{X}_i)\mathbf{B}_{\mathbf{x}}^T(\mathbf{X}_i-\mathbf{x})(\mathbf{X}_i-\mathbf{x})^T\mathbf{B}_{\mathbf{x}}\Bigg]\nonumber\\
	\leq\,&\frac{c_{h,q}(L)^2h^{q+4}}{n}\int_{0}^{2h^{-2}} L^2\left(r\right)r^{\frac{q}{2}+1}(2-rh^2)^{\frac{q}{2}+1} \int_{\Omega_{q-1}} f\left(\mathbf{x}+\boldsymbol\alpha_{\mathbf{x},\boldsymbol\xi}\right)\left(\boldsymbol\xi\boldsymbol\xi^T\right)^2\,\omega_{q-1}(d\boldsymbol\xi)\,dt\nonumber\\
	=\,&\mathcal{O}\left(\frac{h^4}{nh^q}\mathbf{1}\mathbf{1}^T\right).\label{lem:1:7}
	\end{align}
	The desired result now holds by (\ref{lem:1:6}) and (\ref{lem:1:7}), as $\mathcal{O}_\mathbb{P}\Big(\frac{h^2}{\sqrt{nh^q}}\mathbf{1}\mathbf{1}^T\Big)=\mathpzc{o}_\mathbb{P}\big(h^2\mathbf{1}\mathbf{1}^T\big)$ by \ref{assump:a4}.
\end{proof}

\begin{lem}[Equivalent kernel]
	\label{coro:equiv}
	Under \ref{assump:a1}--\ref{assump:a4}, the projected local estimator $\hat m_{h,p}(\mathbf{x})=\linebreak\sum_{i=1}^n W_n^p\left(\mathbf{x},\mathbf{X}_i\right)Y_i$ for $p=0,1$ satisfies uniformly in $\mathbf{x}\in\Omega_q$:
	\[
	\hat m_{h,p}(\mathbf{x})=\sum_{i=1}^n \tilde{L}_h(\mathbf{x},\mathbf{X}_i)Y_i\left(1+\mathpzc{o}_\mathbb{P}\left(1\right)\right),\quad \tilde{L}_h\left(\mathbf{x},\mathbf{X}_i\right)=\frac{1}{nh^q\lambda_q(L)f(\mathbf{x})}L\left(\frac{1-\mathbf{x}^T\mathbf{X}_i}{h^2}\right).
	\]
\end{lem}

\begin{proof}[Proof of Lemma \ref{coro:equiv}]
	Note that $W_n^p\left(\mathbf{x},\mathbf{X}_i\right)=\mathbf{e}_1^T\big(\boldsymbol{\mathcal{X}}_{\mathbf{x},p}^T\boldsymbol{\mathcal{W}}_{\mathbf{x}}\boldsymbol{\mathcal{X}}_{\mathbf{x},p}\big)^{-1}\big(1, \delta_{p,1}(\mathbf{X}_i-\mathbf{x})^T\mathbf{B}_{\mathbf{x}}\big)^T$ $L_h(\mathbf{x},\mathbf{X}_i)$. The matrix $\boldsymbol{\mathcal{X}}_{\mathbf{x},p}^T\boldsymbol{\mathcal{W}}_{\mathbf{x}}\boldsymbol{\mathcal{X}}_{\mathbf{x},p}$ follows by \ref{lem:1:d0}, \ref{lem:1:d1b} and \ref{lem:1:d2b} of Lemma \ref{lem:1}:
	\begin{align*}
	n^{-1}\boldsymbol{\mathcal{X}}_{\mathbf{x},p}^T&\boldsymbol{\mathcal{W}}_{\mathbf{x}}\boldsymbol{\mathcal{X}}_{\mathbf{x},p}\nonumber\\
	=\,&\frac{1}{n} \sum_{i=1}^n \left(\begin{array}{cc}
	L_h(\mathbf{x},\mathbf{X}_i) &  L_h(\mathbf{x},\mathbf{X}_i)(\mathbf{X}_i-\mathbf{x})^T\mathbf{B}_{\mathbf{x}}\nonumber\\
	L_h(\mathbf{x},\mathbf{X}_i)\mathbf{B}_{\mathbf{x}}^T(\mathbf{X}_i-\mathbf{x}) &  L_h(\mathbf{x},\mathbf{X}_i)\mathbf{B}_{\mathbf{x}}^T(\mathbf{X}_i-\mathbf{x})(\mathbf{X}_i-\mathbf{x})^T\mathbf{B}_{\mathbf{x}}
	\end{array}\right)\nonumber\\
	=\,&\left(\begin{array}{cc}
	f(\mathbf{x}) & \frac{2b_q(L)}{q}\boldsymbol\nabla f(\mathbf{x})^T\mathbf{B}_{\mathbf{x}}h^2 \\
	\frac{2b_q(L)}{q}\mathbf{B}_{\mathbf{x}}^T\boldsymbol\nabla f(\mathbf{x})h^2 & \frac{2b_q(L)}{q}\mathbf{I}_q f(\mathbf{x})h^2
	\end{array}\right)+\mathpzc{o}_\mathbb{P}\left(\mathbf{1}\mathbf{1}^T\right).
	\end{align*}
	This matrix can be inverted by the inversion formula of a block matrix, resulting in
	\begin{align}
	\!\!\left(n^{-1}\boldsymbol{\mathcal{X}}_{\mathbf{x},p}^T\boldsymbol{\mathcal{W}}_{\mathbf{x}}\boldsymbol{\mathcal{X}}_{\mathbf{x},p}\right)^{-1}=\left(\begin{array}{cc}
	f(\mathbf{x})^{-1} & -f(\mathbf{x})^{-2}\boldsymbol\nabla f(\mathbf{x})^T\mathbf{B}_{\mathbf{x}} \\
	-f(\mathbf{x})^{-2}\mathbf{B}_{\mathbf{x}}^T\boldsymbol\nabla f(\mathbf{x}) & \left(\frac{2b_q(L)}{q}f(\mathbf{x})h^2\right)^{-1}\mathbf{I}_{q}
	\end{array}\right)+\mathpzc{o}_\mathbb{P}\left(\mathbf{1}\mathbf{1}^T\right).\label{theo:biasvar:4}
	\end{align}	
	Then, by expression (\ref{theo:biasvar:4}), uniformly in $\mathbf{x}\in\Omega_{q}$ it follows that
	\begin{align*}
	\hat m_{h,p}(\mathbf{x}) =&\,\frac{1}{nf(\mathbf{x})}\sum_{i=1}^nL_h(\mathbf{x},\mathbf{X}_i)Y_i(1+\mathpzc{o}_\mathbb{P}\left(1\right))\\
	&+\frac{\delta_{p,1}\boldsymbol\nabla f(\mathbf{x})^T\mathbf{B}_{\mathbf{x}}}{f(\mathbf{x})^2}\frac{1}{n}\sum_{i=1}^nL_h(\mathbf{x},\mathbf{X}_i)\mathbf{B}_{\mathbf{x}}^T(\mathbf{X}_i-\mathbf{x})Y_i(1+\mathpzc{o}_\mathbb{P}\left(1\right)).
	\end{align*}
	By (\ref{KDE}) and (\ref{chq}), the first addend is $\frac{1}{nh^q\lambda_q(L)f(\mathbf{x})}\sum_{i=1}^nL_h(\mathbf{x},\mathbf{X}_i)Y_i(1+\mathpzc{o}_\mathbb{P}\left(1\right))$. The second term is $\mathpzc{o}_\mathbb{P}\left(1\right)$ (see \ref{lem:1:d1c} in Lemma \ref{lem:1}) and negligible in comparison with the first one, which is $\mathcal{O}_\mathbb{P}\left(1\right)$. Then, it can be absorbed inside the factor $\left(1+\mathpzc{o}_\mathbb{P}\left(1\right)\right)$, proving the lemma.
\end{proof}

\begin{lem}
	\label{lem:5}
	Under \ref{assump:a1}--\ref{assump:a4} and \ref{assump:a7n}, for a random sample $\left\{(\mathbf{X}_i,Y_i)\right\}_{i=1}^n$ the following statements~hold: 
	\begin{enumerate}[label=\roman{*}., ref=\textit{\roman{*}}]
		\item  $\int_{\Omega_q}\left(\hat m_{h,p}(\mathbf{x})-\mathcal{L}_{h,p}m(\mathbf{x})\right)f(\mathbf{x})w(\mathbf{x})\,\omega_q(d\mathbf{x})=\mathcal{O}_\mathbb{P}\big(n^{-\frac{1}{2}}\big)$. \label{lem:5:4}
		
		\item $\int_{\Omega_q}\left(\sum_{i=1}^n \tilde{L}_h\left(\mathbf{x},\mathbf{X}_i\right)g(\mathbf{X}_i) \right)^2f(\mathbf{x})w(\mathbf{x})\,\omega_q(d\mathbf{x})=\int_{\Omega_q}g(\mathbf{x})^2 f(\mathbf{x})w(\mathbf{x})\,\omega_q(d\mathbf{x})(1+\mathpzc{o}\left(1\right))$\\ $+\mathcal{O}_\mathbb{P}\big((nh^q)^{-1}+n^{-\frac{1}{2}}\big)$.\label{lem:5:1}
		
		\item $\int_{\Omega_q}\sum_{i=1}^n\sum_{j=1}^n \tilde{L}_h\left(\mathbf{x},\mathbf{X}_i\right)\tilde{L}_h\left(\mathbf{x},\mathbf{X}_j\right)\sigma(\mathbf{X}_i)\varepsilon_ig(\mathbf{X}_j) f(\mathbf{x})w(\mathbf{x})\,\omega_q(d\mathbf{x})=\mathcal{O}_\mathbb{P}\big((nh^\frac{q}{2})^{-1}\big)$.\label{lem:5:2}
		
		\item $\int_{\Omega_q}\sum_{i=1}^n \left(\tilde{L}_h\left(\mathbf{x},\mathbf{X}_i\right)\sigma(\mathbf{X}_i)\varepsilon_i\right)^2f(\mathbf{x})w(\mathbf{x})\,\omega_q(d\mathbf{x})\!=\!\frac{\lambda_q(L^2)\lambda_q(L)^{-2}}{nh^{q}}\int_{\Omega_q}\!\sigma^2(\mathbf{x})w(\mathbf{x})\,\omega_q(d\mathbf{x})$\\$\times(1+\mathpzc{o}\left(1\right))+\mathcal{O}_\mathbb{P}\big((n^\frac{3}{2}h^q)^{-1}\big)$. \label{lem:5:3}
		
		\item $\mathbb{E}\big[W_{ijn}^2\big]=n^{-2}\nu^2\left(1+\mathpzc{o}\left(1\right)\right)$, $\mathbb{E}\left[W_{ijn}W_{jkn}W_{kln}W_{lin}\right]=\mathcal{O}\left(n^{-4}h^{2q}\right)$,  $\mathbb{E}\big[W_{ijn}^4\big]=\mathcal{O}\left((n^4h^{q})^{-1}\right)$, $\mathbb{E}\left[W_{ijn}W_{ikn}^2W_{jkn}\right]=\mathcal{O}\left(n^{-4}\right)$, where $\nu^2\equiv\nu_{\boldsymbol{\theta}_0}^2$ is given in Theorem \ref{theo:limdis}.\label{lem:5:5}
	\end{enumerate}
\end{lem}

\begin{proof}[Proof of Lemma \ref{lem:5}]
	
	\mbox{}\\\textit{Proof of \ref{lem:5:4}}. By Corollary \ref{coro:equiv},
	\begin{align*}
	\int_{\Omega_q}(\hat m_{h,p}(\mathbf{x})-&\mathcal{L}_{h,p}m(\mathbf{x}))f(\mathbf{x})w(\mathbf{x})\,\omega_q(d\mathbf{x})\\
	=&\,\int_{\Omega_q}\sum_{i=1}^n\tilde{L}_h(\mathbf{x},\mathbf{X}_i)(Y_i-m(\mathbf{X}_i))f(\mathbf{x})w(\mathbf{x})\,\omega_q(d\mathbf{x})\left(1+\mathpzc{o}_\mathbb{P}\left(1\right)\right).
	\end{align*}
	Using the properties of the conditional expectation, Fubini, relation (\ref{chq}) and Lemma \ref{lem:4}:
	\begin{align*}
	\mathbb{E}\Bigg[\int_{\Omega_q}&\sum_{i=1}^n\tilde{L}_h(\mathbf{x},\mathbf{X}_i)(Y_i-m(\mathbf{X}_i))f(\mathbf{x})w(\mathbf{x})\,\omega_q(d\mathbf{x})\Bigg]\\
	=&\,0,\\
	\mathbb{V}\mathrm{ar}\Bigg[\int_{\Omega_q}&\sum_{i=1}^n\tilde{L}_h(\mathbf{x},\mathbf{X}_i)(Y_i-m(\mathbf{X}_i))f(\mathbf{x})w(\mathbf{x})\,\omega_q(d\mathbf{x})\Bigg]\\
	=&\,\frac{1}{nh^{q}\lambda_q(L)}\int_{\Omega_q}\int_{\Omega_q}L\left(\frac{1-\mathbf{y}^T\mathbf{x}}{h^2}\right)\sigma^2\left(\mathbf{x}\right)w(\mathbf{x})w(\mathbf{y})\,\omega_q(d\mathbf{x})\,\omega_q(d\mathbf{y})\left(1+\mathpzc{o}\left(1\right)\right)\\
	=&\,\frac{1}{n}\int_{\Omega_q}\sigma^2\left(\mathbf{y}\right)w(\mathbf{y})^2\,\omega_q(d\mathbf{y})\left(1+\mathpzc{o}\left(1\right)\right)\\
	=&\,\mathcal{O}\left(n^{-1}\right).
	\end{align*}
	Then $\int_{\Omega_q}(\hat m_{h,p}(\mathbf{x})-\mathcal{L}_{h,p}m(\mathbf{x}))f(\mathbf{x})w(\mathbf{x})\,\omega_q(d\mathbf{x})=\mathcal{O}_\mathbb{P}\big(n^{-\frac{1}{2}}\big)(1+\mathpzc{o}_\mathbb{P}\left(1\right))=\mathcal{O}_\mathbb{P}\big(n^{-\frac{1}{2}}\big)$.\\
	
	\textit{Proof of \ref{lem:5:1}}. The integral can be split in two addends:
	\begin{align*}
	\int_{\Omega_q}\bigg(&\sum_{i=1}^n \tilde{L}_h\left(\mathbf{x},\mathbf{X}_i\right)g(\mathbf{X}_i) \bigg)^2f(\mathbf{x})w(\mathbf{x})\,\omega_q(d\mathbf{x})\\
	=&\,\frac{1}{n^2h^{2q}\lambda_q(L)^2}\sum_{i=1}^n\int_{\Omega_q} L^2\left(\frac{1-\mathbf{x}^T\mathbf{X}_i}{h^2}\right)\frac{g(\mathbf{X}_i)^2 w(\mathbf{x})}{f(\mathbf{x})}\,\omega_q(d\mathbf{x})\\
	&+\frac{1}{n^2h^{2q}\lambda_q(L)^2}\sum_{i\neq j}\int_{\Omega_q} L\left(\frac{1-\mathbf{x}^T\mathbf{X}_i}{h^2}\right)L\left(\frac{1-\mathbf{x}^T\mathbf{X}_j}{h^2}\right)\frac{g(\mathbf{X}_i)g(\mathbf{X}_j)w(\mathbf{x})}{f(\mathbf{x})} \,\omega_q(d\mathbf{x})\\
	=&\,I_{1}+I_{2}.
	\end{align*}
	Now, by applying Fubini, (\ref{chq}) and Lemma \ref{lem:4},
	\begin{align*}
	\mathbb{E}\left[I_1\right]=&\,\frac{1}{nh^{2q}\lambda_q(L)^2}\int_{\Omega_q}\mathbb{E}\left[L^2\left(\frac{1-\mathbf{x}^T\mathbf{X}}{h^2}\right)g(\mathbf{X})^2\right] \frac{w(\mathbf{x})}{f(\mathbf{x})}\,\omega_q(d\mathbf{x})\\
	=&\,\frac{1}{nh^{2q}\lambda_q(L)^2}\int_{\Omega_q}\int_{\Omega_q} L^2\left(\frac{1-\mathbf{x}^T\mathbf{y}}{h^2}\right)\frac{g(\mathbf{y})^2 w(\mathbf{x})}{f(\mathbf{x})}f(\mathbf{y})\,\omega_q(d\mathbf{y})\,\omega_q(d\mathbf{x})\\
	=&\,\frac{\lambda_q(L^2)}{nh^{q}\lambda_q(L)^2}\int_{\Omega_q}g(\mathbf{x})^2 w(\mathbf{x})\,\omega_q(d\mathbf{x})(1+\mathpzc{o}\left(1\right))\\
	=&\,\mathcal{O}\left((nh^q)^{-1}\right),\\
	\mathbb{V}\mathrm{ar}\left[I_1\right]\leq&\,\frac{1}{n^3h^{4q}\lambda_q(L)^4}\mathbb{E}\left[\bigg(\int_{\Omega_q}L^2\left(\frac{1-\mathbf{x}^T\mathbf{X}}{h^2}\right)\frac{g(\mathbf{X})^2w(\mathbf{x})}{f(\mathbf{x})}\,\omega_q(d\mathbf{x})\bigg)^2\right]\\
	=&\,\frac{\lambda_q(L^2)^2}{n^3h^{2q}\lambda_q(L)^4}\int_{\Omega_q}\frac{g(\mathbf{y})^2w(\mathbf{y})^2 }{f(\mathbf{y})}\,\omega_q(d\mathbf{y})(1+\mathpzc{o}\left(1\right))\\
	=&\,\mathcal{O}\left((n^3h^{2q})^{-1}\right)
	\end{align*}
	and therefore $I_1=\mathcal{O}_\mathbb{P}\left((nh^q)^{-1}\right)$. On the other hand, by Lemma \ref{lem:4} and the independence of $\mathbf{X}_i$ and $\mathbf{X}_j$ if $i\neq j$:
	\begin{align*}
	\mathbb{E}\left[I_2\right]=&\,\frac{1-n^{-1}}{h^{2q}\lambda_q(L)^2}\int_{\Omega_q}\mathbb{E}\left[L\left(\frac{1-\mathbf{x}^T\mathbf{X}}{h^2}\right)g(\mathbf{X})\right]^2 \frac{w(\mathbf{x})}{f(\mathbf{x})}\,\omega_q(d\mathbf{x})\\
	=&\,\left(1-n^{-1}\right)\int_{\Omega_q}g(\mathbf{x})^2 f(\mathbf{x})w(\mathbf{x})\,\omega_q(d\mathbf{x})(1+\mathpzc{o}\left(1\right))\\
	=&\,\int_{\Omega_q}g(\mathbf{x})^2 f(\mathbf{x})w(\mathbf{x})\,\omega_q(d\mathbf{x})(1+\mathpzc{o}\left(1\right)),\\
	\mathbb{E}\left[I_2^2\right]=&\, \frac{1}{n^4h^{4q}\lambda_q(L)^4}\sum_{i\neq j}\sum_{k\neq l} \int_{\Omega_q}\int_{\Omega_q}\mathbb{E}\Bigg[ L\left(\frac{1-\mathbf{x}^T\mathbf{X}_i}{h^2}\right)L\left(\frac{1-\mathbf{x}^T\mathbf{X}_j}{h^2}\right)L\left(\frac{1-\mathbf{y}^T\mathbf{X}_k}{h^2}\right)\\
	&\times L\left(\frac{1-\mathbf{y}^T\mathbf{X}_l}{h^2}\right) g(\mathbf{X}_i)g(\mathbf{X}_j)g(\mathbf{X}_k)g(\mathbf{X}_l) \Bigg] \frac{w(\mathbf{x})w(\mathbf{y})}{f(\mathbf{x})f(\mathbf{y})} \,\omega_q(d\mathbf{x})\,\omega_q(d\mathbf{y})\\
	=&\,\mathcal{O}\left((n^2h^{2q})^{-1}\right) \int_{\Omega_q}\int_{\Omega_q}L^2\left(\frac{1-\mathbf{y}^T\mathbf{x}}{h^2}\right)g(\mathbf{x})^4f(\mathbf{x}) \frac{w(\mathbf{x})w(\mathbf{y})}{f(\mathbf{y})}\,\omega_q(d\mathbf{x})\,\omega_q(d\mathbf{y})\\
	&+\mathcal{O}\left((nh^{q})^{-1}\right)\int_{\Omega_q}\int_{\Omega_q}L\left(\frac{1-\mathbf{y}^T\mathbf{x}}{h^2}\right)g(\mathbf{x})^3g(\mathbf{y})f(\mathbf{x})w(\mathbf{x})w(\mathbf{y})\,\omega_q(d\mathbf{x})\,\omega_q(d\mathbf{y})\\ 
	&+\left(1-\mathcal{O}\left(n^{-1}\right)\right) \mathbb{E}\left[I_2\right]^2\\
	=&\, \mathcal{O}\left((n^2h^{q})^{-1}\right)+\mathcal{O}\left(n^{-1}\right)+\left(1-\mathcal{O}\left(n^{-1}\right)\right) \mathbb{E}\left[I_2\right]^2.
	\end{align*}
	Then $\mathbb{V}\mathrm{ar}\left[I_2\right]=\mathbb{E}\left[I_2^2\right]-\mathbb{E}\left[I_2\right]^2=\mathcal{O}\left(n^{-1}\right)$ and $I_2=\int_{\Omega_q}g(\mathbf{x})^2 f(\mathbf{x})w(\mathbf{x})\,\omega_q(d\mathbf{x})(1+\mathpzc{o}\left(1\right))+\mathcal{O}_\mathbb{P}\big(n^{-\frac{1}{2}}\big)$. Finally, 
	\[
	I_1+I_2=\int_{\Omega_q}g(\mathbf{x})^2 f(\mathbf{x})w(\mathbf{x})\,\omega_q(d\mathbf{x})(1+\mathpzc{o}\left(1\right))+\mathcal{O}_\mathbb{P}\left((nh^q)^{-1}+n^{-\frac{1}{2}}\right).
	\]
	
	\textit{Proof of \ref{lem:5:2}}. By the tower property of the conditional expectation and $\mathbb{E}\left[\varepsilon|\mathbf{X}\right]=0$, the expectation is zero. By the independence between $\varepsilon$'s and $\mathbb{E}\left[\varepsilon^2|\mathbf{X}\right]=1$, the variance is
	\begin{align*}
	\mathbb{V}\mathrm{ar}\Bigg[\int_{\Omega_q}&\sum_{i=1}^n\sum_{j=1}^n \tilde{L}_h\left(\mathbf{x},\mathbf{X}_i\right)\tilde{L}_h\left(\mathbf{x},\mathbf{X}_j\right)\varepsilon_ig(\mathbf{X}_j) f(\mathbf{x})w(\mathbf{x})\,\omega_q(d\mathbf{x})\Bigg]\\
	=&\,\frac{1}{n^4h^{4q}\lambda_q(L)^4}\sum_{i,j,l=1}^n\int_{\Omega_q}\int_{\Omega_q}\mathbb{E}\Bigg[ L\left(\frac{1-\mathbf{x}^T\mathbf{X}_i}{h^2}\right)L\left(\frac{1-\mathbf{x}^T\mathbf{X}_j}{h^2}\right) L\left(\frac{1-\mathbf{y}^T\mathbf{X}_i}{h^2}\right)\\
	&\times L\left(\frac{1-\mathbf{y}^T\mathbf{X}_l}{h^2}\right)
	g(\mathbf{X}_j)g(\mathbf{X}_l)\Bigg] \frac{w(\mathbf{x})w(\mathbf{y})}{f(\mathbf{x})f(\mathbf{y})}\,\omega_q(d\mathbf{x})\,\omega_q(d\mathbf{y})\\
	=&\,\frac{1}{n^4h^{4q}\lambda_q(L)^4}\left\{I_1+I_2+I_3+I_4\right\},
	\end{align*}
	where, by repeated use of Lemma \ref{lem:4}: $I_1=\mathcal{O}\left(nh^{2q}\right)$, $I_2=\mathcal{O}\left(n^2h^{2q}\right)$, $I_3=\mathcal{O}\left(n^2h^{3q}\right)$ and $I_4=\mathcal{O}\left(n^3h^{4q}\right)$.
	Because $\mathcal{O}\left((n^3h^{2q})^{-1}+n^{-2}+(n^2h^{q})^{-1}+n^{-1}\right)=\mathcal{O}\left((n^2h^{q})^{-1}\right)$ by \ref{assump:a4}, it follows that
	\[
	\int_{\Omega_q}\sum_{i=1}^n\sum_{j=1}^n \tilde{L}_h\left(\mathbf{x},\mathbf{X}_i\right)\tilde{L}_h\left(\mathbf{x},\mathbf{X}_j\right)\sigma(\mathbf{X}_i)\varepsilon_ig(\mathbf{X}_j) f(\mathbf{x})w(\mathbf{x})\,\omega_q(d\mathbf{x})=\mathcal{O}_\mathbb{P}\Big(\big(nh^\frac{q}{2}\big)^{-1}\Big).
	\]
	
	\textit{Proof of \ref{lem:5:3}}. Let us denote $I=\int_{\Omega_q}\sum_{i=1}^n \left(\tilde{L}_h\left(\mathbf{x},\mathbf{X}_i\right)\sigma(\mathbf{X}_i)\varepsilon_i\right)^2f(\mathbf{x})w(\mathbf{x})\,\omega_q(d\mathbf{x})$. By the unit conditional variance of $\varepsilon$ and the boundedness of $\mathbb{E}\left[\varepsilon^4|\mathbf{X}\right]$,
	\begin{align*}
	\mathbb{E}[I]=&\,\sum_{i=1}^n\int_{\Omega_q}\mathbb{E}\left[\left(\tilde{L}_h\left(\mathbf{x},\mathbf{X}_i\right)\sigma(\mathbf{X}_i)\right)^2\mathbb{E}\left[\varepsilon_i^2|\mathbf{X}_i\right]\right]f(\mathbf{x})w(\mathbf{x})\,\omega_q(d\mathbf{x})\\
	=&\,\frac{\lambda_q(L^2)\lambda_q(L)^{-2}}{nh^{q}}\int_{\Omega_q} \sigma^2(\mathbf{x})w(\mathbf{x})\,\omega_q(d\mathbf{x})(1+\mathpzc{o}\left(1\right)),\\
	\mathbb{E}\left[I^2\right]=&\,\sum_{i=1}^n\sum_{j=1}^n\int_{\Omega_q}\int_{\Omega_q}\mathbb{E}\left[\left(\tilde{L}_h\left(\mathbf{x},\mathbf{X}_i\right)\sigma(\mathbf{X}_i)\tilde{L}_h\left(\mathbf{y},\mathbf{X}_j\right)\sigma(\mathbf{X}_j)\right)^2\mathbb{E}\left[\varepsilon_i^2\varepsilon_j^2|\mathbf{X}_i,\mathbf{X}_j\right]\right]\\
	&\times f(\mathbf{x})f(\mathbf{y})w(\mathbf{x})w(\mathbf{y})\,\omega_q(d\mathbf{x})\,\omega_q(d\mathbf{y})\\
	=&\,\mathcal{O}\left((n^3h^{2q})^{-1}\right)\int_{\Omega_q}\frac{\sigma^4(\mathbf{x})w(\mathbf{x})^2}{f(\mathbf{x})}\,\omega_q(d\mathbf{x})+\left(1-\mathcal{O}\left(n^{-1}\right)\right)\mathbb{E}\left[I\right]^2\\
	=&\,\mathcal{O}\left((n^3h^{2q})^{-1}\right)+\left(1-\mathcal{O}\left(n^{-1}\right)\right)\mathbb{E}\left[I\right]^2.
	\end{align*}
	Then $\mathbb{V}\mathrm{ar}\left[I\right]=\mathcal{O}\left((n^3h^{2q})^{-1}\right)-\mathcal{O}\left(n^{-1}\right)\mathbb{E}\left[I\right]^2=\mathcal{O}\left((n^3h^{2q})^{-1}\right)$ and as a consequence
	\[
	I=\frac{\lambda_q(L^2)\lambda_q(L)^{-2}}{nh^{q}}\int_{\Omega_q} \sigma^2(\mathbf{x})w(\mathbf{x})\,\omega_q(d\mathbf{x})(1+\mathpzc{o}\left(1\right))+\mathcal{O}_\mathbb{P}\Big(\big(n^\frac{3}{2}h^q\big)^{-1}\Big).
	\]
	
	\textit{Proof of \ref{lem:5:5}}. The computation of
	\begin{align}
	\mathbb{E}\left[W_{ijn}^2\right] =&\,\frac{1}{n^2h^{3q}\lambda_q(L)^4}\int_{\Omega_q}\int_{\Omega_q}\left[\int_{\Omega_q} L\left(\frac{1-\mathbf{x}^T\mathbf{z}}{h^2}\right)L\left(\frac{1-\mathbf{y}^T\mathbf{z}}{h^2}\right)\sigma^2(\mathbf{z})f(\mathbf{z})\,\omega_q(d\mathbf{z})\right]^2\nonumber\\
	&\times\frac{w(\mathbf{x})w(\mathbf{y})}{f(\mathbf{x})f(\mathbf{y})}\,\omega_q(d\mathbf{x})\,\omega_q(d\mathbf{y})\label{lem:5:6:1}
	\end{align}
	is split in the cases where $q\geq 2$ and $q=1$. For the first one, the usual change of variables given by Lemma \ref{lem:2} is applied:
	\begin{align}
	\mathbf{y}=s\mathbf{x}+(1-s^2)^\frac{1}{2}\mathbf{B}_{\mathbf{x}}\boldsymbol\xi,\quad\omega_{q}(d\mathbf{y})=(1-s^2)^{\frac{q}{2}-1}\,\omega_{q-1}(d\boldsymbol\xi)\,ds.\label{lem:5:6:2}
	\end{align}
	Because $q\geq2$, it is possible also to consider an extra change of variables:
	\begin{align}
	\begin{split}
	\mathbf{z}&=t\mathbf{x}+\tau \mathbf{B}_{\mathbf{x}}\boldsymbol\xi+(1-t^2-\tau^2)^\frac{1}{2}\mathbf{B}_{\mathbf{x}}\mathbf{A}_{\boldsymbol\xi}\boldsymbol\eta,\\
	\omega_{q}(d\mathbf{z})&=(1-t^2-\tau^2)^\frac{q-3}{2}\,\omega_{q-2}(d\boldsymbol\eta)\,dt\,d\tau,
	\end{split}
	\label{lem:5:6:3}
	\end{align}
	where $t,\tau\in(-1,1)$, $t^2+\tau^2<1$, $\boldsymbol\eta\in\Omega_{q-2}$ and $\mathbf{A}_{\boldsymbol\xi}=(\mathbf{a}_1,\ldots,\mathbf{a}_q)_{q\times (q-1)}$ is the semi-orthonormal matrix resulting from the completion of $\boldsymbol\xi$ to the orthonormal basis $\left\{\boldsymbol\xi,\mathbf{a}_1,\ldots,\mathbf{a}_{q-1}\right\}$ of $\mathbb{R}^{q}$. This change of variables is obtained by a recursive use of Lemma \ref{lem:2}:
	\begin{align*}
	\int_{\Omega_q}f(\mathbf{z})\,\omega_q(d\mathbf{z})=&\,\int_{-1}^{1}\int_{\Omega_{q-1}} f\left(t\mathbf{x}+(1-t^2)^\frac{1}{2}\mathbf{B}_{\mathbf{x}}\boldsymbol\xi'\right) (1-t^2)^{\frac{q}{2}-1}\,\omega_{q-1}(d\boldsymbol\xi')\,dt\\
	=&\,\int_{-1}^{1}\int_{-1}^{1}\int_{\Omega_{q-2}} f\left(t\mathbf{x}+(1-t^2)^\frac{1}{2}\mathbf{B}_{\mathbf{x}}\left(s\boldsymbol\xi+(1-s^2)^\frac{1}{2}\mathbf{A}_{\boldsymbol\xi}\boldsymbol\eta\right)\right)\\ &\times(1-s^2)^{\frac{q-3}{2}}(1-t^2)^{\frac{q}{2}-1}\omega_{q-2}(d\boldsymbol\eta)\,ds\,dt\\
	=&\,\iint_{t^2+\tau^2<1}\int_{\Omega_{q-2}} f\left(t\mathbf{x}+\tau\mathbf{B}_{\mathbf{x}}\boldsymbol\xi+(1-t^2-\tau^2)^{\frac{1}{2}}\mathbf{B}_{\mathbf{x}}\mathbf{A}_{\boldsymbol\xi}\boldsymbol\eta\right)\\
	&\times \left(1-\tau^2(1-t^2)^{-1}\right)^{\frac{q-3}{2}}(1-t^2)^{\frac{q-3}{2}}\omega_{q-2}(d\boldsymbol\eta)\,d\tau\,dt\\
	=&\,\iint_{t^2+\tau^2<1}\int_{\Omega_{q-2}} f\left(t\mathbf{x}+\tau\mathbf{B}_{\mathbf{x}}\boldsymbol\xi+(1-t^2-\tau^2)^\frac{1}{2}\mathbf{B}_{\mathbf{x}}\mathbf{A}_{\boldsymbol\xi}\boldsymbol\eta\right)\\
	&\times (1-t^2-\tau^2)^{\frac{q-3}{2}}\,\omega_{q-2}(d\boldsymbol\eta)\,d\tau\,dt,
	\end{align*}
	where in the third equality a change of variables $\tau=(1-t^2)^\frac{1}{2}s$ is used. The matrix $\mathbf{B}_{\mathbf{x}}\mathbf{A}_{\boldsymbol\xi}$ of dimension $(q+1)\times(q-1)$ can be interpreted as the one formed by the column vectors that complete the orthonormal set $\left\{\mathbf{x},\mathbf{B}_{\mathbf{x}}\boldsymbol\xi\right\}$ to an orthonormal basis in $\mathbb{R}^{q+1}$.\\
	
	If the changes of variables (\ref{lem:5:6:2}) and (\ref{lem:5:6:3}) is applied first, after that the changes $r=\frac{1-s}{h^2}$ and 
	\[
	\left\{\begin{array}{l}
	\rho=\frac{1-t}{h^2}, \\
	\theta=\tau\left[h\left(\rho(2-h^2\rho)\right)^\frac{1}{2}\right]^{-1},
	\end{array}\right.
	\quad \left|\frac{\partial(t,\tau)}{\partial(\rho,\theta)}\right|=h^3\left[\rho(2-h^2\rho)\right]^\frac{1}{2}
	\]
	are used and, denoting
	\begin{align*}
	\boldsymbol\alpha_{\mathbf{x},\boldsymbol\xi}&=-rh^2\mathbf{x}+\left[rh^2(2-rh^2)\right]^\frac{1}{2}\mathbf{B}_{\mathbf{x}}\boldsymbol\xi,\\
	\boldsymbol\beta_{\mathbf{x},\boldsymbol\xi}&=-h^2\rho\mathbf{x}+h\left[\rho(2-h^2\rho)\right]^\frac{1}{2}\left[\theta \mathbf{B}_{\mathbf{x}}\boldsymbol\xi+(1-\theta^2)^\frac{1}{2}\mathbf{B}_{\mathbf{x}}\mathbf{A}_{\boldsymbol\xi}\boldsymbol\eta\right],
	\end{align*}
	then the following result is obtained employing the DCT (see Lemma 4 of \cite{Garcia-Portugues:clt} for technical details in a similar situation):
	\begin{align*}
	(\mathrm{\ref{lem:5:6:1}})=&\,\frac{1}{n^2h^{3q}\lambda_q(L)^4}\int_{\Omega_q}\int_{\Omega_q}\Bigg[\iint_{t^2+\tau^2<1}\int_{\Omega_{q-2}} \nonumber\\
	&\times L\left(\frac{1-t}{h^2}\right)L\Bigg(\frac{1-\mathbf{y}^T\big(t\mathbf{x}+\tau \mathbf{B}_{\mathbf{x}}\boldsymbol\xi+(1-t^2-\tau^2)^\frac{1}{2}\mathbf{B}_{\mathbf{x}}\mathbf{A}_{\boldsymbol\xi}\boldsymbol\eta\big)}{h^2}\Bigg)\nonumber\\
	&\times\sigma^2\left(t\mathbf{x}+\tau \mathbf{B}_{\mathbf{x}}\boldsymbol\xi+(1-t^2-\tau^2)^\frac{1}{2}\mathbf{B}_{\mathbf{x}}\mathbf{A}_{\boldsymbol\xi}\boldsymbol\eta\right)\nonumber\\
	&\times f\left(t\mathbf{x}+\tau \mathbf{B}_{\mathbf{x}}\boldsymbol\xi+(1-t^2-\tau^2)^\frac{1}{2}\mathbf{B}_{\mathbf{x}}\mathbf{A}_{\boldsymbol\xi}\boldsymbol\eta\right)(1-t^2-\tau^2)^\frac{q-3}{2}\,\omega_{q-2}(d\boldsymbol\eta)\,dt\,d\tau\Bigg]^2\nonumber\\
	&\times\frac{w(\mathbf{x})w(\mathbf{y})}{f(\mathbf{x})f(\mathbf{y})}\,\omega_q(d\mathbf{x})\,\omega_q(d\mathbf{y})\nonumber\\
	=&\,\frac{1}{n^2h^{3q}\lambda_q(L)^4}\int_{-1}^1\int_{\Omega_{q-1}}\int_{\Omega_q}\Bigg[\iint_{t^2+\tau^2<1}\int_{\Omega_{q-2}} L\left(\frac{1-t}{h^2}\right)L\left(\frac{1-st-\tau(1-s^2)^\frac{1}{2}}{h^2}\right)\nonumber\\
	&\times\sigma^2\left(t\mathbf{x}+\tau \mathbf{B}_{\mathbf{x}}\boldsymbol\xi+(1-t^2-\tau^2)^\frac{1}{2}\mathbf{B}_{\mathbf{x}}\mathbf{A}_{\boldsymbol\xi}\boldsymbol\eta\right)\nonumber\\
	&\times f\left(t\mathbf{x}+\tau \mathbf{B}_{\mathbf{x}}\boldsymbol\xi+(1-t^2-\tau^2)^\frac{1}{2}\mathbf{B}_{\mathbf{x}}\mathbf{A}_{\boldsymbol\xi}\boldsymbol\eta\right)(1-t^2-\tau^2)^\frac{q-3}{2}\,\omega_{q-2}(d\boldsymbol\eta)\,dt\,d\tau\Bigg]^2\nonumber\\
	&\times\frac{w(\mathbf{x})w\Big(s\mathbf{x}+(1-s^2)^\frac{1}{2}\mathbf{B}_{\mathbf{x}}\boldsymbol\xi\Big)}{f(\mathbf{x})f\Big(s\mathbf{x}+(1-s^2)^\frac{1}{2}\mathbf{B}_{\mathbf{x}}\boldsymbol\xi\Big)}\,\omega_q(d\mathbf{x})(1-s^2)^{\frac{q}{2}-1}\,\omega_{q-1}(d\boldsymbol\xi)\,ds \nonumber\\
	=&\,\frac{1}{n^2\lambda_q(L)^4}\int_{0}^{2h^{-2}}\int_{\Omega_{q-1}}\int_{\Omega_q}\Bigg[\int_0^{2h^{-2}}\int_{-1}^1\int_{\Omega_{q-2}} L\left(\rho\right)\nonumber\\
	&\times L\left(r+\rho-h^2r\rho-\theta\left[r\rho(2-h^2r)(2-h^2\rho)\right]^\frac{1}{2}\right)\sigma^2\left(\mathbf{x}+\boldsymbol\beta_{\mathbf{x},\boldsymbol\xi,\boldsymbol\eta}\right)f\left(\mathbf{x}+\boldsymbol\beta_{\mathbf{x},\boldsymbol\xi,\boldsymbol\eta}\right)\nonumber\\
	&\times(1-\theta^2)^\frac{q-3}{2}\rho^{\frac{q}{2}-1}(2-h^2\rho)^{\frac{q}{2}-1} \,\omega_{q-2}(d\boldsymbol\eta)\,dt\,d\tau\Bigg]^2\frac{w(\mathbf{x})w\big(\mathbf{x}+\boldsymbol\alpha_{\mathbf{x},\boldsymbol\xi}\big)}{f(\mathbf{x})f\big(\mathbf{x}+\boldsymbol\alpha_{\mathbf{x},\boldsymbol\xi}\big)}\nonumber\\
	&\times \,\omega_q(d\mathbf{x})\,r^{\frac{q}{2}-1}(2-h^2r)^{\frac{q}{2}-1}\,\omega_{q-1}(d\boldsymbol\xi)\,dr\nonumber\\
	=&\,\frac{(1+\mathpzc{o}\left(1\right))}{n^2\lambda_q(L)^4}\int_{0}^{\infty}\int_{\Omega_{q-1}}\int_{\Omega_q}\Bigg[\int_0^{\infty}\int_{-1}^1\int_{\Omega_{q-2}} L\left(\rho\right)L\left(r+\rho-2\theta(r\rho)^\frac{1}{2}\right)\sigma^2\left(\mathbf{x}\right)f(\mathbf{x})\\
	&\times(1-\theta^2)^\frac{q-3}{2}\rho^{\frac{q}{2}-1}2^{\frac{q}{2}-1}\omega_{q-2}(d\boldsymbol\eta)\,dt\,d\tau\Bigg]^2\frac{w\left(\mathbf{x}\right)^2}{f(\mathbf{x})^2}\,\omega_q(d\mathbf{x})\,r^{\frac{q}{2}-1}2^{\frac{q}{2}-1}\,\omega_{q-1}(d\boldsymbol\xi)\,dr\\
	=&\,\left(1+\mathpzc{o}\left(1\right)\right)\frac{\omega_{q-1}\omega_{q-2}^22^{\frac{3q}{2}-3}}{n^2\lambda_q(L)^4}\int_{\Omega_q}\sigma^4\left(\mathbf{x}\right)w(\mathbf{x})^2\,\omega_q(d\mathbf{x})\\
	&\times\int_{0}^{\infty}r^{\frac{q}{2}-1}\left\{\int_0^{\infty} \rho^{\frac{q}{2}-1}L\left(\rho\right)\int_{-1}^1 (1-\theta^2)^\frac{q-3}{2}L\left(r+\rho-2\theta(r\rho)^\frac{1}{2}\right)\,d\theta\,d\rho\right\}^2\,dr\\
	=&\,n^{-2}\nu^2\left(1+\mathpzc{o}\left(1\right)\right).
	\end{align*}
	
	For $q=1$, define the change of variables:
	\begin{align*}
	\mathbf{y}=&\,s\mathbf{x}+(1-s^2)^\frac{1}{2}\mathbf{B}_{\mathbf{x}}\boldsymbol\xi,\quad\omega_{1}(d\mathbf{y})=(1-s^2)^{-\frac{1}{2}}\,\omega_{0}(d\boldsymbol\xi)\,ds,\\
	\mathbf{z}=&\,t\mathbf{x}+(1-t^2)^\frac{1}{2}\mathbf{B}_{\mathbf{x}}\boldsymbol\eta,\quad\omega_{1}(d\mathbf{z})=(1-t^2)^{-\frac{1}{2}}\,\omega_{0}(d\boldsymbol\eta)\,dt,
	\end{align*}
	where $\boldsymbol\xi,\boldsymbol\eta\in\Omega_{0}=\left\{-1,1\right\}$. Note that as $q=1$ and $\mathbf{x}^T (\mathbf{B}_{\mathbf{x}}\boldsymbol\xi)=\mathbf{x}^T (\mathbf{B}_{\mathbf{x}}\boldsymbol\eta)=0$, then necessarily $\mathbf{B}_{\mathbf{x}}\boldsymbol\xi=\mathbf{B}_{\mathbf{x}}\boldsymbol\eta$ or $\mathbf{B}_{\mathbf{x}}\boldsymbol\xi=-\mathbf{B}_{\mathbf{x}}\boldsymbol\eta$. These changes of variables are applied first, later $\rho=\frac{1-t}{h^2}$ and finally $r=\frac{1-s}{h^2}$, using that:
	\begin{align*}
	&\frac{1-st-(1-s^2)^\frac{1}{2}(1-t^2)^\frac{1}{2}(\mathbf{B}_{\mathbf{x}}\boldsymbol\xi)^T\mathbf{B}_{\mathbf{x}}\boldsymbol\eta}{h^2}\\
	&\qquad\qquad\qquad\qquad=r+\rho-h^2r\rho-\left(r\rho(2-h^2r)(2-h^2\rho)\right)^\frac{1}{2}(\mathbf{B}_{\mathbf{x}}\boldsymbol\xi)^T\mathbf{B}_{\mathbf{x}}\boldsymbol\eta.
	\end{align*}
	Finally, considering
	\begin{align*}
	\boldsymbol\alpha_{\mathbf{x},\boldsymbol\xi}=-rh^2\mathbf{x}+\left[rh^2(2-rh^2)\right]^\frac{1}{2}\mathbf{B}_{\mathbf{x}}\boldsymbol\xi,\quad\boldsymbol\beta_{\mathbf{x},\boldsymbol\eta}=-\rho h^2\mathbf{x}+\left[\rho h^2(2-\rho h^2)\right]^\frac{1}{2}\mathbf{B}_{\mathbf{x}}\boldsymbol\eta,
	\end{align*}
	it follows by the use of the DCT:
	\begin{align*}
	(\mathrm{\ref{lem:5:6:1}})=&\,\frac{1}{n^2h^{3}\lambda_q(L)^4}\int_{\Omega_{1}}\int_{\Omega_{1}}\Bigg[\int_{-1}^1\int_{\Omega_{0}} L\left(\frac{1-t}{h^2}\right)L\Bigg(\frac{1-\mathbf{y}^T\big(t\mathbf{x}+(1-t^2)^\frac{1}{2}\mathbf{B}_{\mathbf{x}}\boldsymbol\eta\big)}{h^2}\Bigg)\nonumber\\
	&\times\sigma^2\left(t\mathbf{x}+(1-t^2)^\frac{1}{2}\mathbf{B}_{\mathbf{x}}\boldsymbol\eta\right) f\left(t\mathbf{x}+(1-t^2)^\frac{1}{2}\mathbf{B}_{\mathbf{x}}\boldsymbol\eta\right)(1-t^2)^{-\frac{1}{2}}\,\omega_{0}(d\boldsymbol\eta)\,dt\Bigg]^2\nonumber\\
	&\times\frac{w(\mathbf{x})w(\mathbf{y})}{f(\mathbf{x})f(\mathbf{y})}\,\omega_{1}(d\mathbf{x})\,\omega_{1}(d\mathbf{y})\nonumber\\
	=&\,\frac{1}{n^2h^{3}\lambda_q(L)^4}\int_{-1}^1\int_{\Omega_{0}}\int_{\Omega_{1}}\Bigg[\int_{-1}^1\int_{\Omega_{0}} \nonumber\\
	&\times L\left(\frac{1-t}{h^2}\right)L\left(\frac{1-st-(1-t^2)^\frac{1}{2}(1-s^2)^\frac{1}{2}(\mathbf{B}_{\mathbf{x}}\boldsymbol\xi)^T(\mathbf{B}_{\mathbf{x}}\boldsymbol\eta)}{h^2}\right)\nonumber\\
	&\times \sigma^2\left(t\mathbf{x}+(1-t^2)^\frac{1}{2}\mathbf{B}_{\mathbf{x}}\boldsymbol\xi\right)f\left(t\mathbf{x}+(1-t^2)^\frac{1}{2}\mathbf{B}_{\mathbf{x}}\boldsymbol\xi\right)(1-t^2)^{-\frac{1}{2}}\,\omega_{0}(d\boldsymbol\eta)\,dt\Bigg]^2\nonumber\\
	&\times\frac{w(\mathbf{x})w\Big(s\mathbf{x}+(1-s^2)^\frac{1}{2}\mathbf{B}_{\mathbf{x}}\boldsymbol\xi\Big)}{f(\mathbf{x})f\Big(s\mathbf{x}+(1-s^2)^\frac{1}{2}\mathbf{B}_{\mathbf{x}}\boldsymbol\xi\Big)} \,\omega_{1}(d\mathbf{x})\,(1-s^2)^{-\frac{1}{2}}\,\omega_{0}(d\boldsymbol\xi)\,ds\nonumber\\
	=&\,\frac{1}{n^2\lambda_q(L)^4}\int_{0}^{2h^{-2}}\int_{\Omega_{0}}\int_{\Omega_{1}}\Bigg[\int_{0}^{2h^{-2}}\int_{\Omega_{0}}\nonumber\\
	&\times L\left(\rho\right)L\left(r+\rho-h^2r\rho-\left(r\rho(2-h^2r)(2-h^2\rho)\right)^\frac{1}{2}(\mathbf{B}_{\mathbf{x}}\boldsymbol\xi)^T\mathbf{B}_{\mathbf{x}}\boldsymbol\eta\right)\nonumber\\
	&\times \sigma^2\left(\mathbf{x}+\boldsymbol\beta_{\mathbf{x},\boldsymbol\eta}\right)f\left(\mathbf{x}+\boldsymbol\beta_{\mathbf{x},\boldsymbol\eta}\right)\rho^{-\frac{1}{2}}(2-h^2\rho)^{-\frac{1}{2}}\,\omega_{0}(d\boldsymbol\eta)\,d\rho\Bigg]^2\frac{w(\mathbf{x})w\big(\mathbf{x}+\boldsymbol\alpha_{\mathbf{x},\boldsymbol\xi}\big)}{f(\mathbf{x})f\big(\mathbf{x}+\boldsymbol\alpha_{\mathbf{x},\boldsymbol\xi}\big)}\nonumber\\
	&\times \,\omega_{1}(d\mathbf{x})\,r^{-\frac{1}{2}}(2-h^2r)^{-\frac{1}{2}}\,\omega_{0}(d\boldsymbol\xi)\,dr\nonumber\\
	=&\,\frac{2^{-1}\left(1+\mathpzc{o}\left(1\right)\right)}{n^2\lambda_q(L)^4}\int_{0}^{\infty}\int_{\Omega_{0}}\int_{\Omega_{1}}\Bigg[\int_{0}^{\infty}\int_{\Omega_{0}} L\left(\rho\right)L\left(r+\rho-2\left(r\rho\right)^\frac{1}{2}(\mathbf{B}_{\mathbf{x}}\boldsymbol\xi)^T\mathbf{B}_{\mathbf{x}}\boldsymbol\eta\right)\nonumber\\
	&\times \sigma^2\left(\mathbf{x}\right)f\left(\mathbf{x}\right)\rho^{-\frac{1}{2}}\,\omega_{0}(d\boldsymbol\eta)\,d\rho\Bigg]^2\frac{w(\mathbf{x})^2}{f(\mathbf{x})^2} \,\omega_{1}(d\mathbf{x})\,r^{-\frac{1}{2}}\,\omega_{0}(d\boldsymbol\xi)\,dr\\
	=&\,\frac{\omega_{0}2^{-\frac{3}{2}}\left(1+\mathpzc{o}\left(1\right)\right)}{n^2\lambda_q(L)^4}\int_{\Omega_{1}}\sigma^4\left(\mathbf{x}\right)w(\mathbf{x})^2 \,\omega_{1}(d\mathbf{x})\\
	&\times \int_{0}^{\infty}r^{-\frac{1}{2}}\bigg\{\int_{0}^{\infty} \rho^{-\frac{1}{2}}L\left(\rho\right) \left[L\left(r+\rho-2\left(r\rho\right)^\frac{1}{2}\right)+L\left(r+\rho+2\left(r\rho\right)^\frac{1}{2}\right)\right] \,d\rho\bigg\}^2\,\,dr\\
	=&\,n^{-2}\nu^2\left(1+\mathpzc{o}\left(1\right)\right).
	\end{align*}
	
	The rest of the results are provided by the recursive use of Lemma \ref{lem:4}, bearing in mind that the indexes are pairwise different:
	\begin{align*}
	\mathbb{E}\big[W_{ijn}^4&\big]\\
	=&\,\frac{n^4h^{2q}}{n^8h^{8q}\lambda_q(L)^8}\int_{\Omega_q}\times\stackrel{4}{\cdots}\times\int_{\Omega_q}\mathbb{E}\Bigg[\prod_{k=1}^4L\left(\frac{1-\mathbf{x}_k^T\mathbf{X}}{h^2}\right)\sigma^4(\mathbf{X})\mathbb{E}\left[\varepsilon^4|\mathbf{X}\right]\Bigg]^2\prod_{k=1}^4\frac{w(\mathbf{x}_k)}{f(\mathbf{x}_k)}\,\omega_q(d\mathbf{x}_k)\\
	=&\,\mathcal{O}\left((n^{4}h^{4q})^{-1}\right)\int_{\Omega_q}\times\stackrel{4}{\cdots}\times\int_{\Omega_q}\prod_{k=2}^4L^2\left(\frac{1-\mathbf{x}_k^T\mathbf{X}}{h^2}\right)\sigma^8(\mathbf{x}_1)f(\mathbf{x}_1)\prod_{k=1}^8\frac{w(\mathbf{x}_k)}{f(\mathbf{x}_k)}\,\omega_q(d\mathbf{x}_k)\\
	=&\,\mathcal{O}\left((n^{4}h^{q})^{-1}\right),\\
	\mathbb{E}\Big[W_{ijn}&W_{jkn}W_{kln}W_{lin}\Big]\\
	=&\,\frac{n^4h^{2q}}{n^8h^{8q}\lambda_q(L)^8}\int_{\Omega_q}\times\stackrel{4}{\cdots}\times\int_{\Omega_q}
	\mathbb{E}\left[L\left(\frac{1-\mathbf{x}_1^T\mathbf{X}}{h^2}\right)\!L\left(\frac{1-\mathbf{x}_4^T\mathbf{X}}{h^2}\right)\!\sigma^2(\mathbf{X})\right]\\
	&\times\mathbb{E}\left[L\left(\frac{1-\mathbf{x}_1^T\mathbf{X}}{h^2}\right)\!L\left(\frac{1-\mathbf{x}_2^T\mathbf{X}}{h^2}\right)\!\sigma^2(\mathbf{X})\right]\!\mathbb{E}\left[L\left(\frac{1-\mathbf{x}_2^T\mathbf{X}}{h^2}\right)\!L\left(\frac{1-\mathbf{x}_3^T\mathbf{X}}{h^2}\right)\!\sigma^2(\mathbf{X})\right]\\
	&\times\mathbb{E}\left[L\left(\frac{1-\mathbf{x}_3^T\mathbf{X}}{h^2}\right)\!L\left(\frac{1-\mathbf{x}_4^T\mathbf{X}}{h^2}\right)\!\sigma^2(\mathbf{X})\right]\prod_{k=1}^8\frac{w(\mathbf{x}_k)}{f(\mathbf{x}_k)}\,\omega_q(d\mathbf{x}_k)\\
	=&\,\mathcal{O}\left((n^{4}h^{2q})^{-1}\right)\int_{\Omega_q}\times\stackrel{4}{\cdots}\times\int_{\Omega_q}L\left(\frac{1-\mathbf{x}_4^T\mathbf{x}_1}{h^2}\right)L\left(\frac{1-\mathbf{x}_2^T\mathbf{x}_1}{h^2}\right)L\left(\frac{1-\mathbf{x}_2^T\mathbf{x}_3}{h^2}\right)\\
	&\times L\left(\frac{1-\mathbf{x}_4^T\mathbf{x}_3}{h^2}\right)\sigma^4(\mathbf{x}_1)\sigma^4(\mathbf{x}_3)\frac{f(\mathbf{x}_1)f(\mathbf{x}_3) }{f(\mathbf{x}_2)f(\mathbf{x}_3)}\prod_{k=1}^4w(\mathbf{x}_k)\,\omega_q(d\mathbf{x}_k)\\
	=&\,\mathcal{O}\left(n^{-4}h^{2q}\right),\\
	\mathbb{E}\Big[W_{ijn}&W^2_{ikn}W_{jkn}\Big]\\
	=&\,\frac{n^4h^{2q}}{n^8h^{8q}\lambda_q(L)^8}\int_{\Omega_q}\times\stackrel{4}{\cdots}\times\int_{\Omega_q}
	\mathbb{E}\left[L\left(\frac{1-\mathbf{x}_1^T\mathbf{X}}{h^2}\right)L\left(\frac{1-\mathbf{x}_2^T\mathbf{X}}{h^2}\right)\sigma^2(\mathbf{X})\right]\\
	&\times\mathbb{E}\left[L\left(\frac{1-\mathbf{x}_1^T\mathbf{X}}{h^2}\right)L\left(\frac{1-\mathbf{x}_3^T\mathbf{X}}{h^2}\right)L\left(\frac{1-\mathbf{x}_4^T\mathbf{X}}{h^2}\right)\sigma^3(\mathbf{X})\mathbb{E}\left[\varepsilon^3|\mathbf{X}\right]\right]^2 \prod_{k=1}^4\frac{w(\mathbf{x}_k)}{f(\mathbf{x}_k)}\,\omega_q(d\mathbf{x}_k)\\
	=&\,\mathcal{O}\left((n^{4}h^{3q})^{-1}\right)\int_{\Omega_q}\times\stackrel{4}{\cdots}\times\int_{\Omega_q} L\left(\frac{1-\mathbf{x}_2^T\mathbf{x}_1}{h^2}\right)L^2\left(\frac{1-\mathbf{x}_3^T\mathbf{x}_1}{h^2}\right)L^2\left(\frac{1-\mathbf{x}_4^T\mathbf{x}_1}{h^2}\right)\\
	&\times\sigma^8(\mathbf{x}_1)\frac{f(\mathbf{x}_1)^2}{f(\mathbf{x}_2)f(\mathbf{x}_3)f(\mathbf{x}_4)}\prod_{k=1}^4w(\mathbf{x}_k)\,\omega_q(d\mathbf{x}_k)\\
	=&\,\mathcal{O}\left(n^{-4}\right).
	\end{align*}
\end{proof}

\begin{lem}
	\label{lem:6}
	Under \ref{assump:a1}--\ref{assump:a6n} and \ref{assump:a9n}, for a random sample $\left\{(\mathbf{X}_i,Y_i)\right\}_{i=1}^n$ the following statements~hold: 
	\begin{enumerate}[label=\roman{*}., ref=\textit{\roman{*}}]
		
		\item $\int_{\Omega_q}\sum_{i=1}^n \left(W_n^{p}\left(\mathbf{x},\mathbf{X}_i\right)\hat\varepsilon_iV_i^*\right)^2\hat f_h(\mathbf{x})w(\mathbf{x})\,\omega_q(d\mathbf{x})=\frac{\lambda_q(L^2)\lambda_q(L)^{-2}}{nh^q}\int_{\Omega_q}\sigma_{\boldsymbol{\theta}_1}^2(\mathbf{x})w(\mathbf{x})\,\omega_q(d\mathbf{x})$\\$\times(1+\mathpzc{o}_\mathbb{P}\left(1\right))+\mathcal{O}_\mathbb{P^*}\big((n^\frac{3}{2}h^q)^{-1}\big)$. \label{lem:6:1}
		
		\item $2n^2h^q\sum_{i\neq j} \big[\int_{\Omega_q}W_n^p\left(\mathbf{x},\mathbf{X}_i\right)W_n^p\left(\mathbf{x},\mathbf{X}_j\right)\hat\varepsilon_i\hat\varepsilon_j \hat f_h(\mathbf{x})w(\mathbf{x})\,\omega_q(d\mathbf{x})\big]^2=2\nu_{\boldsymbol{\theta}_1}^2(1+\mathpzc{o}_\mathbb{P}\left(1\right))$,\\ $\mathbb{E}^*\big[W_{ijn}^*W_{jkn}^*W_{kln}^*W_{lin}^*\big]=\mathcal{O}_\mathbb{P}\left(n^{-4}h^{2q}\right)$, $\mathbb{E}^*\big[W_{ijn}^{*4}\big]=\mathcal{O}_\mathbb{P}\left((n^4h^{q})^{-1}\right)$ and $\mathbb{E}^*\big[W_{ijn}^*W_{ikn}^{*2}W_{jkn}^*]$ $=\mathcal{O}_\mathbb{P}\left(n^{-4}\right)$. \label{lem:6:2}
		
	\end{enumerate}
\end{lem}

\begin{proof}[Proof of Lemma \ref{lem:6}]
	\mbox{}\\\textit{Proof of \ref{lem:6:1}}.
	Using that the $V_i^*$'s are iid and independent with respect to the sample, 
	\begin{align}
	\mathbb{E}^*\Bigg[\int_{\Omega_q}\sum_{i=1}^n W_n^{p}&(\mathbf{x},\mathbf{X}_i)^2\hat\varepsilon_i^{\,2}V_i^{*2} \hat f_h(\mathbf{x})w(\mathbf{x})\,\omega_q(d\mathbf{x})\Bigg]\nonumber\\
	=&\,\int_{\Omega_q}\sum_{i=1}^nW_n^{p}(\mathbf{x},\mathbf{X}_i)^2(Y_i-m_{\hat{\boldsymbol{\theta}}}(\mathbf{X}_i))^2\hat f_h(\mathbf{x})w(\mathbf{x})\,\omega_q(d\mathbf{x})\nonumber\\
	=&\,\int_{\Omega_q}\sum_{i=1}^n\tilde{L}_h(\mathbf{x},\mathbf{X}_i)^2(Y_i-m_{\boldsymbol{\theta}_1}(\mathbf{X}_i))^2 f(\mathbf{x})w(\mathbf{x})\,\omega_q(d\mathbf{x})(1+\mathpzc{o}_\mathbb{P}\left(1\right))\label{lem:6:1:1}
	\end{align}
	where the last equality holds because by assumptions \ref{assump:a5n} and \ref{assump:a6n}, $m_{\hat{\boldsymbol{\theta}}}(\mathbf{x})-m_{\boldsymbol{\theta}_1}(\mathbf{x})=\mathcal{O}_\mathbb{P}\big(n^{-\frac{1}{2}}\big)$ uniformly in $\mathbf{x}\in\Omega_q$. By applying the tower property of the conditional expectation as in \ref{lem:1:d1c} from Lemma \ref{lem:1}, it is easy to derive from \ref{lem:5:3} in Lemma \ref{lem:5} that
	\begin{align*}
	(\mathrm{\ref{lem:6:1:1}})=&\,\frac{\lambda_q(L^2)\lambda_q(L)^{-2}}{nh^{q}}\int_{\Omega_q}\sigma^2_{\boldsymbol{\theta}_1}(\mathbf{x})w(\mathbf{x})\,\omega_q(d\mathbf{x})(1+\mathpzc{o}_\mathbb{P}\left(1\right)).
	\end{align*}
	The order of the variance is obtained applying the same idea, i.e., first deriving the variance with respect to the $V_i^*$'s and then applying the order computation given in the proof of \ref{lem:5:3} in Lemma \ref{lem:5} (adapted via the conditional expectation):
	\begin{align*}
	\mathbb{V}\mathrm{ar}^*\Bigg[\int_{\Omega_q}\sum_{i=1}^n&\tilde{L}_h(\mathbf{x},\mathbf{X}_i)^2(Y_i-m_{\hat{\boldsymbol{\theta}}}(\mathbf{X}_i))^2V_i^{*2} f(\mathbf{x})w(\mathbf{x})\,\omega_q(d\mathbf{x})\Bigg]\\
	=&\,\sum_{i=1}^n\bigg(\int_{\Omega_q}\tilde{L}_h(\mathbf{x},\mathbf{X}_i)^2(Y_i-m_{\hat{\boldsymbol{\theta}}}(\mathbf{X}_i))^2f(\mathbf{x})w(\mathbf{x})\,\omega_q(d\mathbf{x})\bigg)^2\mathbb{V}\mathrm{ar}^*\left[V_i^{*2}\right]\\
	=&\,\mathcal{O}\left((n^4h^{4q})^{-1}\right)\sum_{i=1}^n\Bigg(\int_{\Omega_q}L^2\left(\frac{1-\mathbf{x}^T\mathbf{X}_i}{h^2}\right)(Y_i-m_{\boldsymbol{\theta}_1}(\mathbf{X}_i))^2\frac{w(\mathbf{x})}{f(\mathbf{x})}\,\omega_q(d\mathbf{x})\Bigg)^2\\
	=&\,\mathcal{O}_\mathbb{P}\left((n^3h^{2q})^{-1}\right).
	\end{align*}
	The statement holds by Chebychev's inequality with respect to the probability law $\mathbb{P}^*$.\\
	
	\textit{Proof of \ref{lem:6:2}}. 
	First, by Corollary \ref{coro:equiv}, the expansion for the kernel density estimate and the fact $m_{\hat{\boldsymbol{\theta}}}(\mathbf{x})-m_{\boldsymbol{\theta}_1}(\mathbf{x})=\mathcal{O}_\mathbb{P}\big(n^{-\frac{1}{2}}\big)$ uniformly in $\mathbf{x}\in\Omega_q$, 
	\[
	I_n=2n^2h^q\sum_{i\neq j} \left[\int_{\Omega_q} W_n^p\left(\mathbf{x},\mathbf{X}_i\right)W_n^p\left(\mathbf{x},\mathbf{X}_j\right)\hat\varepsilon_i\hat\varepsilon_j \hat f_h(\mathbf{x})w(\mathbf{x})\,\omega_q(d\mathbf{x})\right]^2=2\sum_{i\neq j} I_{ijn}(1+\mathpzc{o}_\mathbb{P}\left(1\right)),
	\]
	where
	\begin{align*}
	I_{ijn}=&\,n^2h^q\int_{\Omega_q}\int_{\Omega_q} \tilde{L}_h\left(\mathbf{x},\mathbf{X}_i\right)\tilde{L}_h\left(\mathbf{x},\mathbf{X}_j\right) \tilde{L}_h\left(\mathbf{y},\mathbf{X}_i\right)\tilde{L}_h\left(\mathbf{y},\mathbf{X}_j\right)\\
	&\times (Y_i-m_{\boldsymbol{\theta}_1}(\mathbf{X}_i))^2(Y_j-m_{\boldsymbol{\theta}_1}(\mathbf{X}_j))^2f(\mathbf{x})f(\mathbf{y})w(\mathbf{x})w(\mathbf{y})\,\omega_q(d\mathbf{x})\,\omega_q(d\mathbf{y}).
	\end{align*}
	By the tower property of the conditional expectation and \ref{lem:5:3} in Lemma \ref{lem:5}, $\mathbb{E}\left[I_{ijn}\right]=\mathbb{E}\big[W^2_{ijn}\big]=n^{-2}\nu_{\boldsymbol{\theta}_1}^2(1+\mathpzc{o}\left(1\right))$ (considering that the $W_{ijn}$'s are defined with respect to $\boldsymbol{\theta}_1$ instead of $\boldsymbol{\theta}_0$). To prove that $I_n\stackrel{p}{\longrightarrow}2\nu_{\boldsymbol{\theta}_1}^2$, consider $\widetilde I_n=2\sum_{i\neq j} I_{ijn}$ and, by (\ref{VWn}) and (\ref{EWn4}), 
	\begin{align*}
	\mathbb{V}\mathrm{ar}\left[\widetilde I_{n}\right]=&\,\mathbb{E}\bigg[\Big(2\sum_{i\neq j}I_{ijn}\Big)^2\bigg]-4n^2(n-1)^2\mathbb{E}\left[I_{ijn}\right]^2\\
	&\,=4\sum_{i\neq j}\sum_{k\neq l}\mathbb{E}\left[W_{ijn}^2W_{kln}^2\right]-4n^2(n-1)^2\mathbb{E}\left[W_{ijn}^2\right]^2\\
	&\,=\frac{1}{3}\mathbb{E}\left[W_n^4\right]-\mathbb{V}\mathrm{ar}\left[W_{n}\right]^2+\mathpzc{o}\left(1\right)\\
	&\,=\mathbb{V}\mathrm{ar}\left[W_{n}\right]^2\left(\frac{1}{3}\mathbb{V}\mathrm{ar}\left[W_{n}\right]^{-2}\mathbb{E}\left[W_n^4\right]-1\right)+\mathpzc{o}\left(1\right)\\
	&\,=2\nu_{\boldsymbol{\theta}_1}^2(1+\mathpzc{o}\left(1\right))\mathpzc{o}\left(1\right)+\mathpzc{o}\left(1\right)\\
	&\,=\mathpzc{o}\left(1\right),
	\end{align*}
	because, as was shown in the proof of Theorem \ref{theo:limdis}, conditions \textit{b)} and \textit{d)} hold. Then, $\widetilde I_n-\mathbb{E}\big[\widetilde I_n\big]$ converges to zero in squared mean, which implies that it converges in probability and therefore
	\[
	I_n=\widetilde I_n(1+\mathpzc{o}_\mathbb{P}\left(1\right))=\left(\widetilde I_n-\mathbb{E}\big[\widetilde I_n\big]+2\nu^2_{\boldsymbol{\theta}_1}+\mathpzc{o}\left(1\right)\right)(1+\mathpzc{o}_\mathbb{P}\left(1\right))=2\nu^2_{\boldsymbol{\theta}_1}+\mathpzc{o}_\mathbb{P}\left(1\right),
	\]
	which proofs the first statement.\\
	
	Second, it follows straightforwardly that $\mathbb{E}^*\big[W_{ijn}^{*4}\big]=\mathcal{O}_\mathbb{P}\big(W_{ijn}^{4}\big)$, $\mathbb{E}^*\big[W_{ijn}^*W_{jkn}^*W_{kln}^*W_{lin}^*\big]=$\linebreak $\mathcal{O}_\mathbb{P}\left(W_{ijn}W_{jkn}W_{kln}W_{lin}\right)$ and $\mathbb{E}^*\big[W_{ijn}^*W_{ikn}^{*2}W_{jkn}^*\big]=\mathcal{O}_\mathbb{P}\left(W_{ijn}W_{ikn}^{2}W_{jkn}\right)$. The idea now is to use that, for a rv $X_n$ and by Markov's inequality, $X_n=\mathbb{E}\left[X_n\right]+\mathcal{O}_\mathbb{P}\left(\mathbb{E}\left[|X_n|\right]\right)$. The expectations of the variables are given in \ref{lem:5:5} from Lemma \ref{lem:5}. The orders of the absolute expectations are the same: in the definition of $W_{ijn}$ the only factor with sign is $\varepsilon_i\varepsilon_j$, which is handled by the assumption of boundedness of $\mathbb{E}\left[|\varepsilon|^3|\mathbf{X}\right]$. Therefore, $W_{ijn}^{4}=\mathcal{O}_\mathbb{P}\left((n^4h^{q})^{-1}\right)$, $W_{ijn}W_{jkn}W_{kln}W_{lin}=\mathcal{O}_\mathbb{P}\left(n^{-4}h^{2q}\right)$ and $W_{ijn}W_{ikn}^{2}W_{jkn}=\mathcal{O}_\mathbb{P}\left(n^{-4}\right)$, so the statement is proved.
\end{proof}

\section{Empirical evidence of the asymptotic distribution}
\label{appendix:empev}

\begin{figure}[b!]
	\centering
	\vspace{-0.25cm}
	\includegraphics[width=0.5\textwidth]{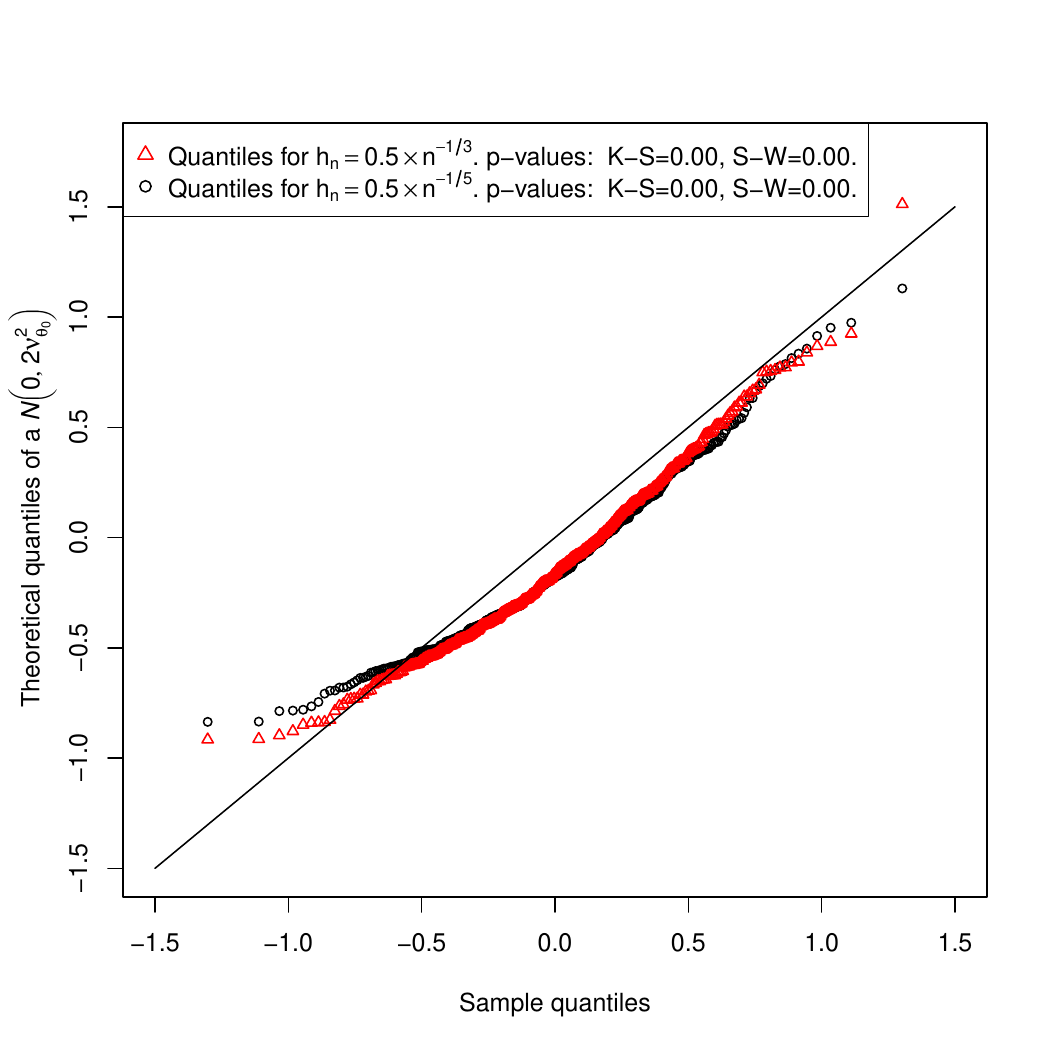}\includegraphics[width=0.5\textwidth]{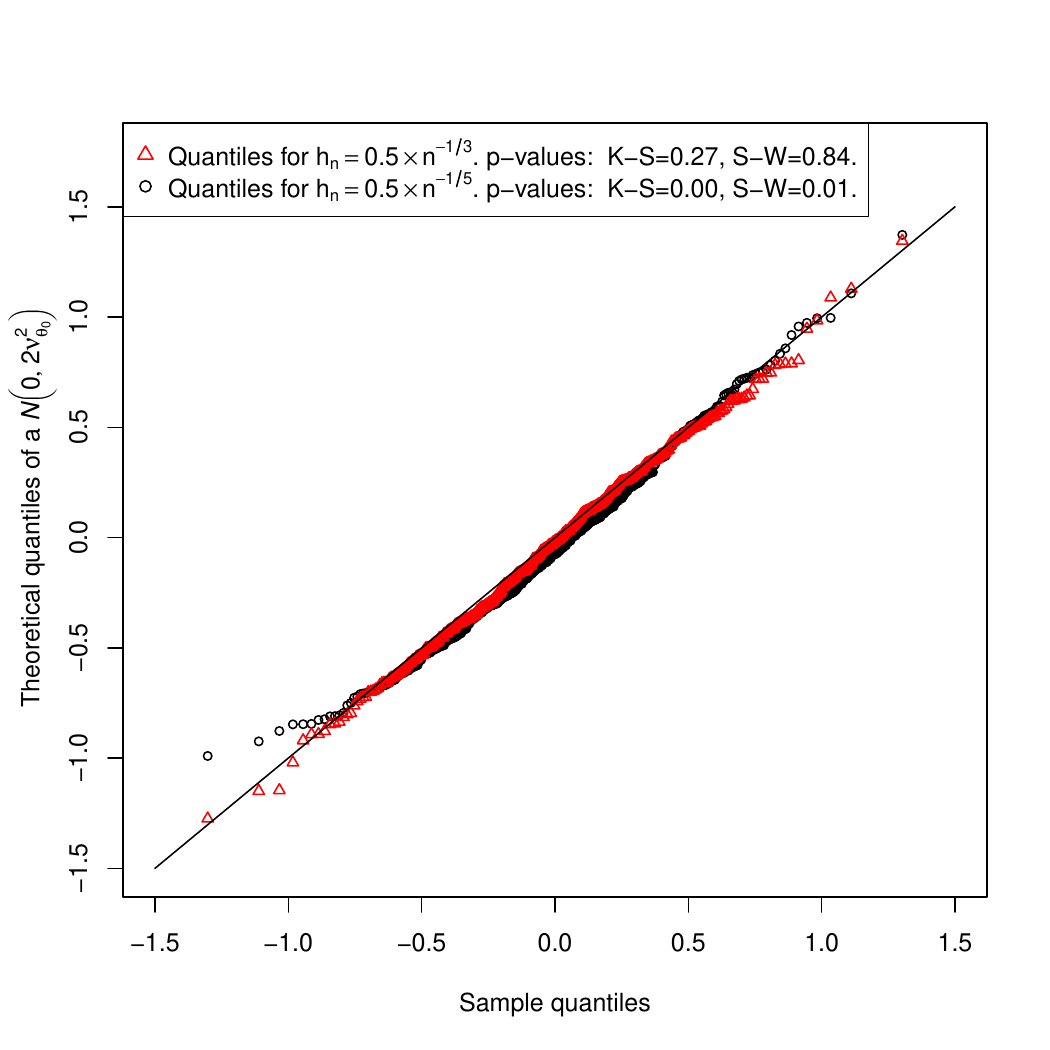}
	\vspace{-0.5cm}
	\caption{\small QQ-plot comparing the quantiles of the asymptotic distribution given by Theorem \ref{theo:limdis} with the sample quantiles for $\big\{nh^{\frac{1}{2}}\big(T^j_n-\frac{\sqrt{\pi}}{4}nh\big)\big\}_{j=1}^{500}$ with $n=10^2$ (left) and $n=5\times 10^5$ (right). }
	\label{fig:asymp}
\end{figure}

To illustrate the effective convergence of the statistic to the asymptotic distribution, a simple numerical experiment is provided. The regression setting is the model $Y=c+{\varepsilon}$, with $c=1$, ${\varepsilon}\sim\mathcal{N}(0,\sigma^2)$, $\sigma^2=\frac{1}{2}$ and $\mathbf{X}$ uniformly distributed on the circle ($q=1$). The composite hypothesis $H_0: m\equiv c$, for $c\in\mathbb{R}$ unknown (test for no effect), is checked using the local constant estimator ($p=0$) with von Mises kernel and considering the weight function $w\equiv1$. Figure \ref{fig:asymp} presents two QQ-plots computed from samples $\big\{nh^{\frac{1}{2}}\big(T^j_n-\frac{\sqrt{\pi}}{4}nh\big)\big\}_{j=1}^{500}$ obtained for different sample sizes $n$. Two bandwidth sequences $h_n=\frac{1}{2}\times n^{-r}$, $r=\frac{1}{3},\frac{1}{5}$ are chosen to illustrate the effect of the bandwidths in the convergence to the asymptotic distribution, and, specifically, that the effect of undersmoothing boosts the convergence since the bias is mitigated. The Kolmogorov-Smirnov (K-S) and Shapiro-Wilk (S-W) tests are applied on to measure how close the empirical distribution of the test statistic is to a $\mathcal{N}\big(0,2\nu_{\boldsymbol{\theta}_0}^2\big)$ and to normality, respectively.

\section{Further information on the simulation study}
\label{appendix:simus}

The densities employed for the directional predictor $\mathbf{X}$ are taken from the models in \cite{Garcia-Portugues:exact} and are included in Table \ref{tab:dens} for the sake of completeness. Their graphical representations are shown in Figure \ref{fig:dens}. These densities aim to capture simple designs like the uniform and more challenging ones with regions of low density in the support.

\begin{table}[H]
	\scriptsize
	\centering
	\setlength{\tabcolsep}{3pt}
	\begin{tabular}{cp{3.8cm}l}
		\toprule\toprule
		Model & Description & Density \\[0.05cm]\midrule
		M1 & Uniform & $\omega_{q}^{-1}$\\[0.05cm]
		M4 & Projected normal, non \phantom{aaa} rotationally symmetric & $\mathrm{PN}((1,\mathbf{0}),2\boldsymbol\Sigma_1)$ \\[0.05cm]
		M12 & Mixture of PN and DC & $\frac{3}{4}\mathrm{PN}((1,\mathbf{0}),\boldsymbol\Sigma_1)+\frac{1}{4}\mathrm{DC}\big(\big(\frac{1}{2},\frac{\sqrt{3}}{2},\mathbf{0}\big),50\big)$\\[0.05cm]
		M16 & Double small circle & $\frac{1}{2}\mathrm{SC}((\mathbf{0},1),10)+\frac{1}{2}\mathrm{SC}((1,\mathbf{0}),10)$\\[0.05cm]
		M20 & Windmill (4 vM)& \\[0.05cm]
		& $q=1$ & $\frac{2}{11}\mathrm{vM}\left((0,1),20\right)+\frac{1}{11}\sum_{i=1}^3\mathrm{vM}\left(\rho_1\left(\frac{2i\pi}{3}\right),15\right)$\\[0.05cm]
		& $q>1$ & $\frac{2}{11}\mathrm{vM}\left((\mathbf{0}_q,1),20\right)+\frac{1}{11}\sum_{i=1}^3\sum\limits_{j\in\{3,5,6\}}\mathrm{vM}\left(\left(\rho_2\left(\frac{2i\pi}{3},\frac{\pi}{j}\right),\mathbf{0}\right),15\right)$\\[0.05cm]
		\bottomrule\bottomrule
	\end{tabular}
	\caption{\small Directional densities considered in the simulation study.  $\rho_1(\theta)=(\cos(\theta),\sin(\theta))$ and  $\rho_2(\theta,\phi)=(\cos(\theta)\sin(\phi),\sin(\theta)\sin(\phi),\cos(\phi))$, with $\theta\in[0,2\pi),\,\phi\in[0,\pi)$. $\boldsymbol\Sigma_1$ is such that the first three elements of $\mathrm{diag}(\boldsymbol\Sigma_1)$ are $\frac{1}{2},\,\frac{1}{4},\,\frac{1}{8}$ and the rest of them are $1$. vM, PN, SC and DC stands for von Mises, Projected Normal, Small Circle and Directional Cauchy, respectively, all densities with a location parameter (first argument) and a concentration/dispersion parameter (second argument). \label{tab:dens}}
\end{table}

\begin{figure}[h!]
	\centering
	\includegraphics[width=0.225\textwidth]{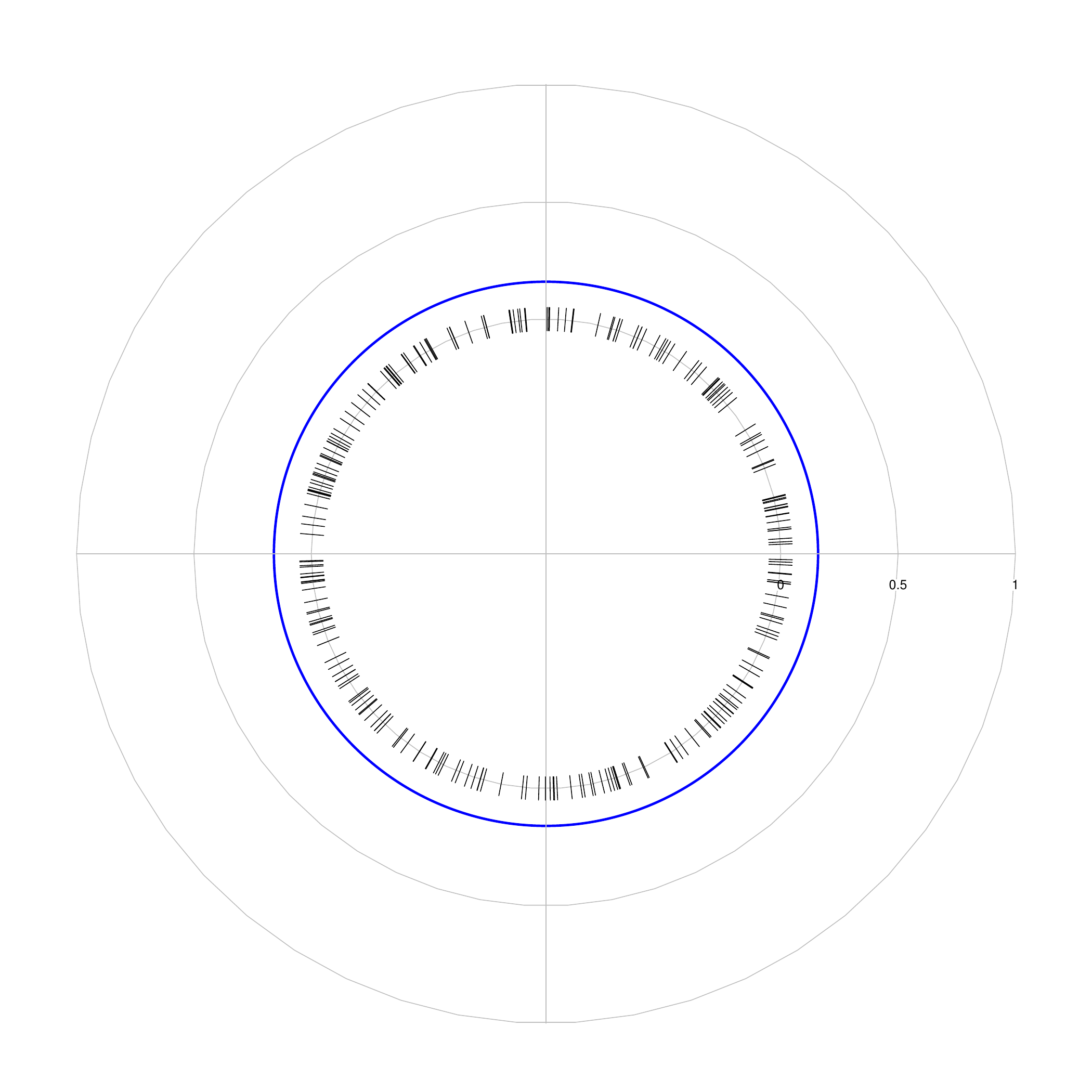}
	\includegraphics[width=0.225\textwidth]{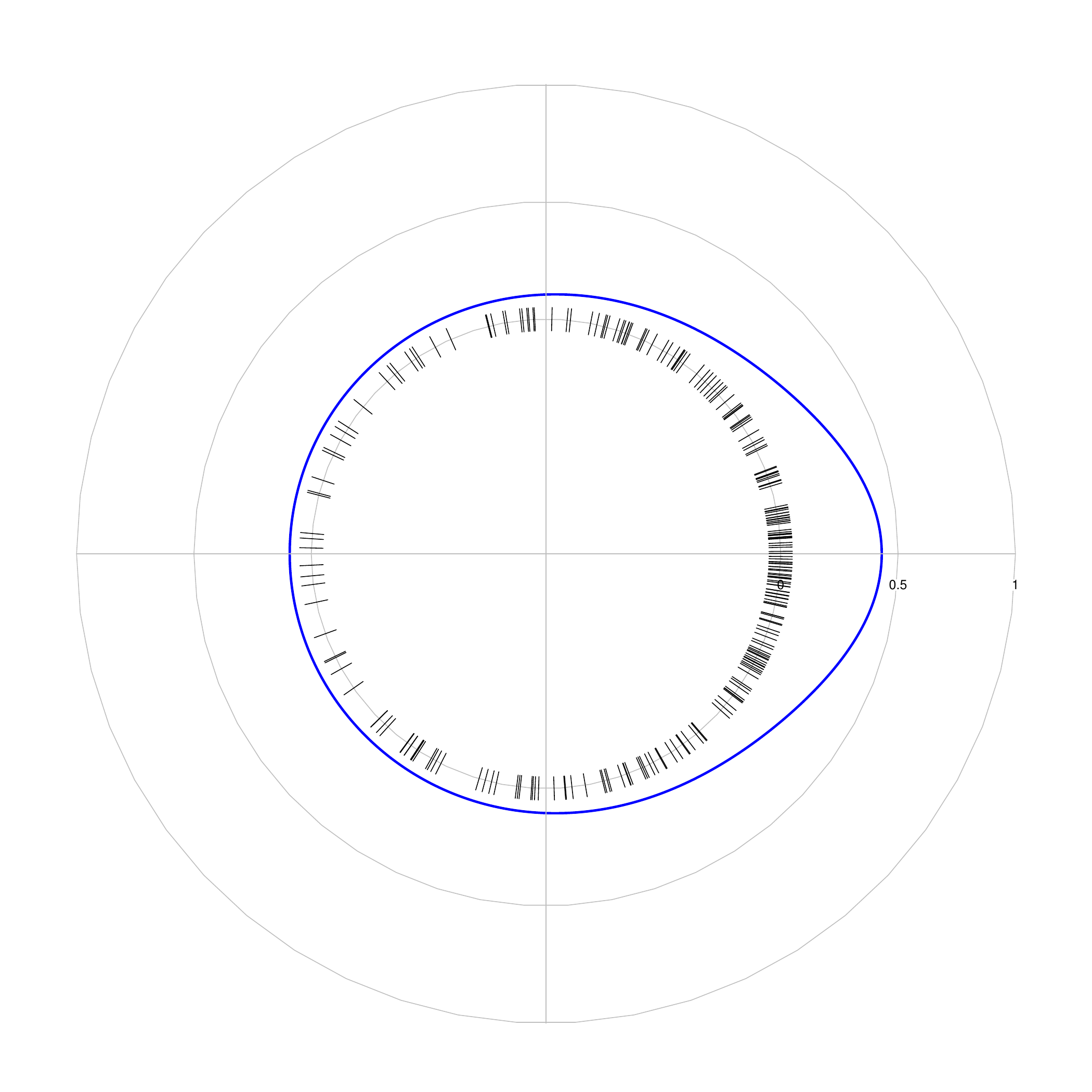}
	\includegraphics[width=0.225\textwidth]{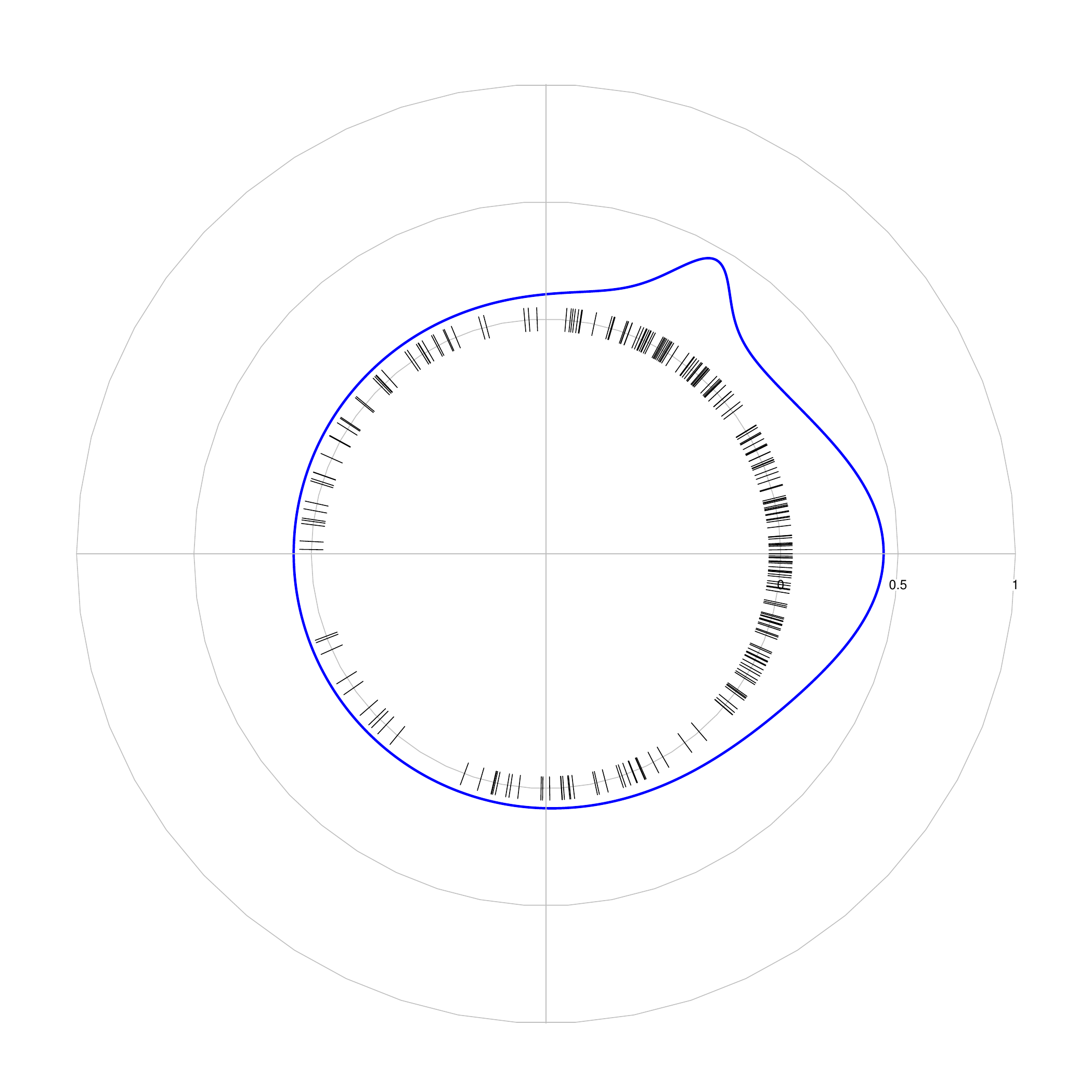}
	\includegraphics[width=0.225\textwidth]{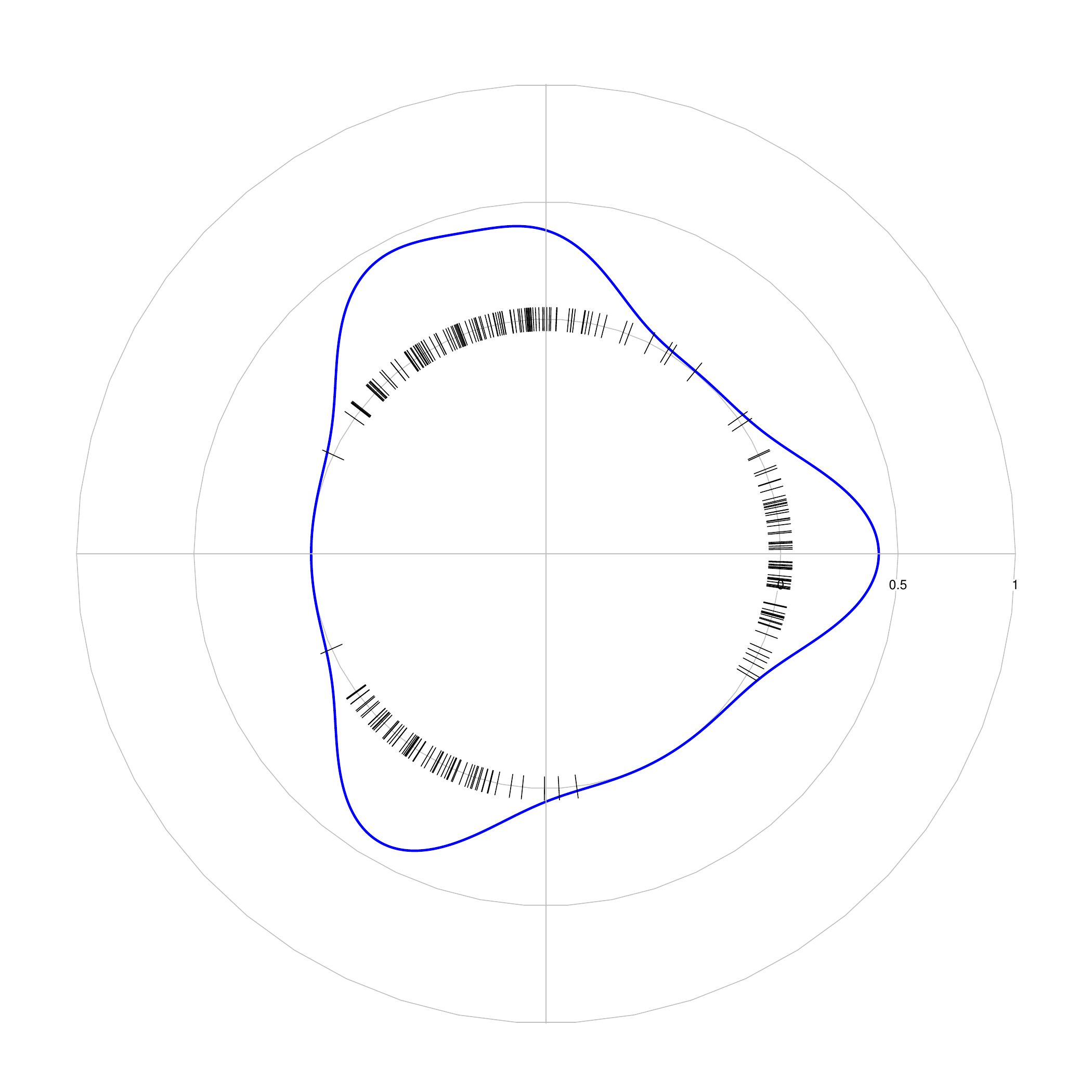}\\
	\includegraphics[width=0.225\textwidth]{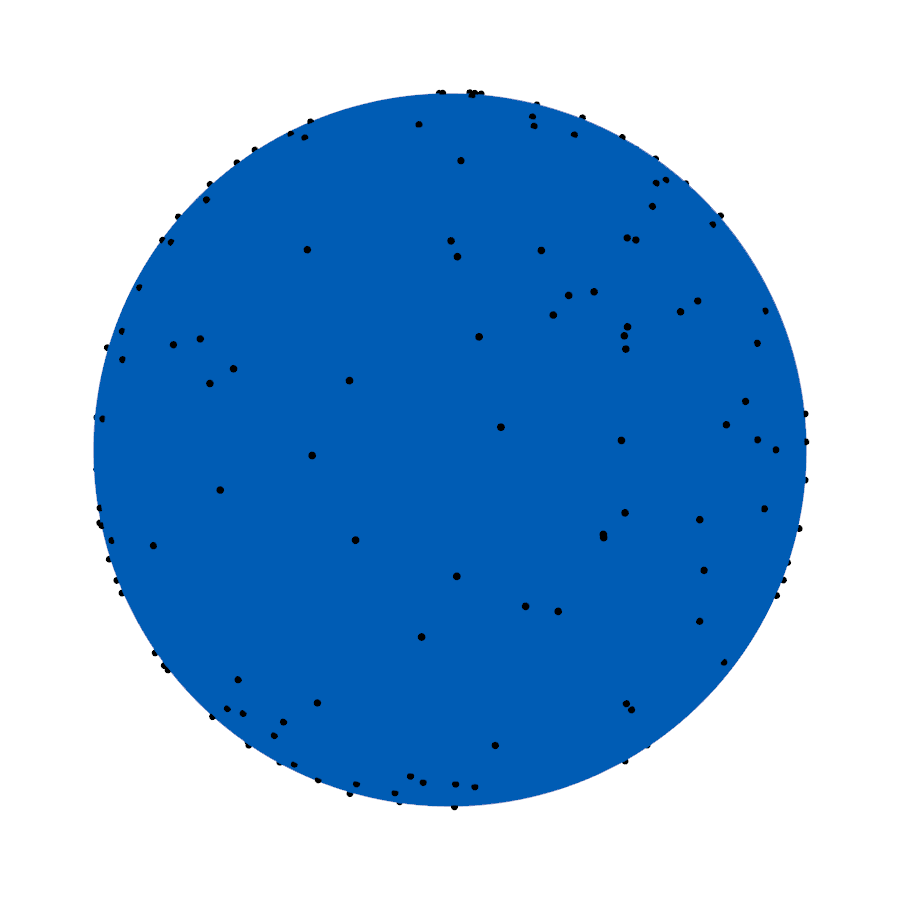}
	\includegraphics[width=0.225\textwidth]{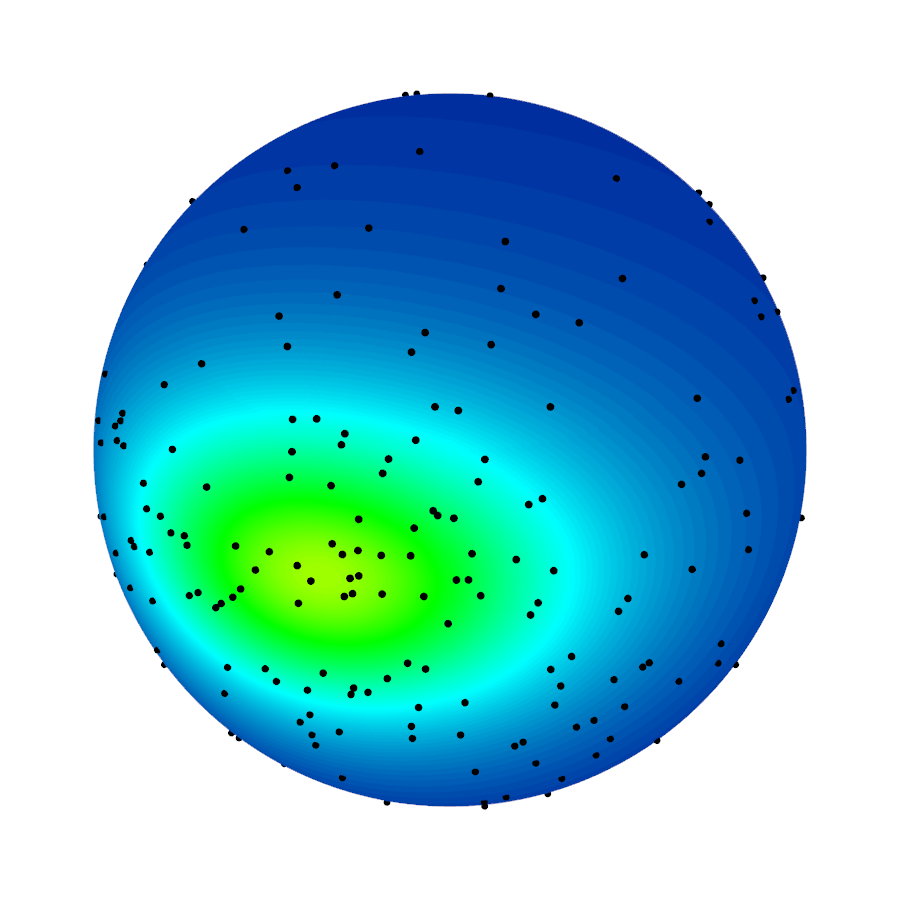}
	\includegraphics[width=0.225\textwidth]{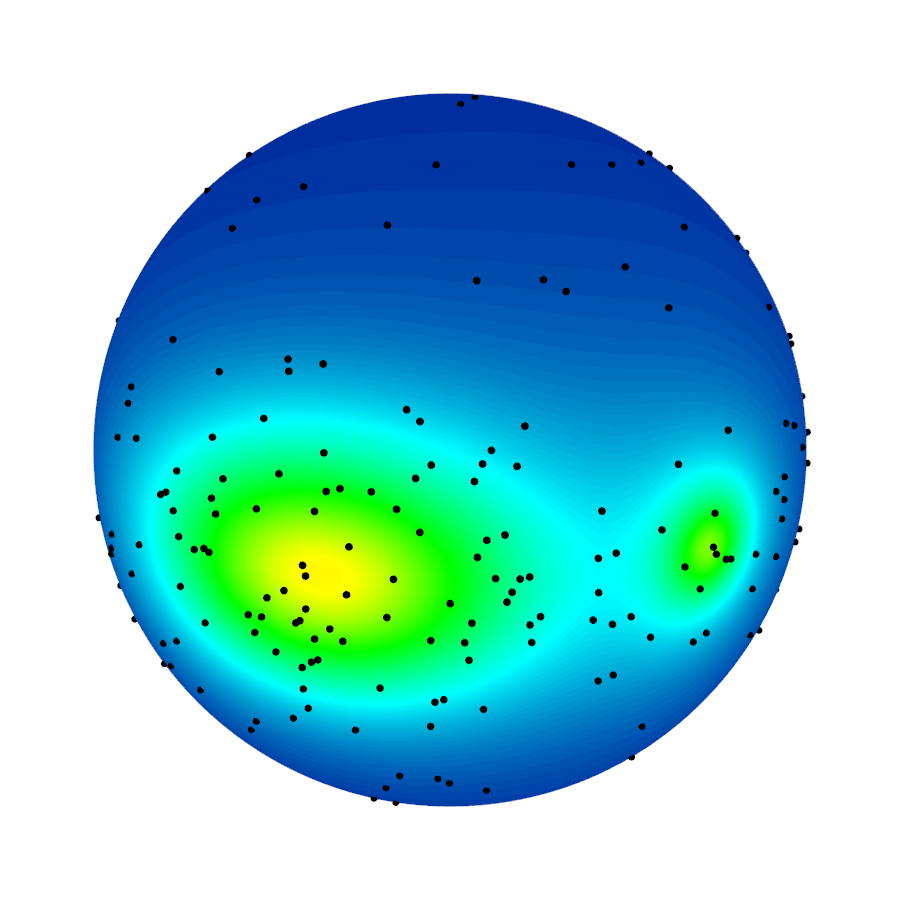}
	\includegraphics[width=0.225\textwidth]{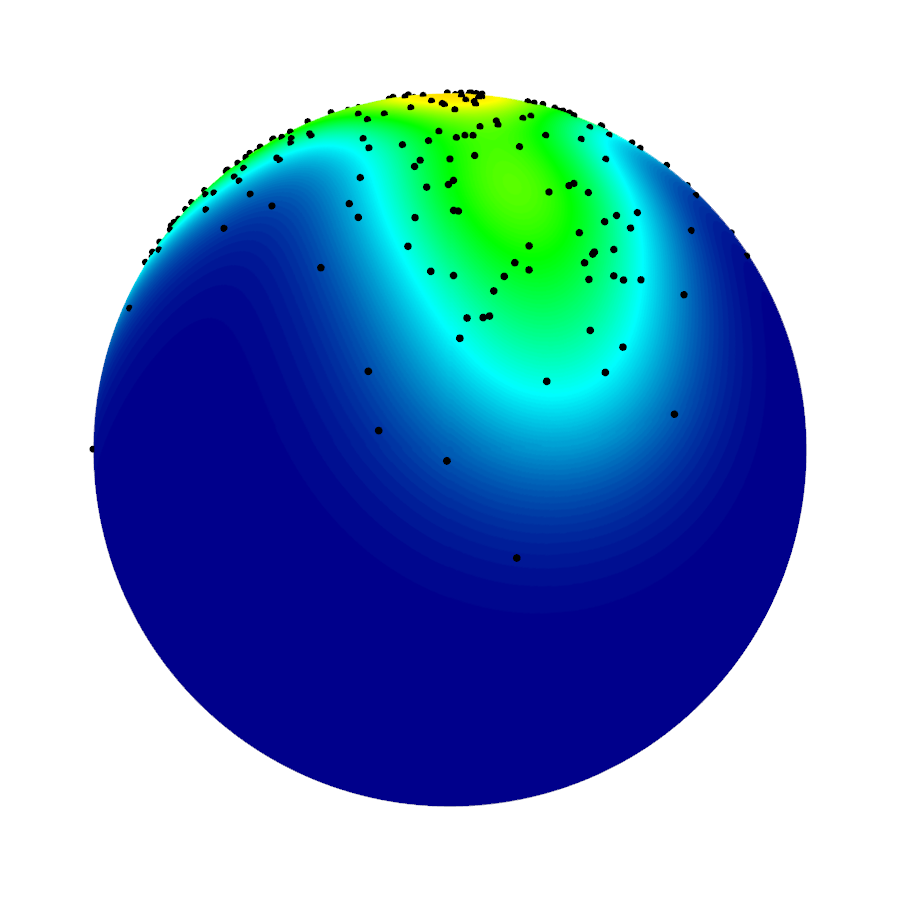}
	\caption{\small From left to right: directional densities for scenarios S1 to S4 for circular and spherical cases. Color shading according to density. \label{fig:dens}}
\end{figure}

The noise considered is $\varepsilon\sim\mathcal{N}(0,1)$ and is combined with two different conditional standard deviations given by $\sigma_1(\mathbf{x})=\frac{1}{2}$ (homocedastic, Hom.) and $\sigma_2(\mathbf{x})=\frac{1}{4}+3f_{\mathrm{M16}}(\mathbf{x})$ (heteroskedastic, Het.), with $f_\mathrm{M16}$ being the density of the M16 model. The alternative hypothesis $H_1$ is built by adding the deviations $\Delta_1(\mathbf{x})=\cos(2\pi x_1)(x_{q+1}^3-1)/\log(2+\left|x_{q+1}\right|)$ and $\Delta_2(\mathbf{x})=\cos(2\pi x_1^2x_2)\exp\left\{x_{q+1}\right\}$ to the true regression function $m_{\boldsymbol{\theta}_0}(\mathbf{x})$. The deviations from the null hypothesis, $\Delta_1$ and $\Delta_2$, are shown in Figure \ref{fig:devs}, jointly with the conditional standard deviation function used to generate data with heteroskedastic noise. The combinations of regression model, density and noise for each simulation scenario are depicted in Table \ref{tab:mods}. 

\begin{figure}[h!]
	\centering
	\includegraphics[width=0.225\textwidth]{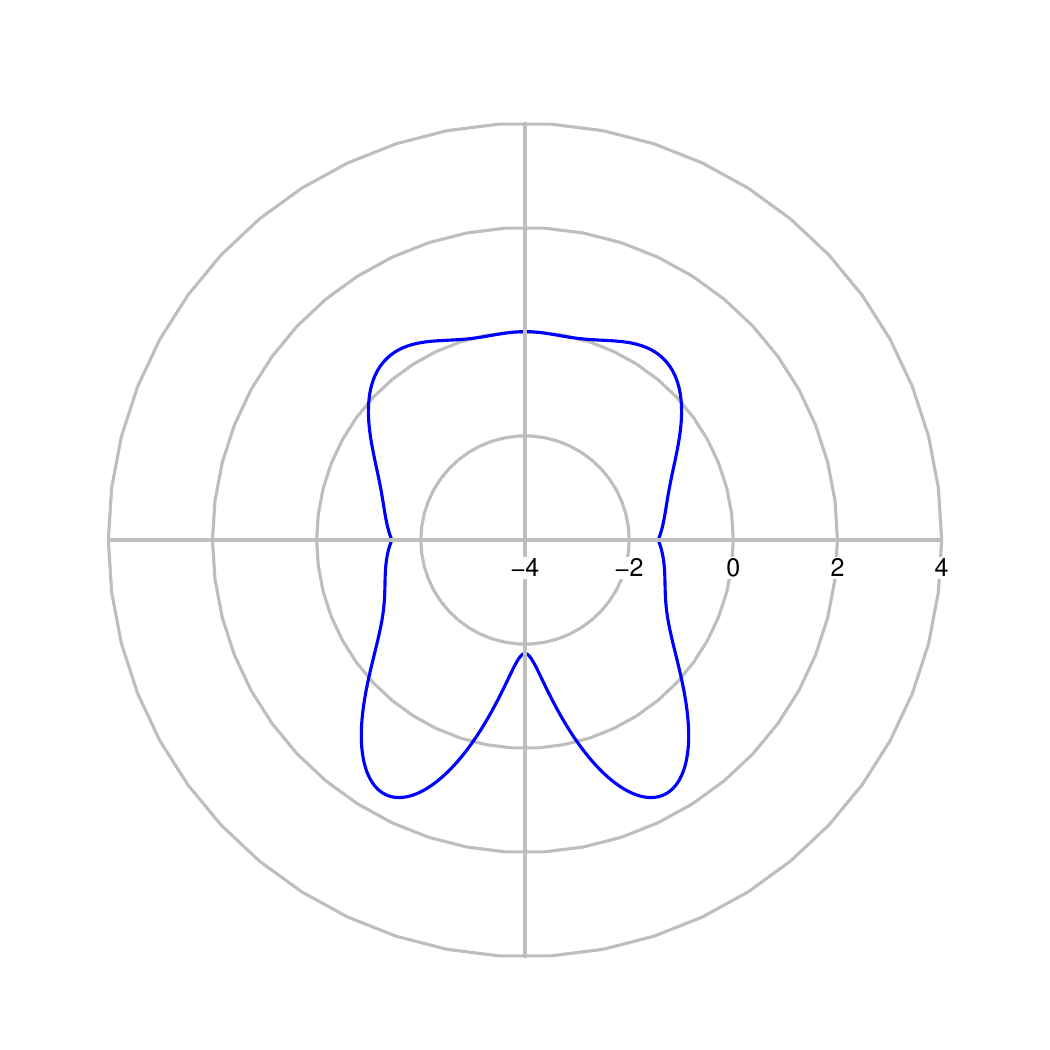}
	\includegraphics[width=0.225\textwidth]{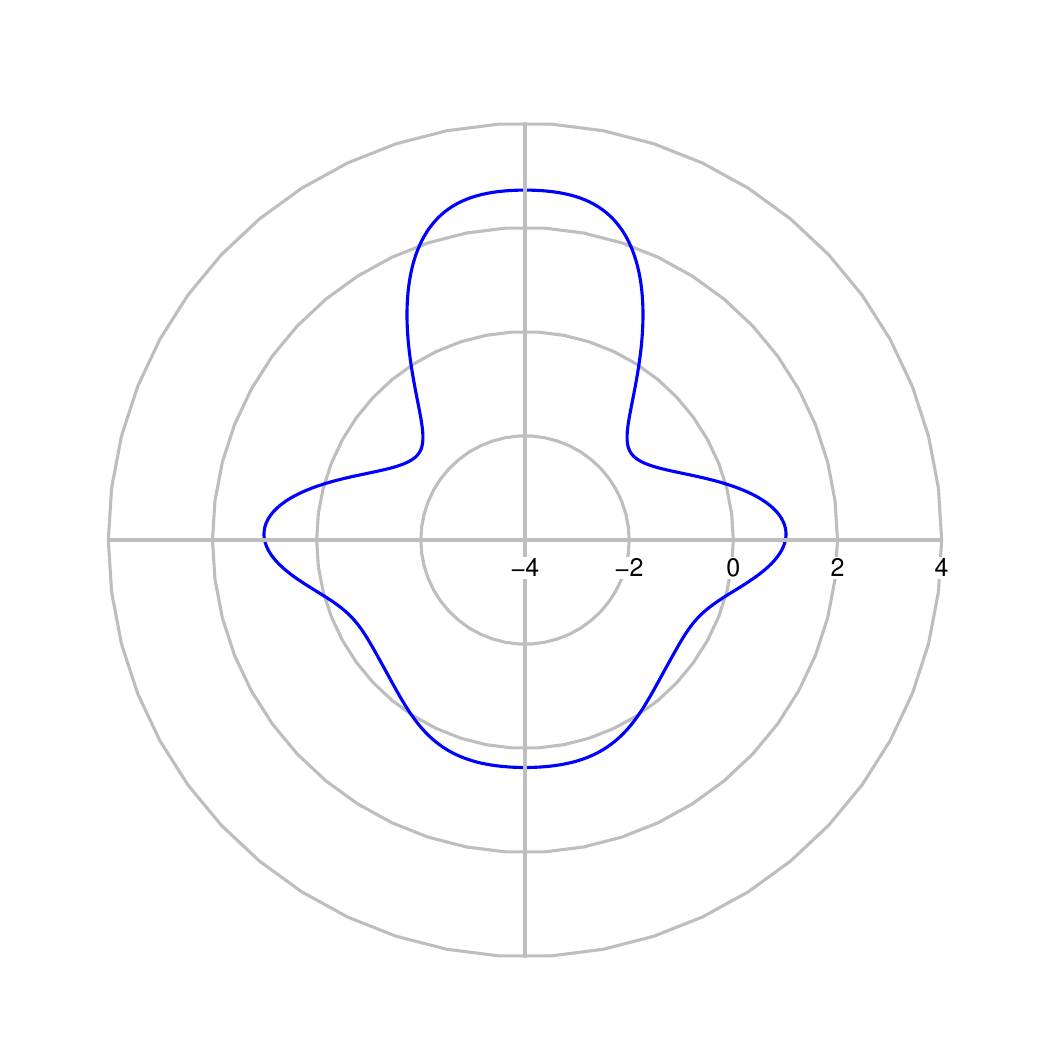}
	\includegraphics[width=0.225\textwidth]{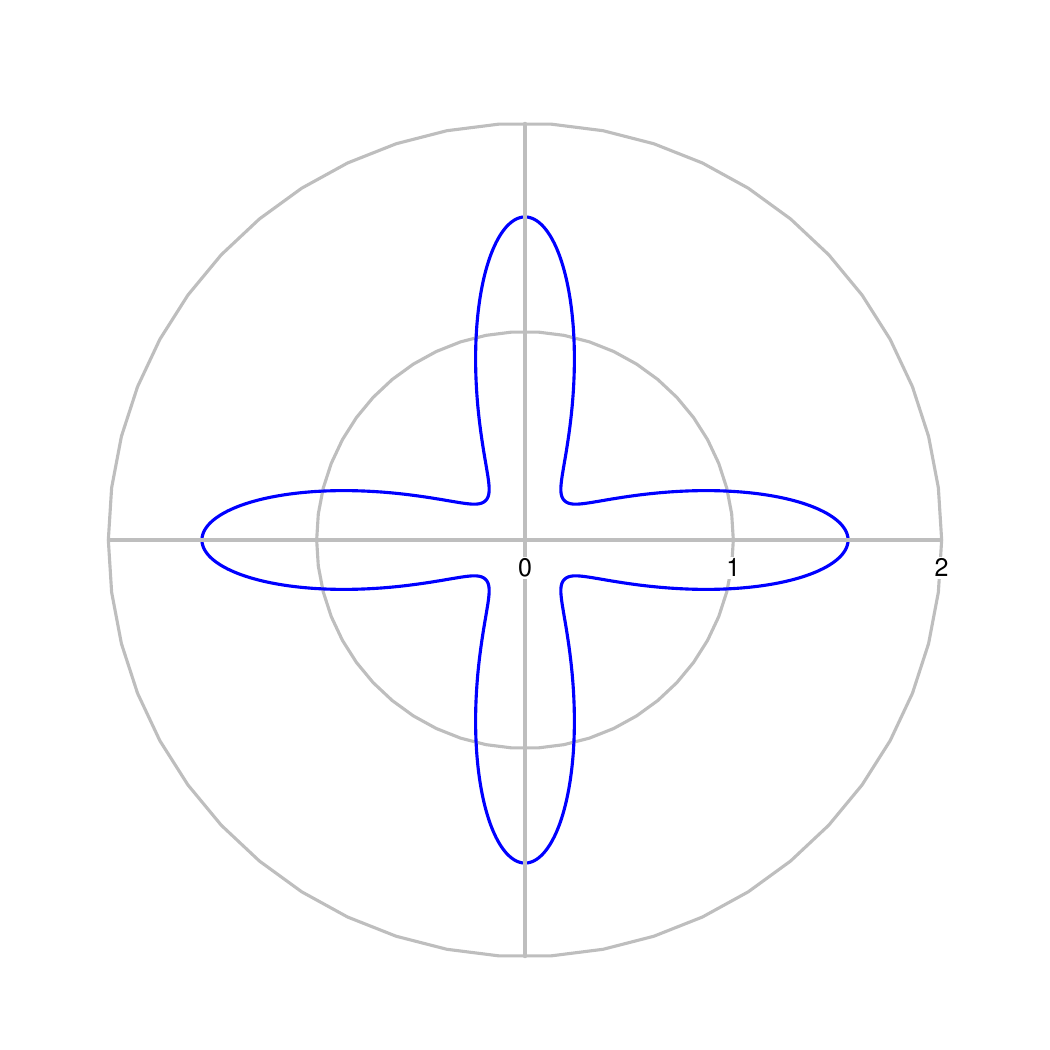}\\
	\includegraphics[width=0.225\textwidth]{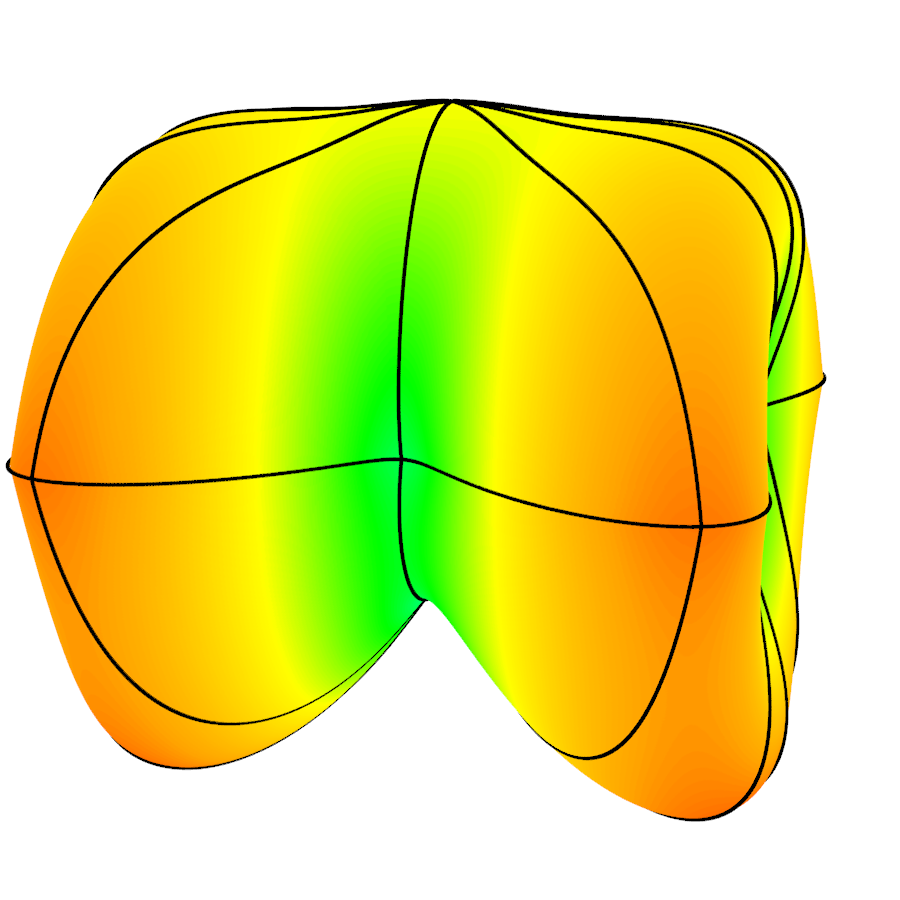}
	\includegraphics[width=0.225\textwidth]{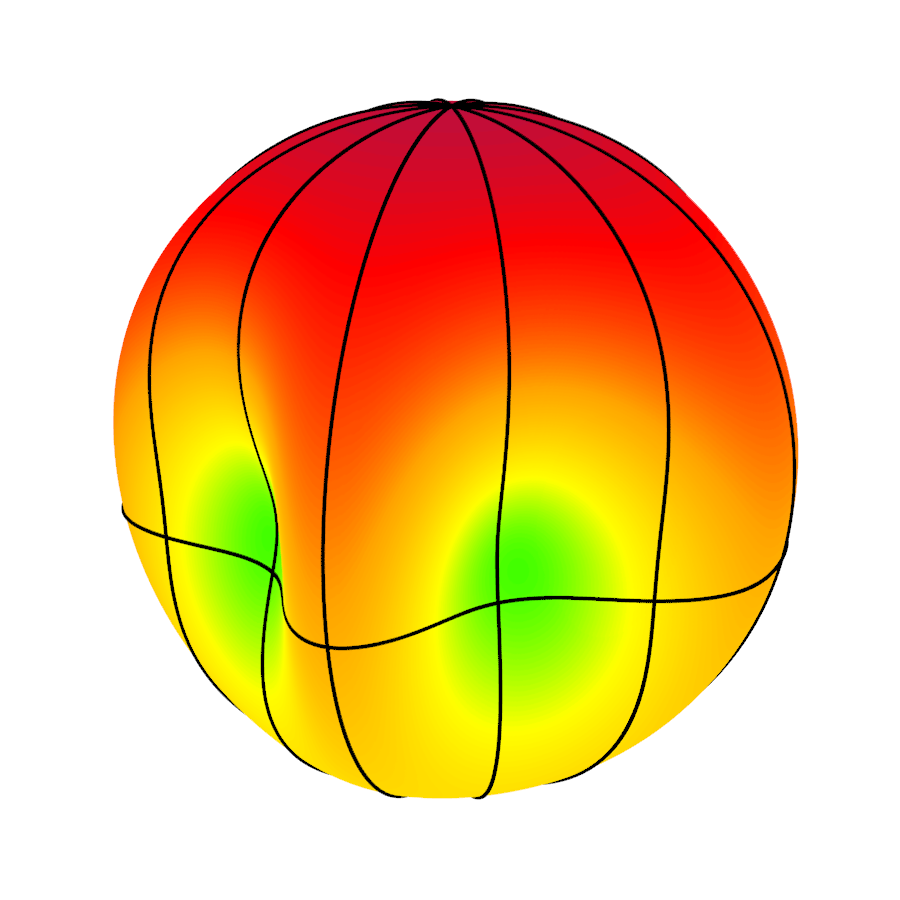}
	\includegraphics[width=0.225\textwidth]{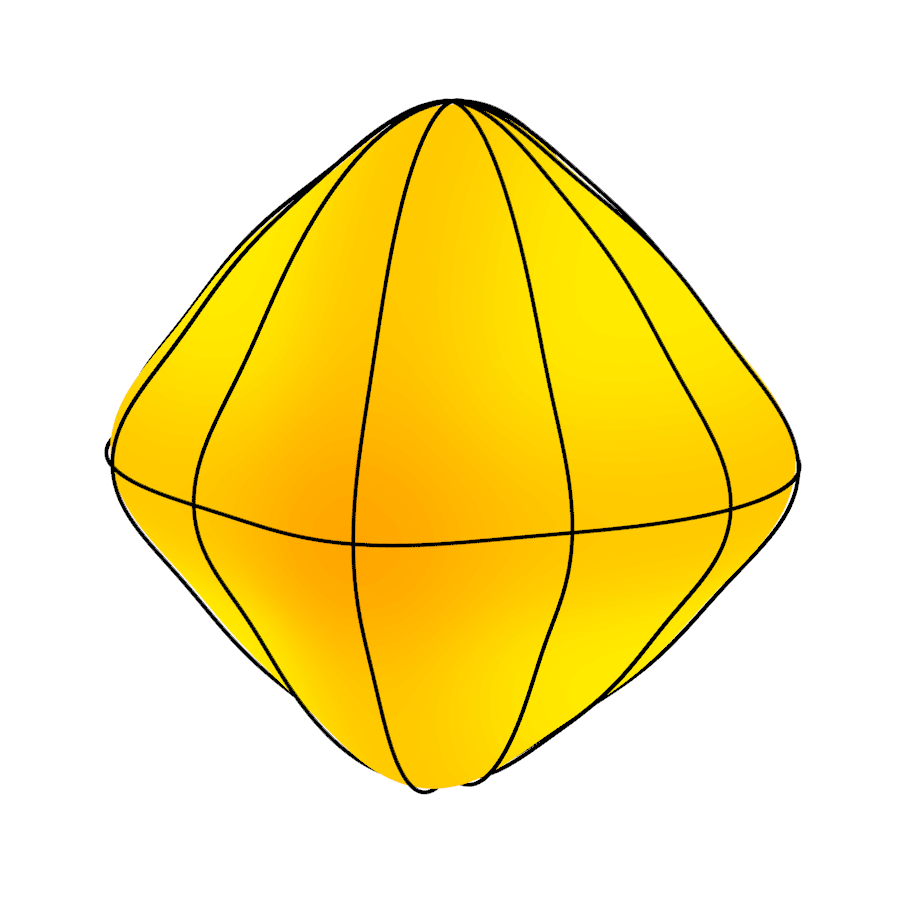}
	\caption{\small From left to right: deviations $\Delta_1$ and $\Delta_2$ and conditional standard deviation function $\sigma_2$ for circular and spherical cases.  \label{fig:devs}}
\end{figure}

\begin{table}[H]
	\centering
	\scriptsize
	\begin{tabular}{c|c|c|c|c|c}
		\toprule\toprule
		Scenario & Regression function & Parameters & Density  & Noise & Deviation  \\
		\midrule
		S1  & $m(\mathbf{x})=c$ & $c=0$ & M1 & Het. & $\phantom{-}\frac{3}{4}\Delta_1(\mathbf{x})$\\\midrule
		S2  & $m(\mathbf{x})=c+\boldsymbol\eta^T\mathbf{x}$ & $c=1$, $\boldsymbol\eta=\big(-\frac{3}{2},\mathbf{\frac{1}{2}}_q\big)$& $\frac{3}{5}\text{M4}+\frac{2}{5}\text{M1}$ & Het. & $-\frac{3}{4}\Delta_1(\mathbf{x})$\\\midrule
		S3  & $m(\mathbf{x})=c+a\sin(2\pi x_2)+b\cos(2\pi x_1)$ & $c=0$, $a=1$, $b=\frac{3}{2}$ & $\frac{3}{5}\text{M12}+\frac{2}{5}\text{M1}$ & Hom. & $\phantom{-}\frac{3}{4}\Delta_2(\mathbf{x})$\\\midrule
		S4  & $m(\mathbf{x})=c+a\sin\big(2\pi b\left(2+x_{q+1}\right)^{-1}\big)$ & $c=0$, $a=3$, $b=4$ & M20 & Hom. & $\phantom{-}\frac{1}{2}\Delta_2(\mathbf{x})$\\
		\bottomrule\bottomrule
	\end{tabular}
	\caption{\small Specification of simulation scenarios.\label{tab:mods}}
\end{table}

The coefficients $\delta$ for obtaining deviations $\delta\Delta_1$ and $\delta\Delta_2$ in each scenario were chosen such that the density of the response $Y=m_{\boldsymbol{\theta}_0}(\mathbf{X})+\delta\Delta(\mathbf{X})+\sigma(\mathbf{X})\varepsilon$ under $H_0$ ($\delta=0$) and under $H_1$ ($\delta\neq0$) were similar. Figure \ref{fig:densdevs} shows the densities of $Y$ under the null and the alternative for the four scenarios and dimensions considered. This is a graphical way of ensuring that the deviation is not trivial to detect and hence is not straightforward to reject $H_0$. \\

Note that, due to the design of the deviations and its pairing with the regression functions, design densities and kind of noises, it is harder to reject $H_0$ in particular situations. This is what happens for example in S4 for $q=2$: due to the design density, most of the observations happen close to the north pole, where the shape of the parametric model and of $\Delta_2$ are similar, resulting in a harder detectable deviation for that dimension. A different situation happens for S1, where the heteroskedastic noise masks the deviation $\Delta_1$ for moderate and large values of the smoothing parameter $h$. These two combinations are provided to check the performance of the test under challenging situations.

\pagebreak

The empirical sizes of the test for significance levels $\alpha=0.01,0.05,0.10$ are given in Figures \ref{fig:size:1}, \ref{fig:size:2} and \ref{fig:size:3}, corresponding to sample sizes $n=100$, $250$ and $500$, respectively. Nominal levels are respected in most scenarios, except for unrealistically small bandwidths. Finally, the empirical powers for $n=100,250$ and $500$ are given in Figure \ref{fig:pow:2} and, as can be seen, the rejection rates increase with $n$. The hardest deviation to detect corresponds to S1, which is hidden for large bandwidths.

\begin{figure}[H]
	\centering
	\includegraphics[width=0.24\textwidth]{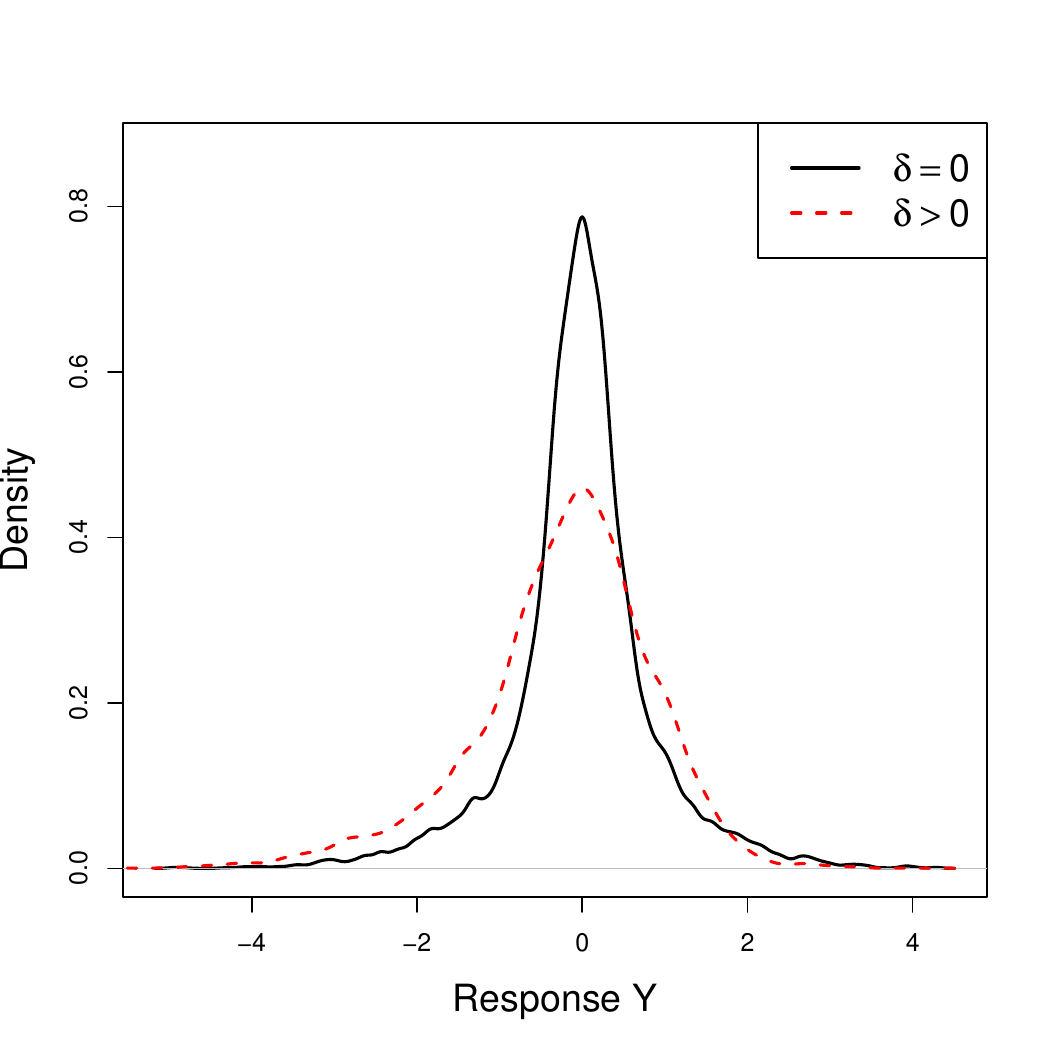}
	\includegraphics[width=0.24\textwidth]{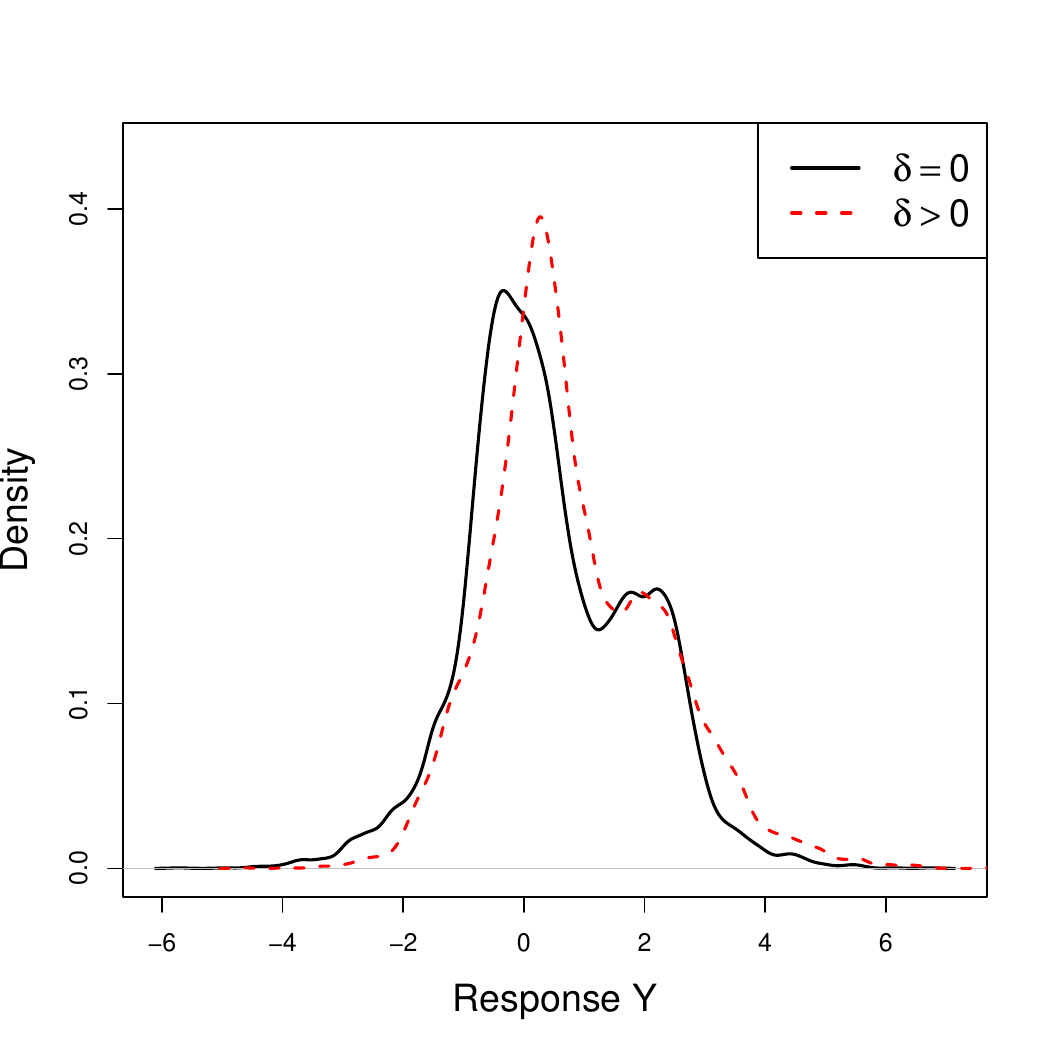}
	\includegraphics[width=0.24\textwidth]{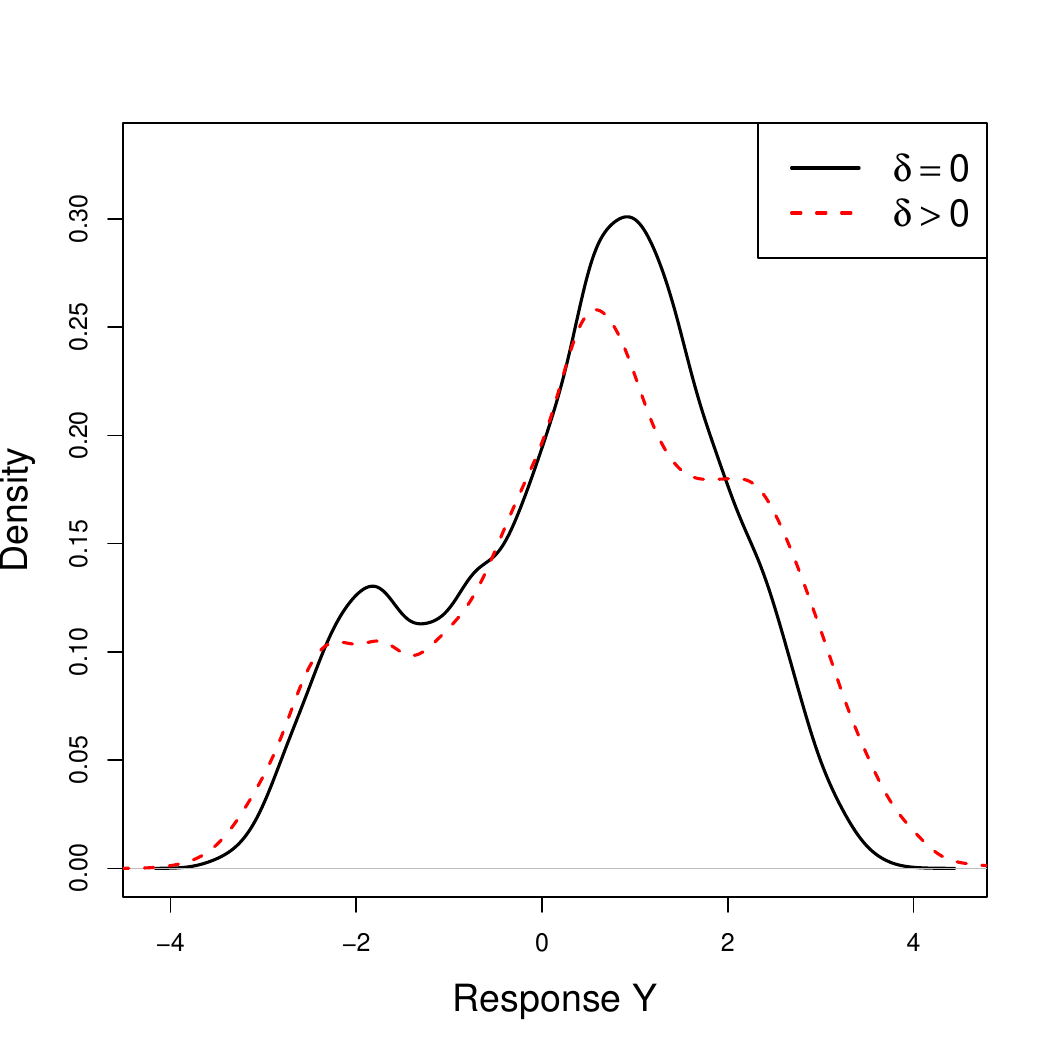}
	\includegraphics[width=0.24\textwidth]{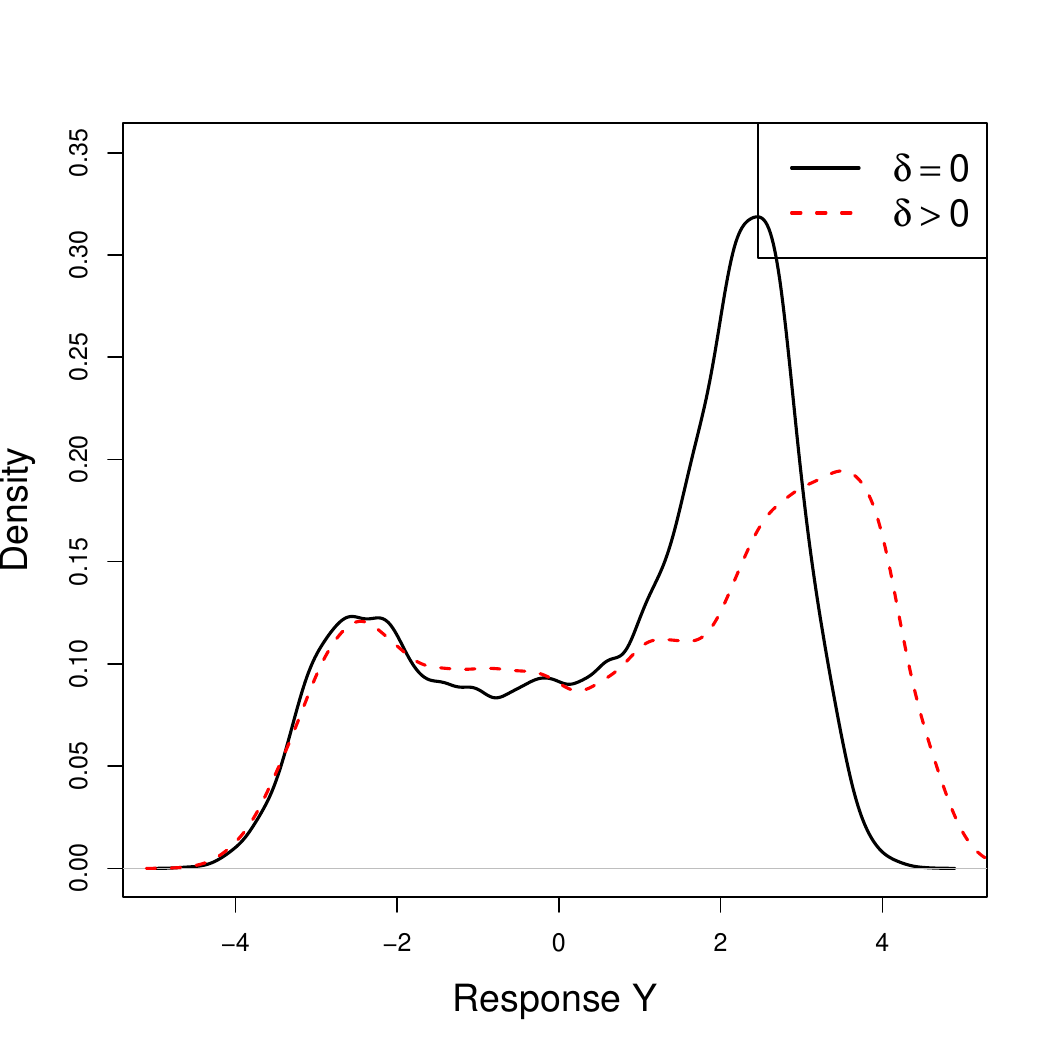}\\
	\includegraphics[width=0.24\textwidth]{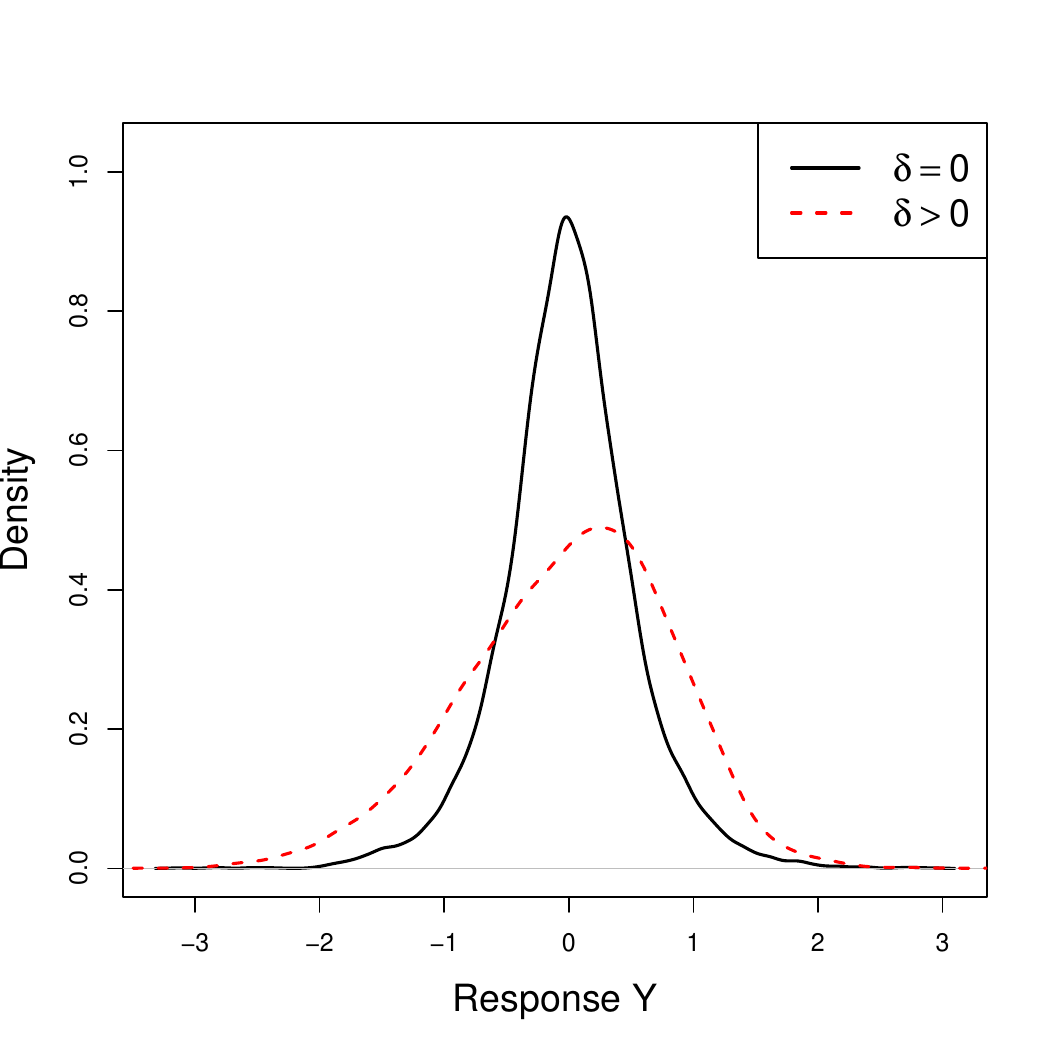}
	\includegraphics[width=0.24\textwidth]{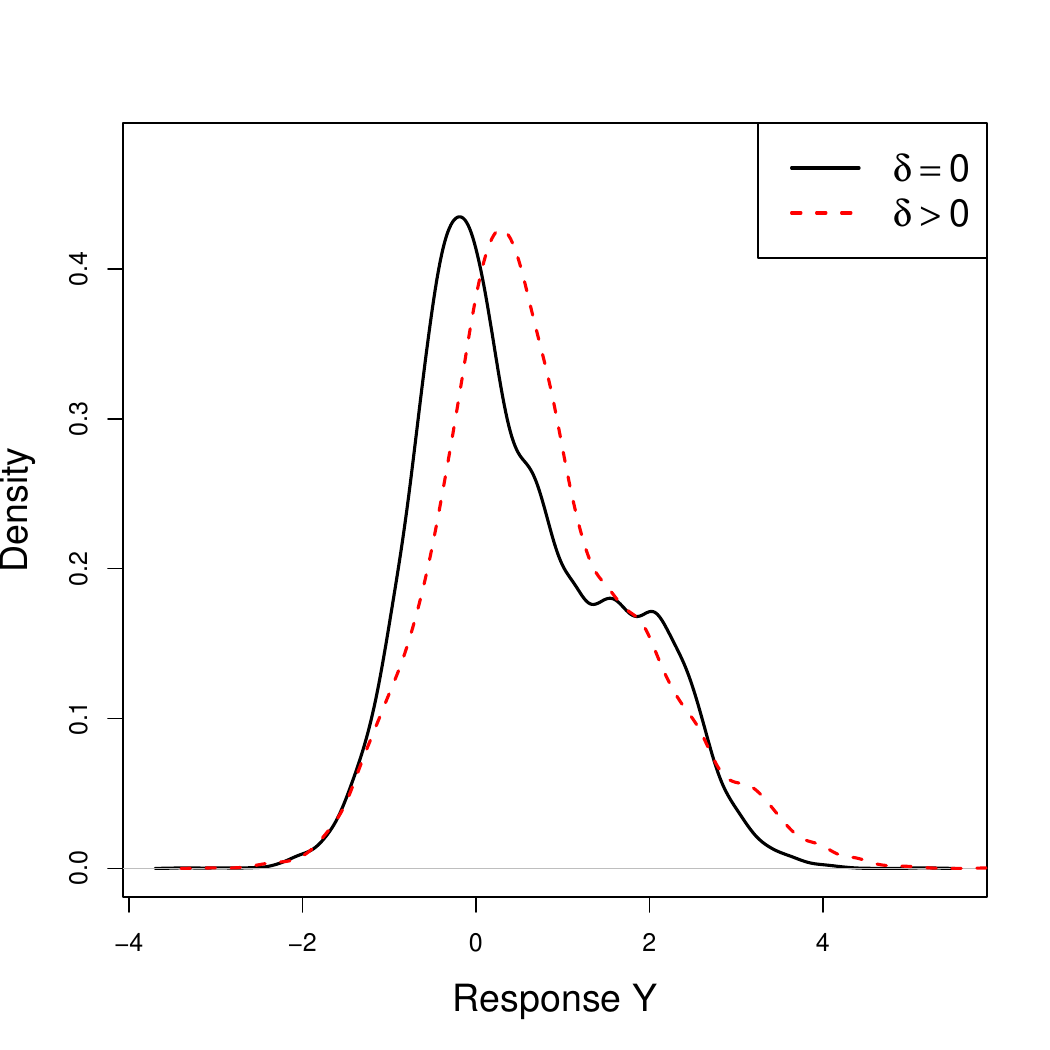}
	\includegraphics[width=0.24\textwidth]{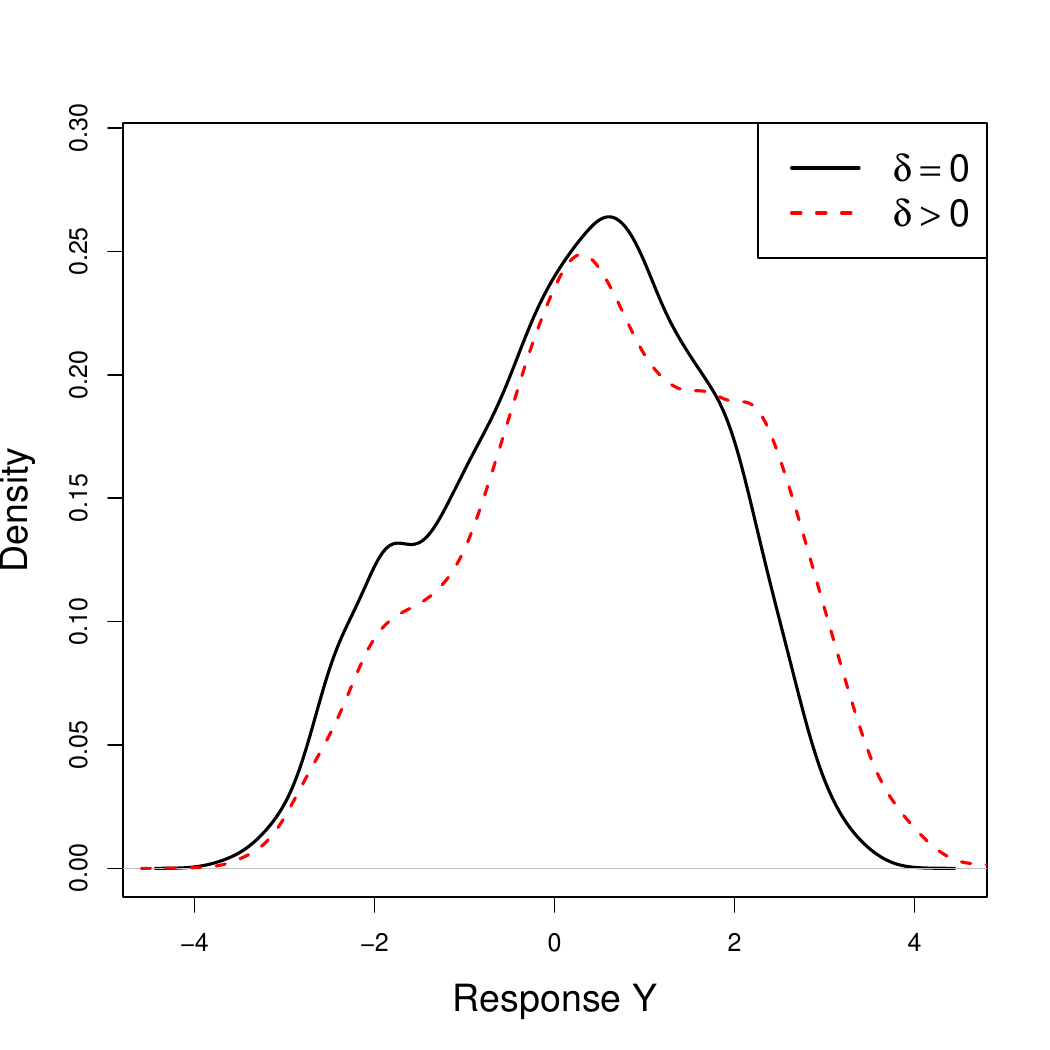}
	\includegraphics[width=0.24\textwidth]{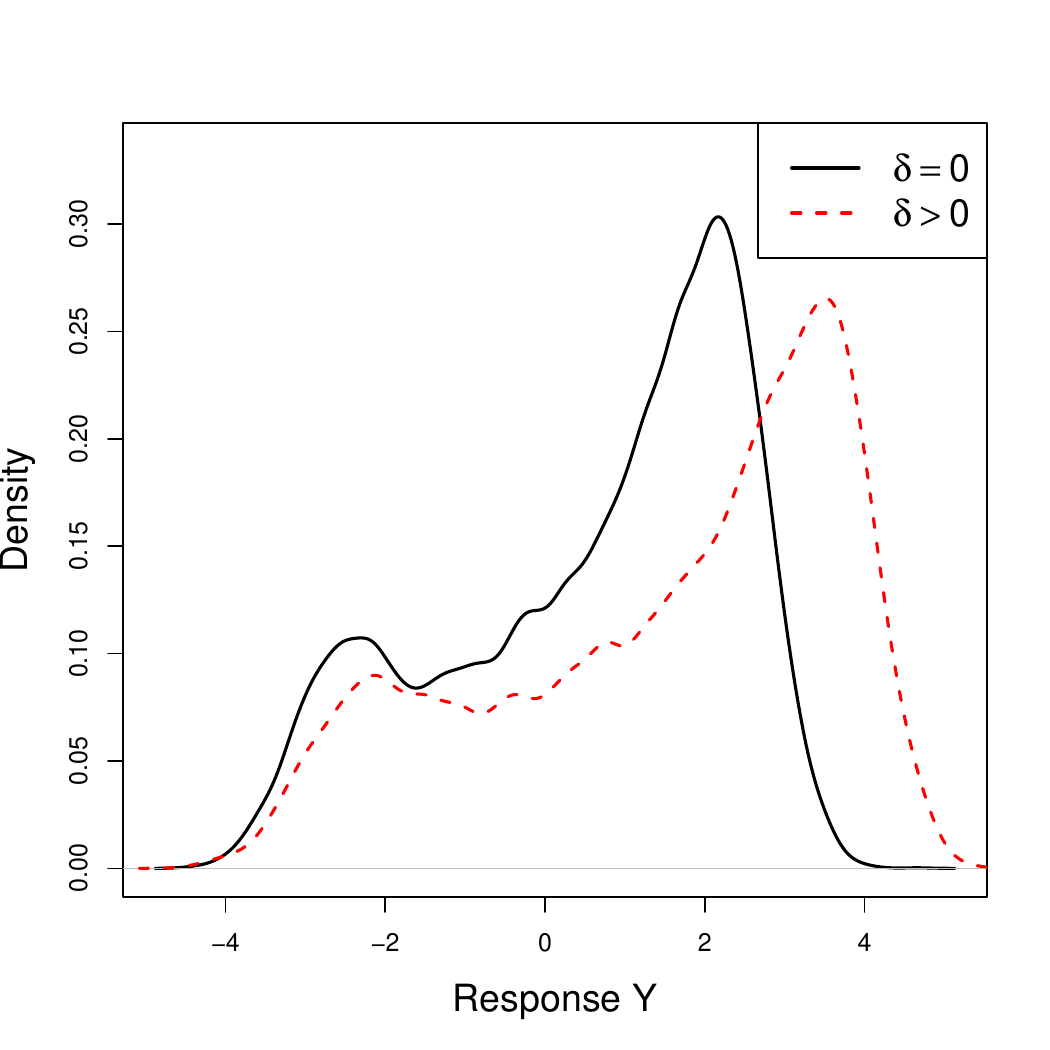}\\
	\includegraphics[width=0.24\textwidth]{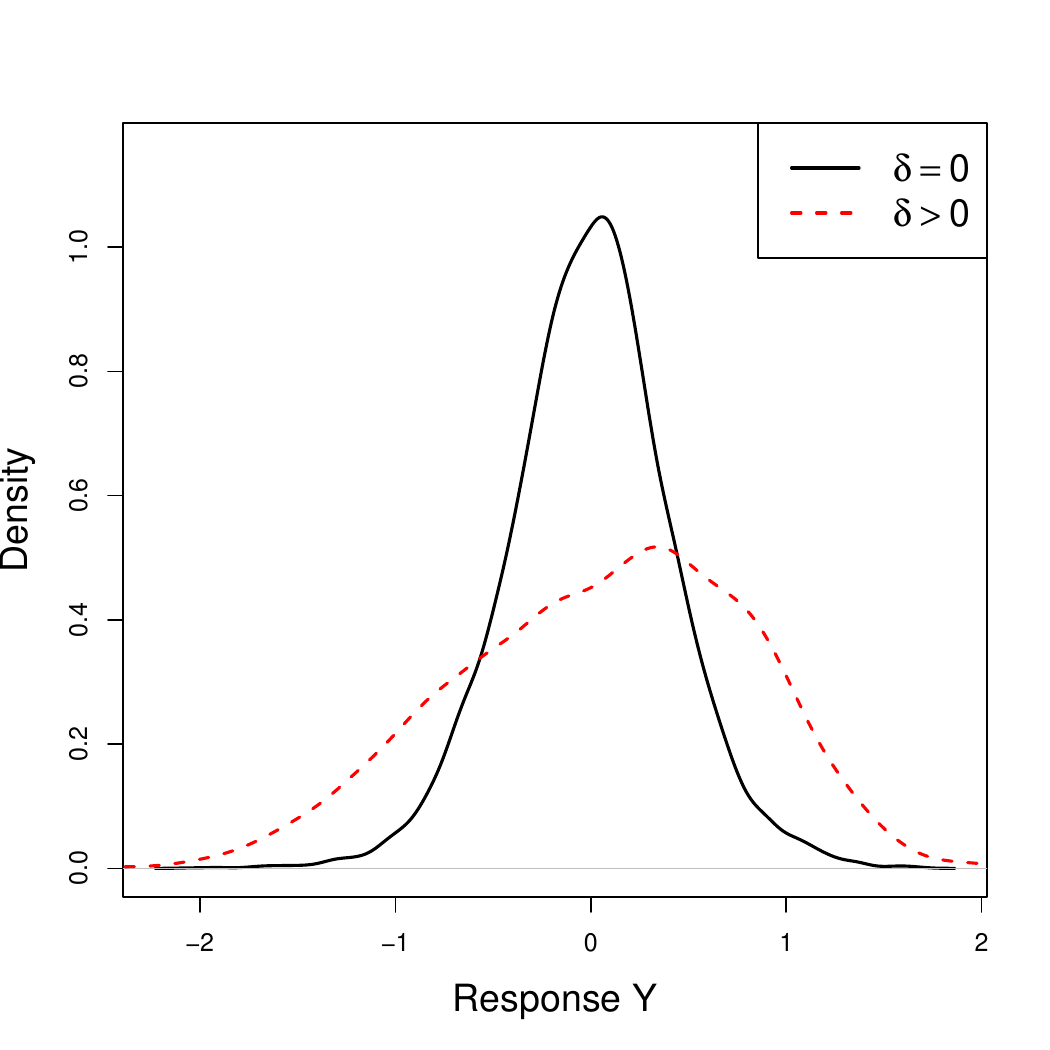}
	\includegraphics[width=0.24\textwidth]{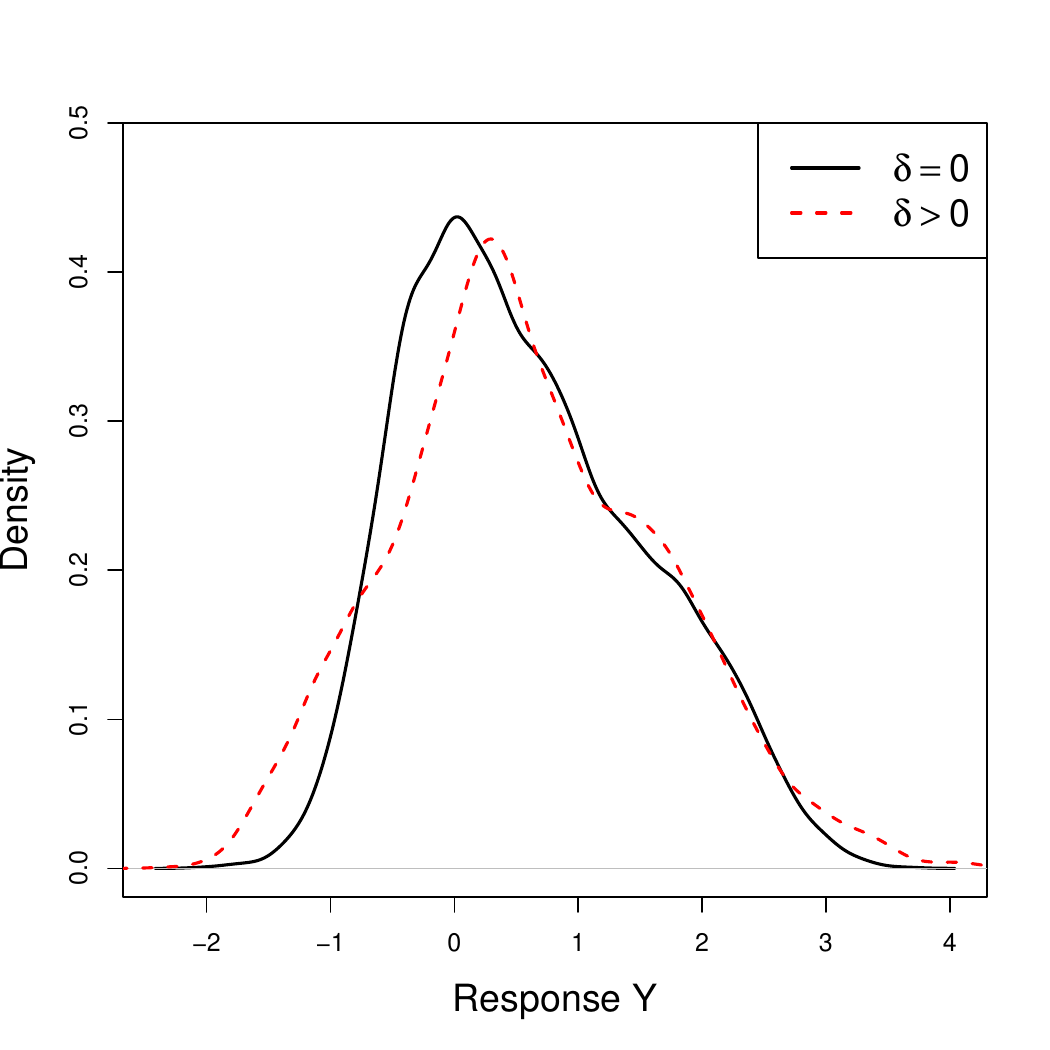}
	\includegraphics[width=0.24\textwidth]{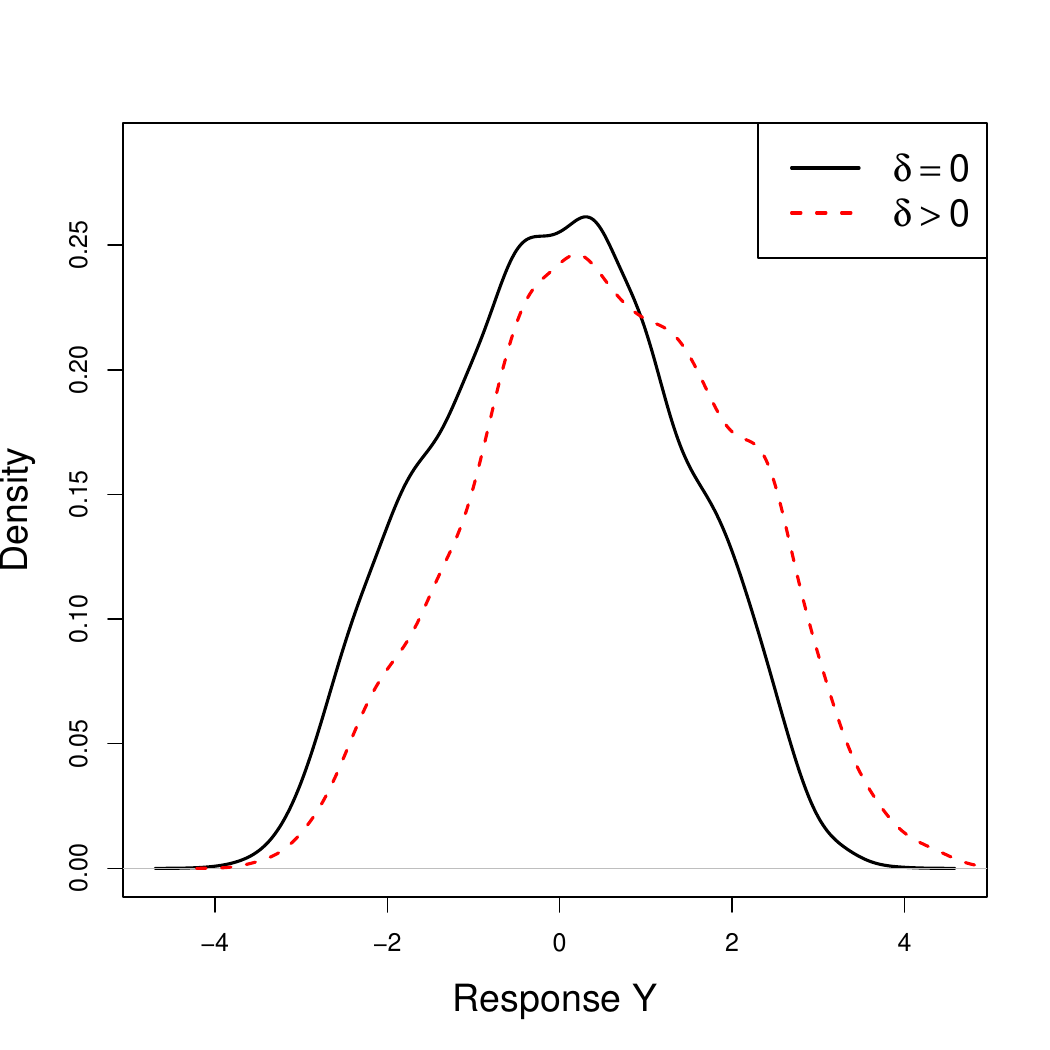}
	\includegraphics[width=0.24\textwidth]{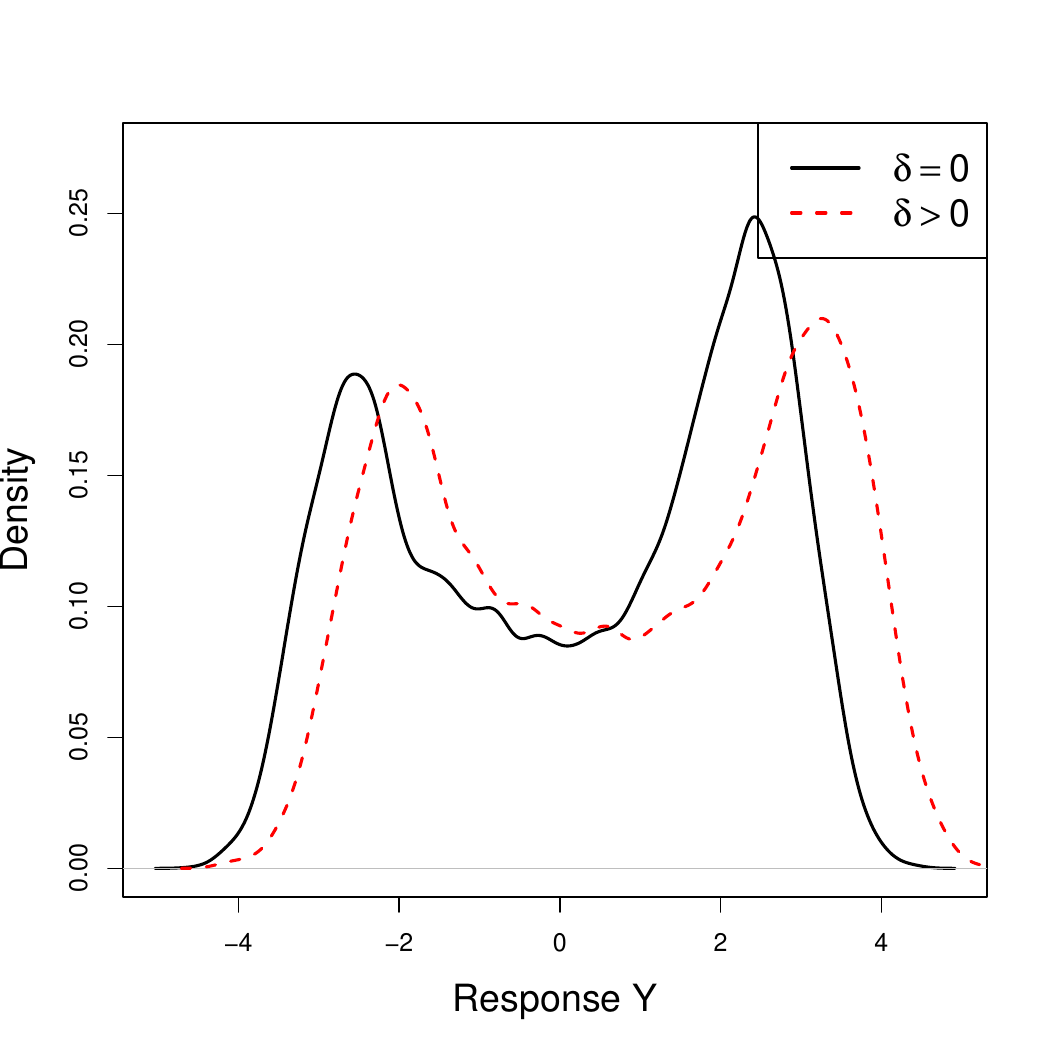}
	\caption{\small Densities of the response $Y$ under the null (solid line) and under the alternative (dashed line) for scenarios S1 to S4 (columns, from left to right) and dimensions $q=1,2,3$ (rows, from top to bottom).\label{fig:densdevs}}
\end{figure}

\vspace*{\fill}

\pagebreak

\vspace*{\fill}

\begin{figure}[H]
	\centering
	\includegraphics[width=0.32\textwidth]{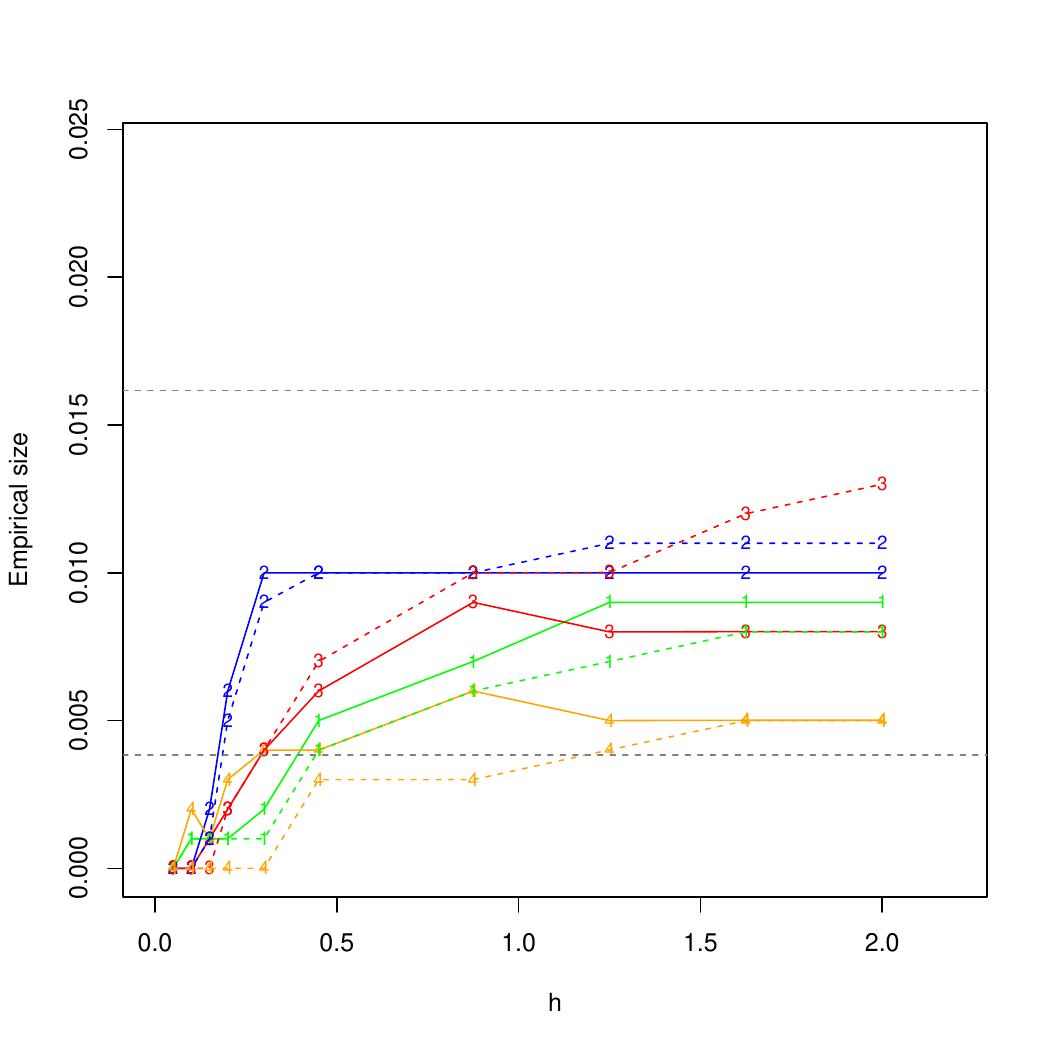}
	\includegraphics[width=0.32\textwidth]{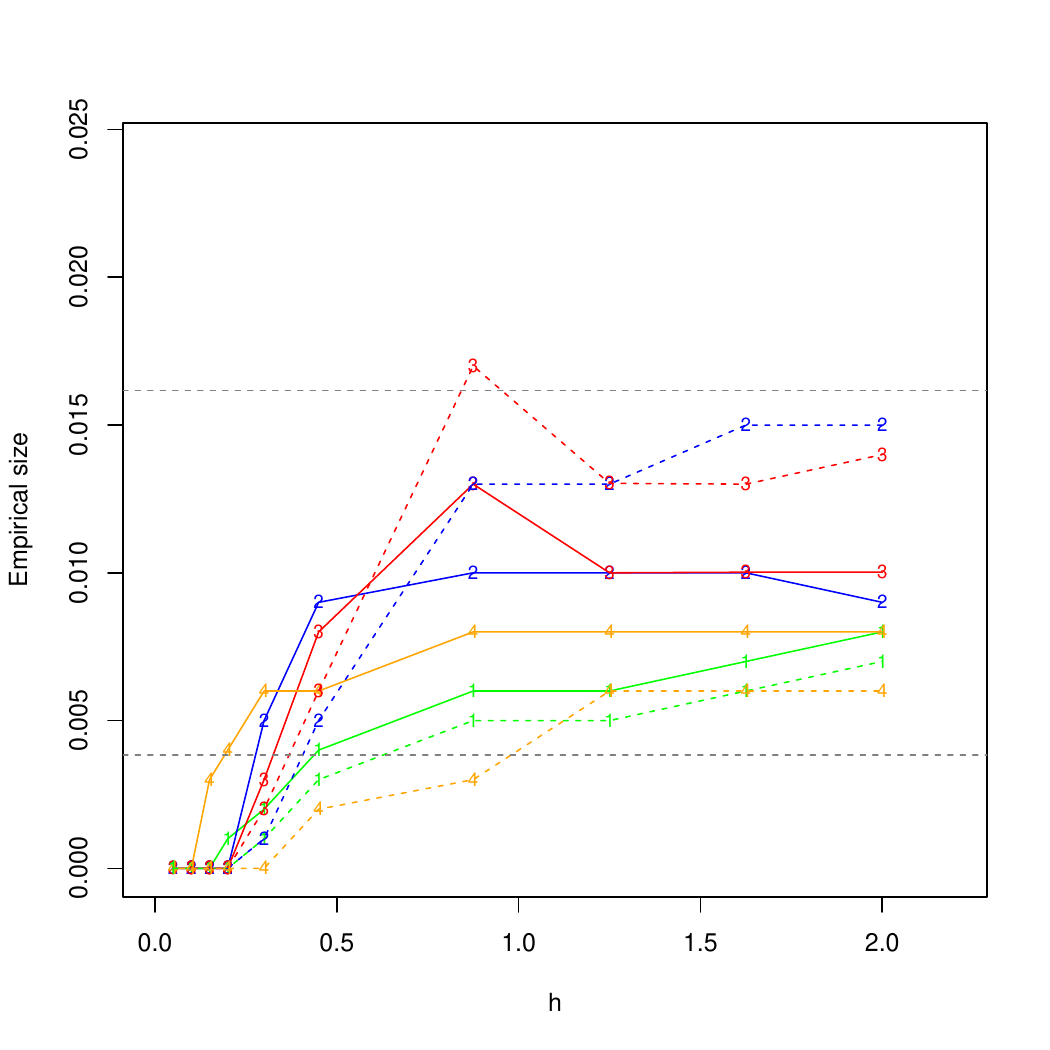}
	\includegraphics[width=0.32\textwidth]{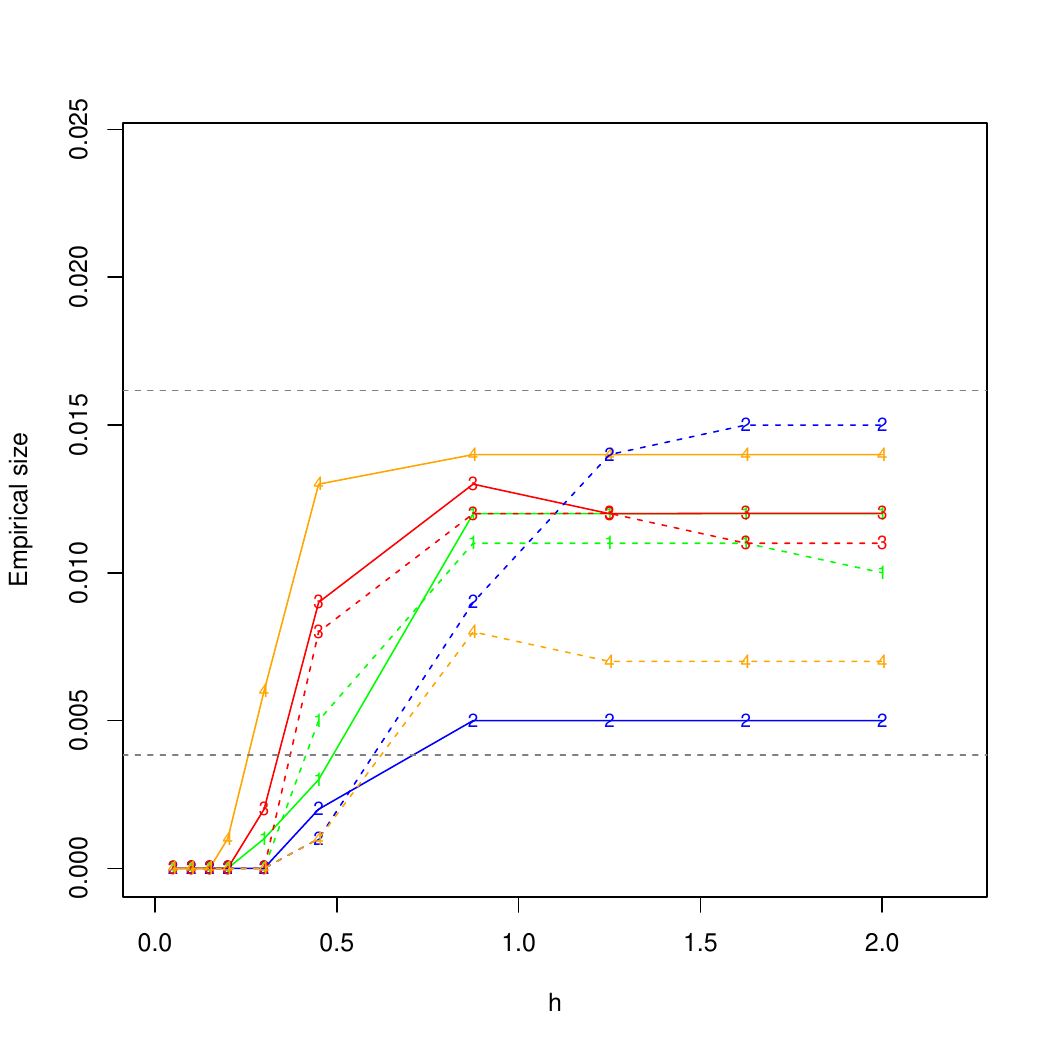}\\
	\includegraphics[width=0.32\textwidth]{img/fig9.pdf}
	\includegraphics[width=0.32\textwidth]{img/fig10.pdf}
	\includegraphics[width=0.32\textwidth]{img/fig11.pdf}\\
	\includegraphics[width=0.32\textwidth]{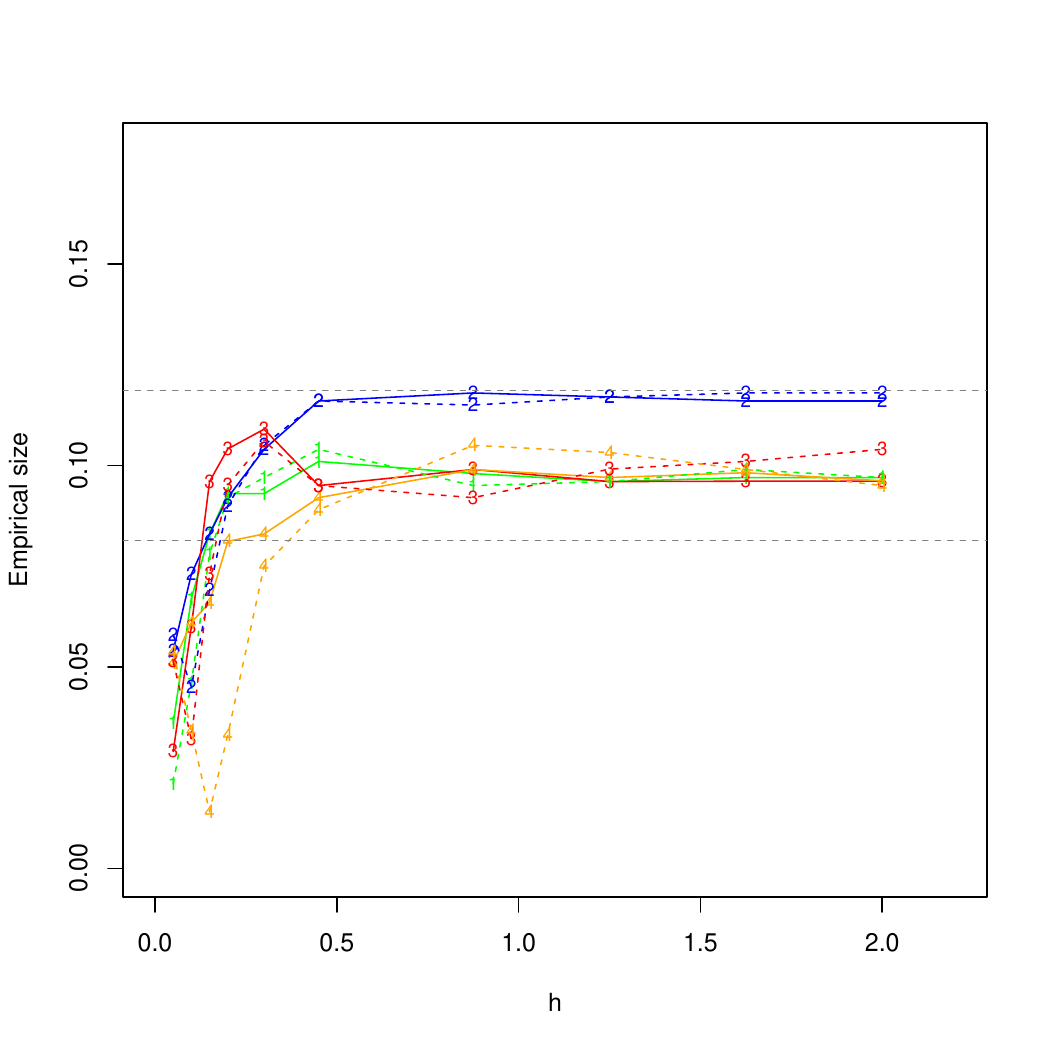}
	\includegraphics[width=0.32\textwidth]{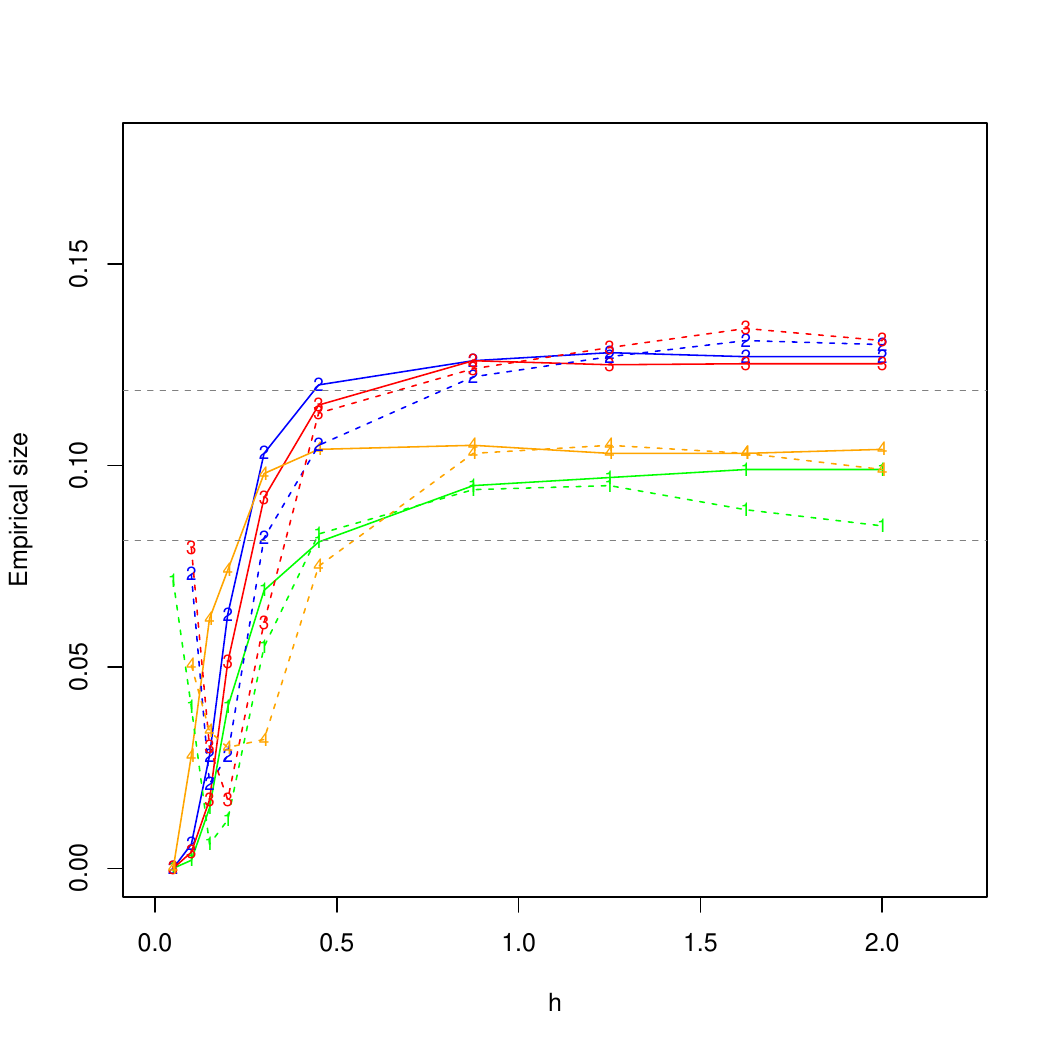}
	\includegraphics[width=0.32\textwidth]{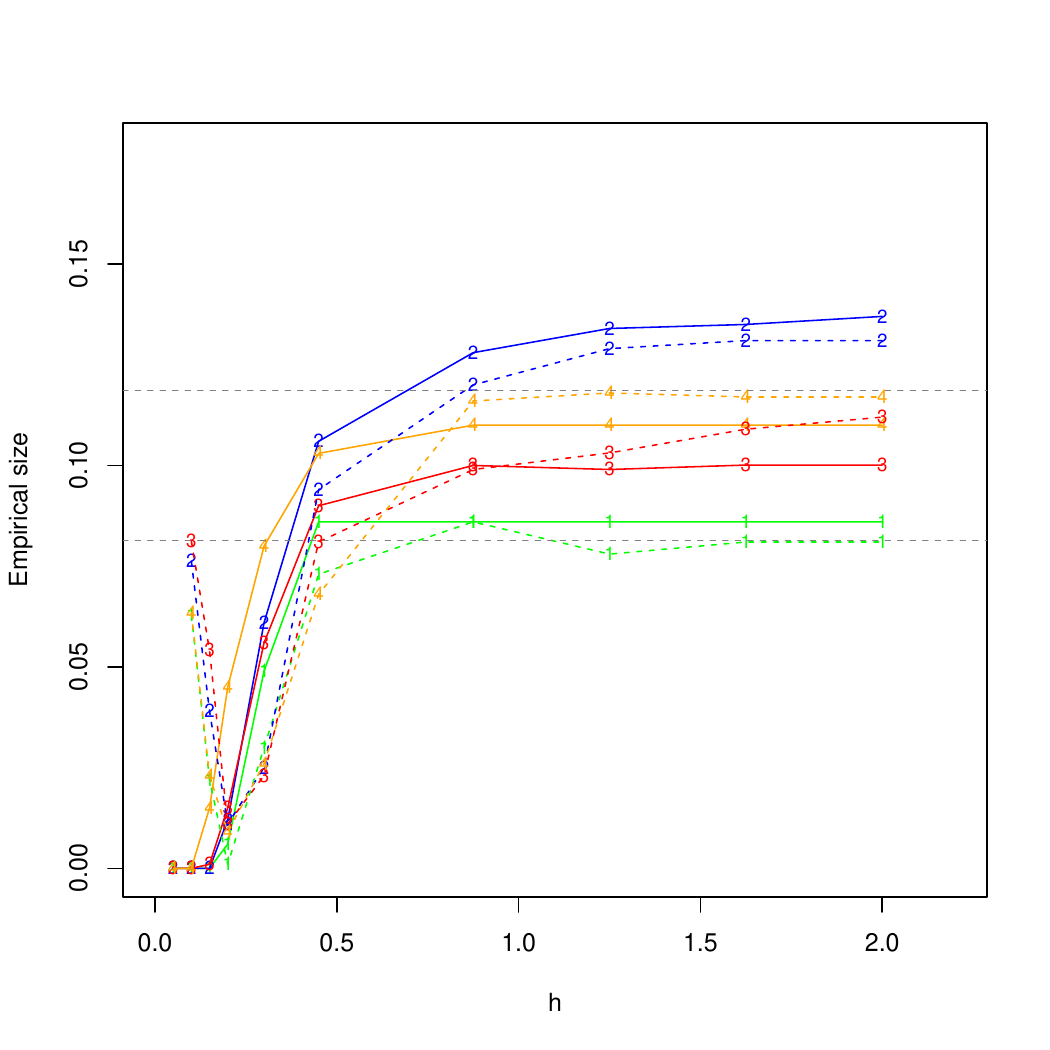}
	\caption{\small Empirical sizes for $\alpha=0.01$ (first row), $\alpha=0.05$ (second row) and $\alpha=0.10$ (third row) for the different scenarios, with $p=0$ (solid line) and $p=1$ (dashed line). From left to right, columns represent dimensions $q=1,2,3$ with sample size $n=100$. Green, blue, red and orange colors correspond to scenarios S1 to S4, respectively. \label{fig:size:1}}
\end{figure}

\vspace*{\fill}

\pagebreak

\vspace*{\fill}

\begin{figure}[H]
	\centering
	\includegraphics[width=0.32\textwidth]{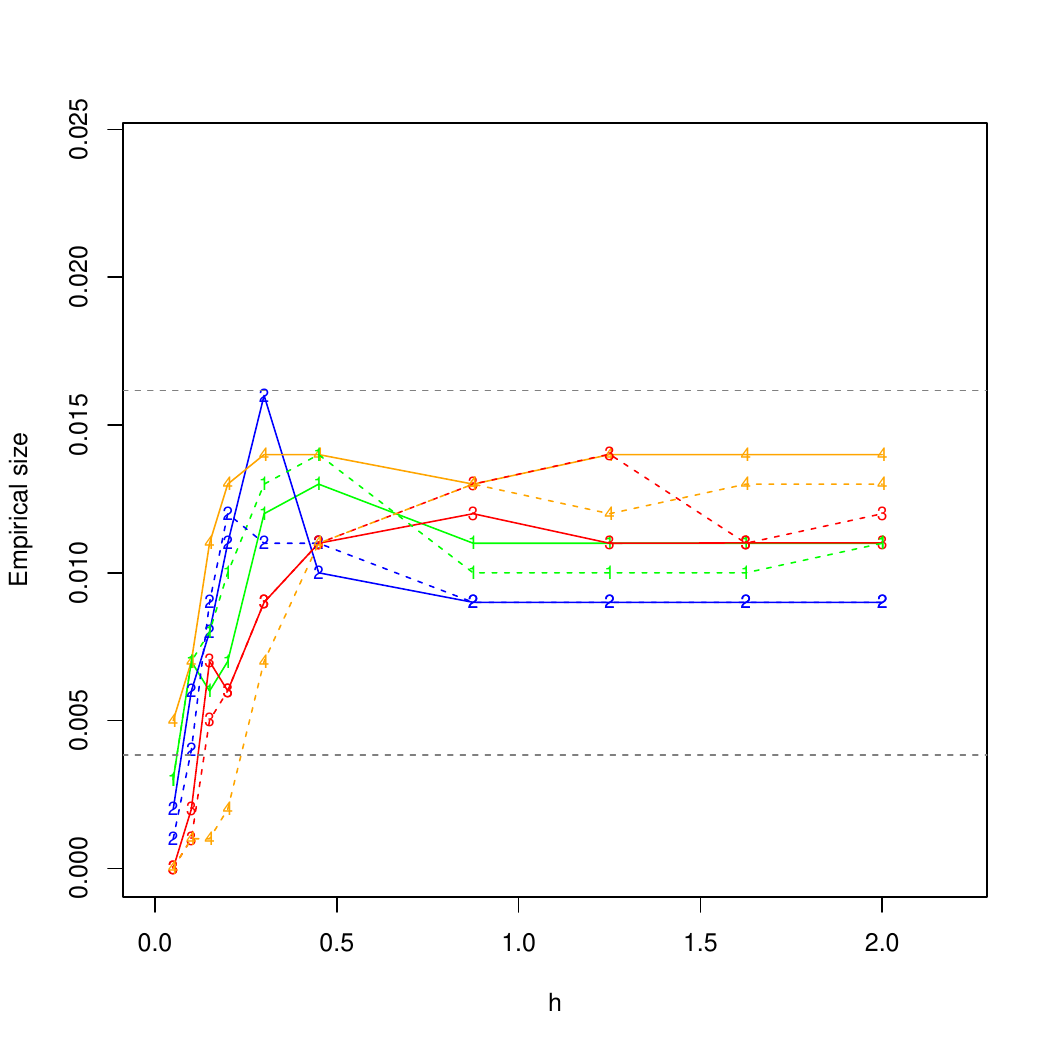}
	\includegraphics[width=0.32\textwidth]{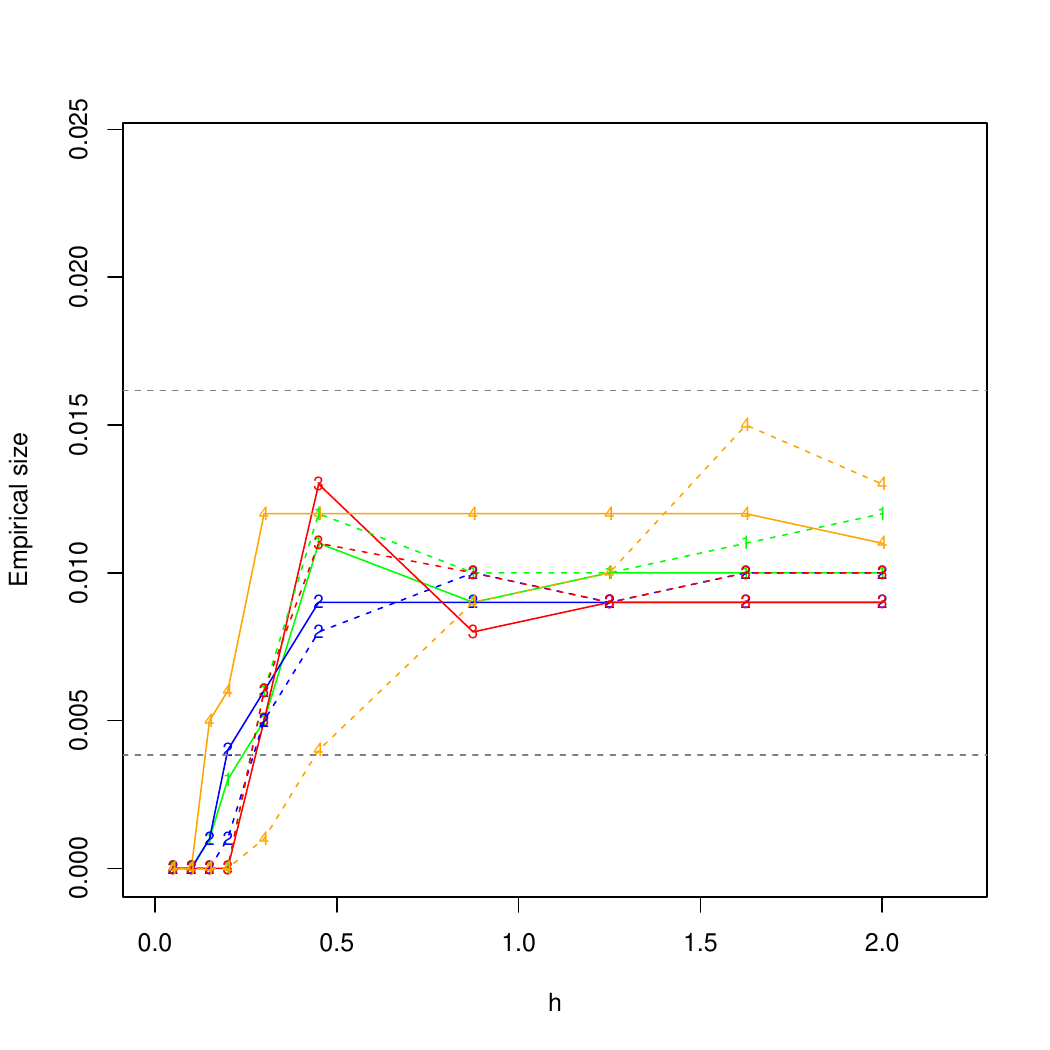}
	\includegraphics[width=0.32\textwidth]{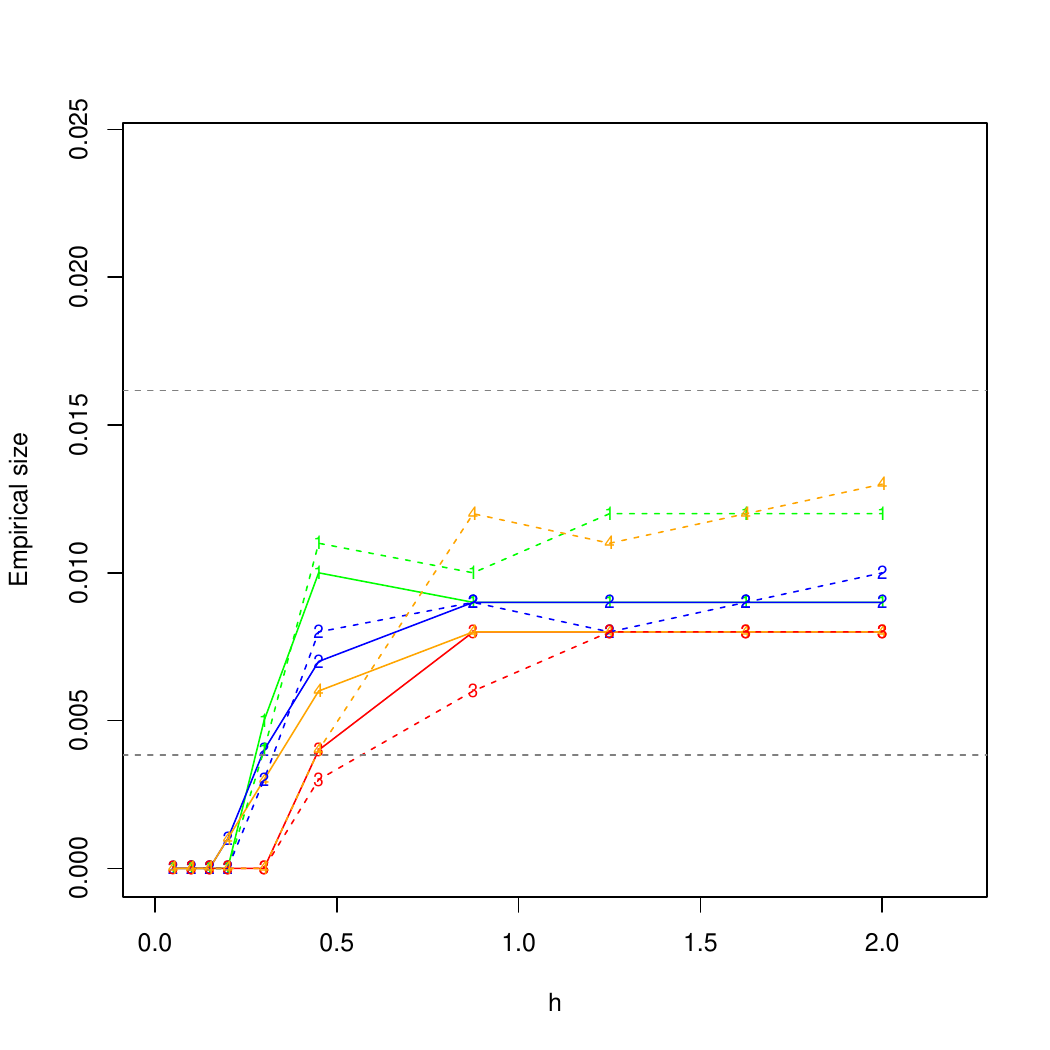}\\
	\includegraphics[width=0.32\textwidth]{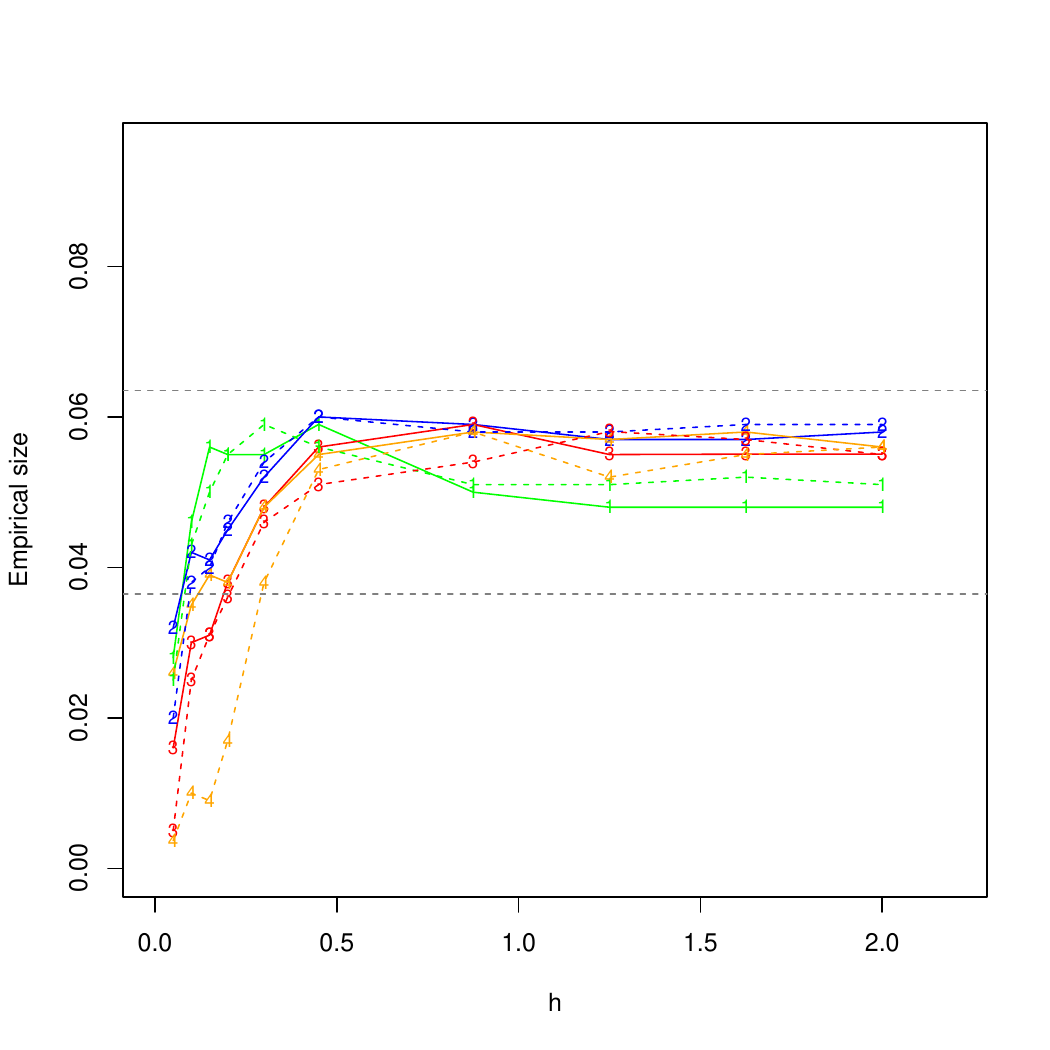}
	\includegraphics[width=0.32\textwidth]{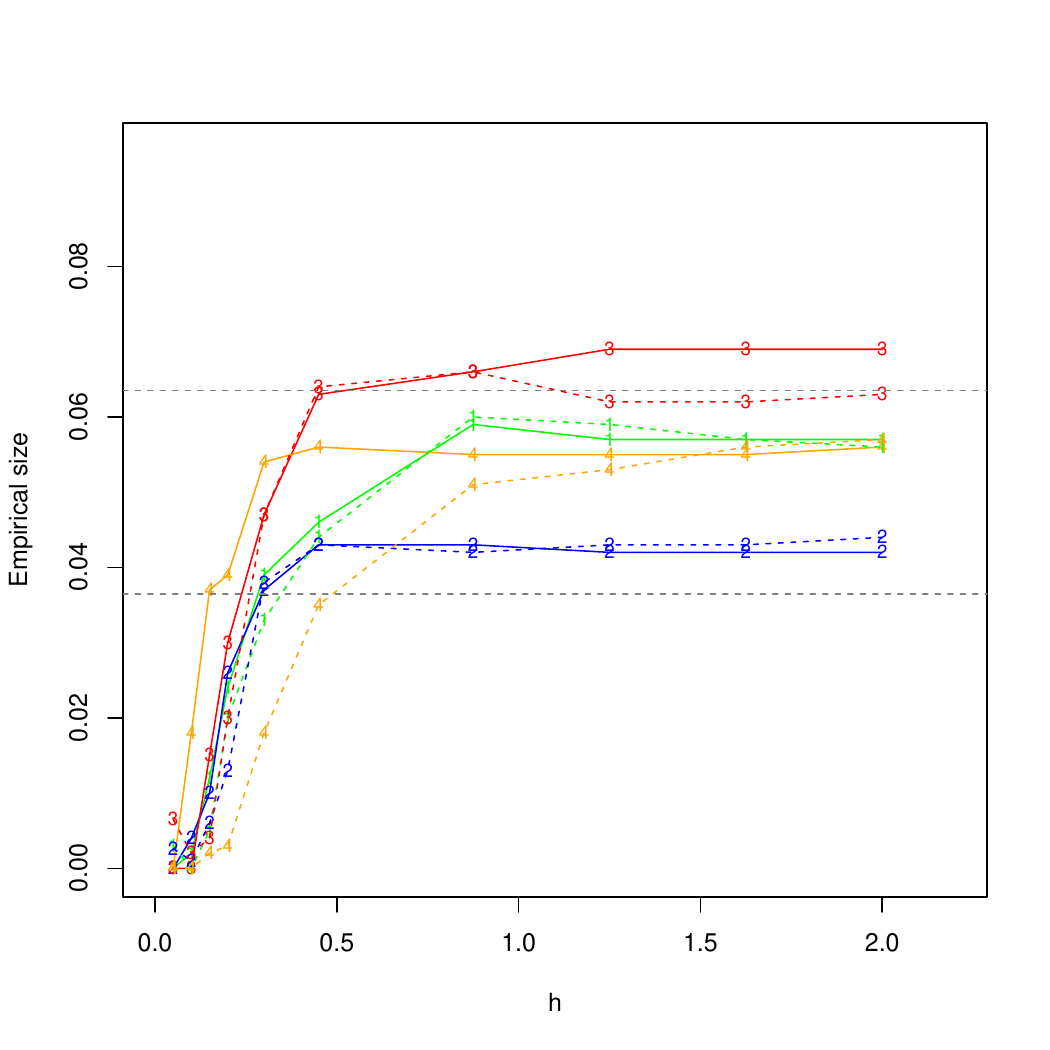}
	\includegraphics[width=0.32\textwidth]{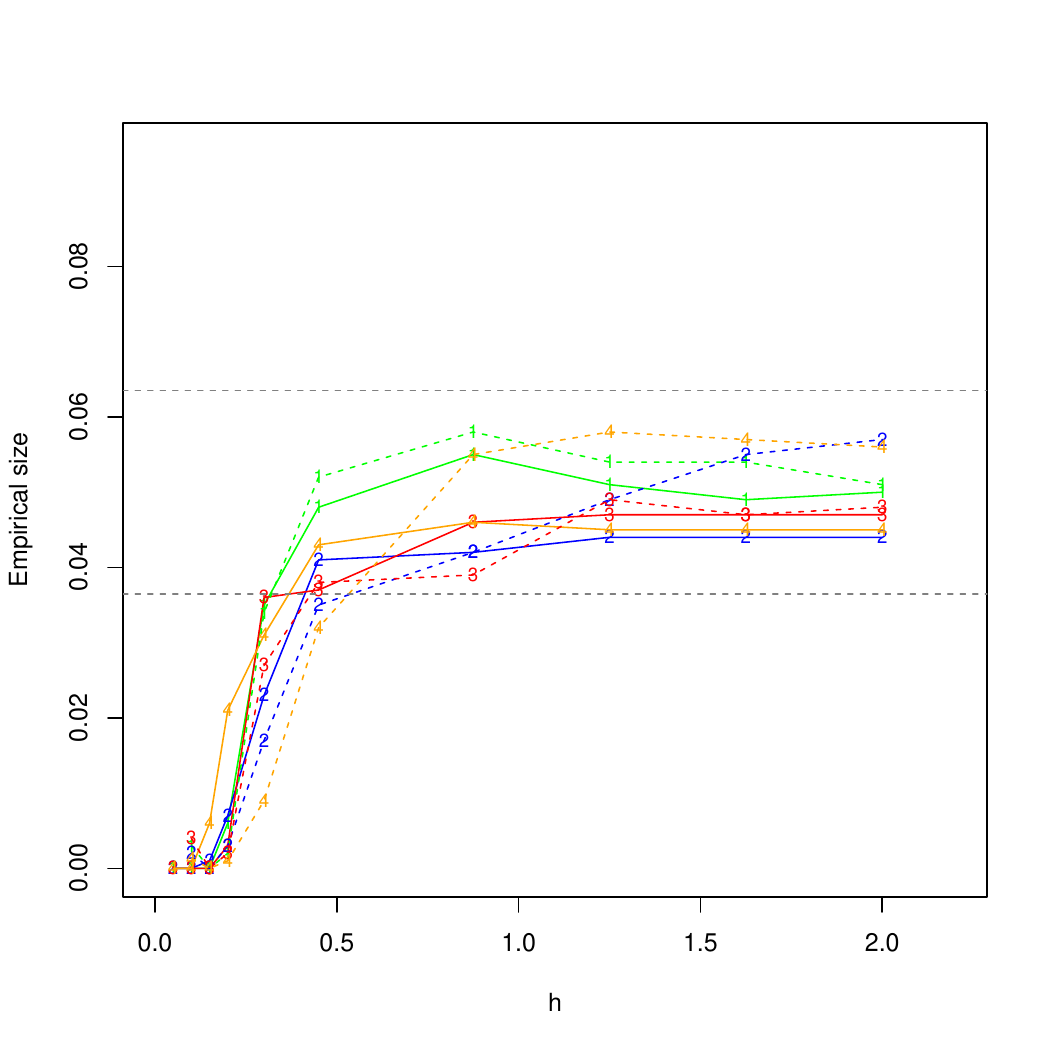}\\
	\includegraphics[width=0.32\textwidth]{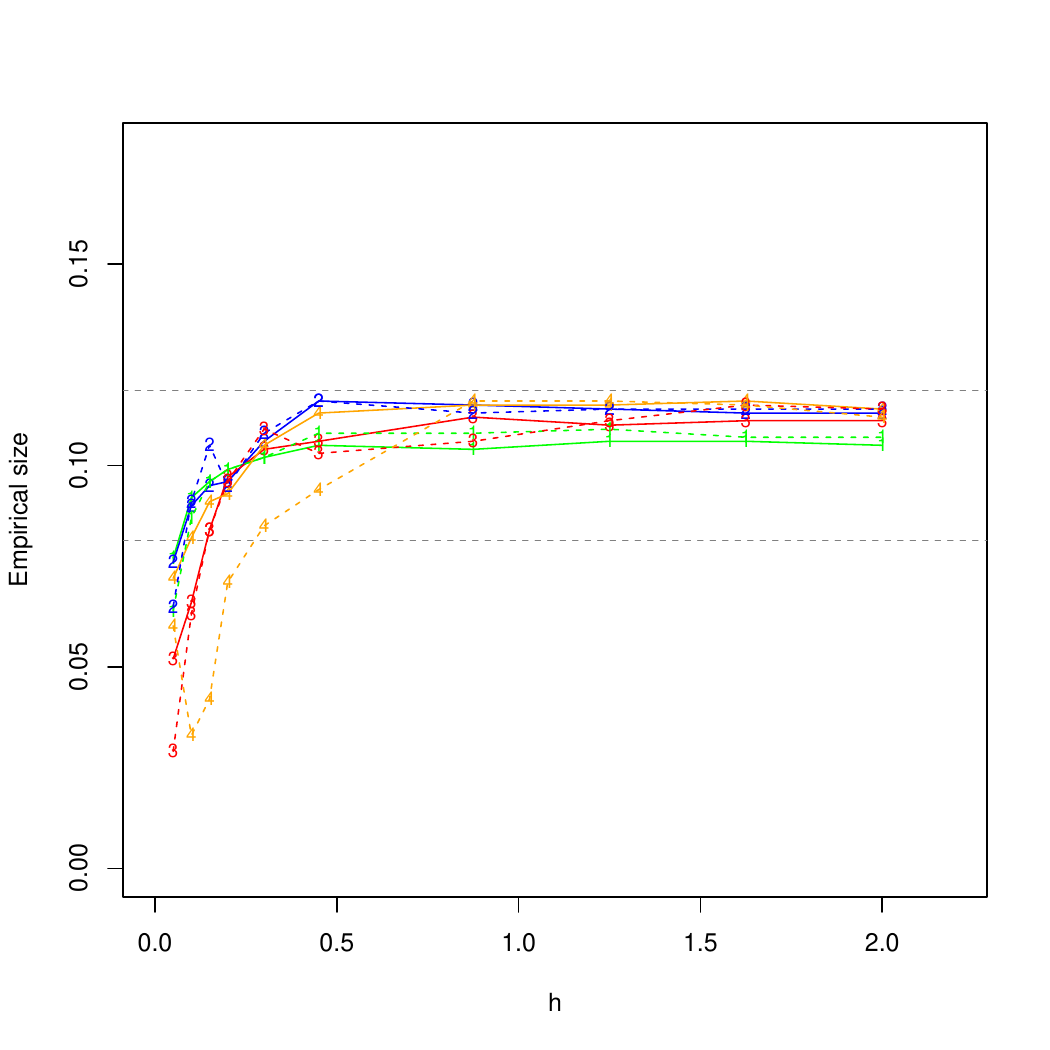}
	\includegraphics[width=0.32\textwidth]{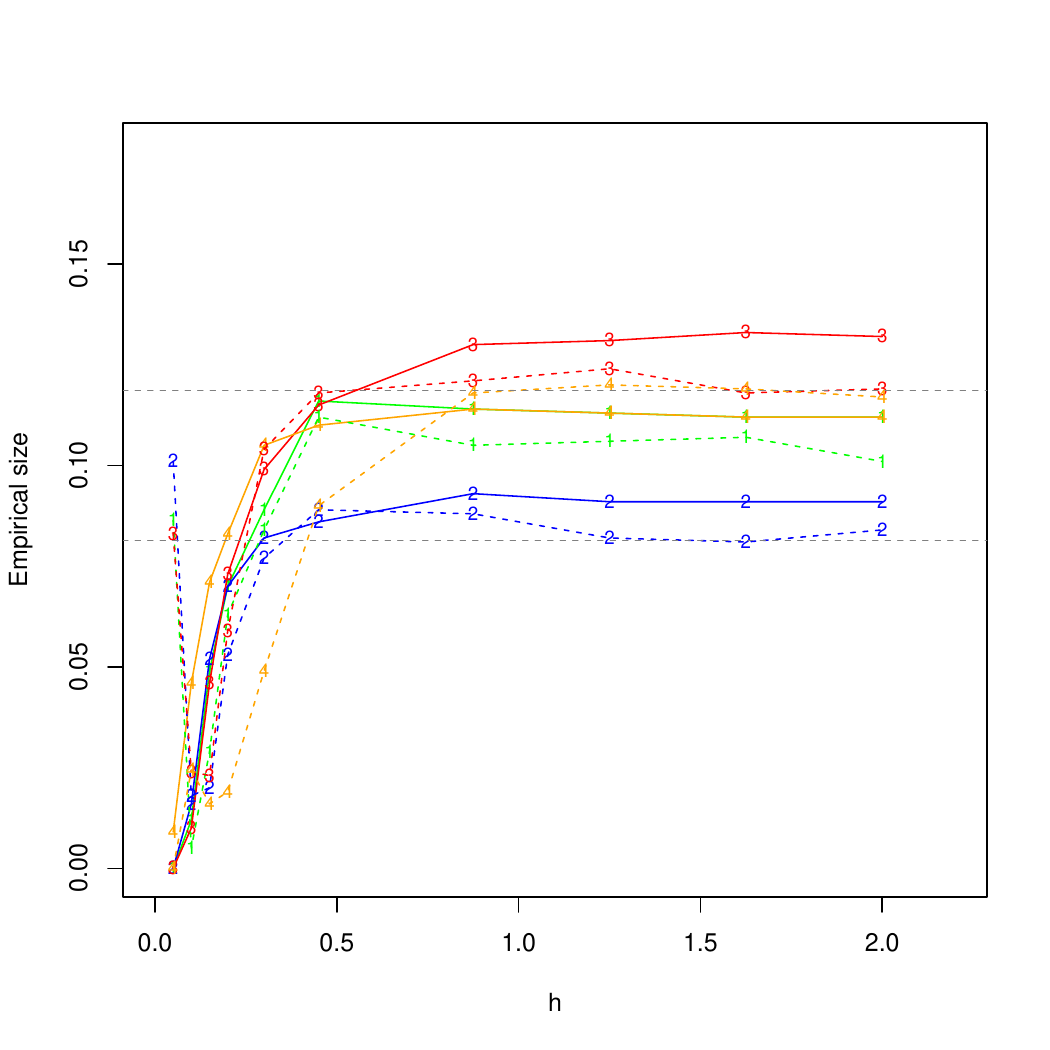}
	\includegraphics[width=0.32\textwidth]{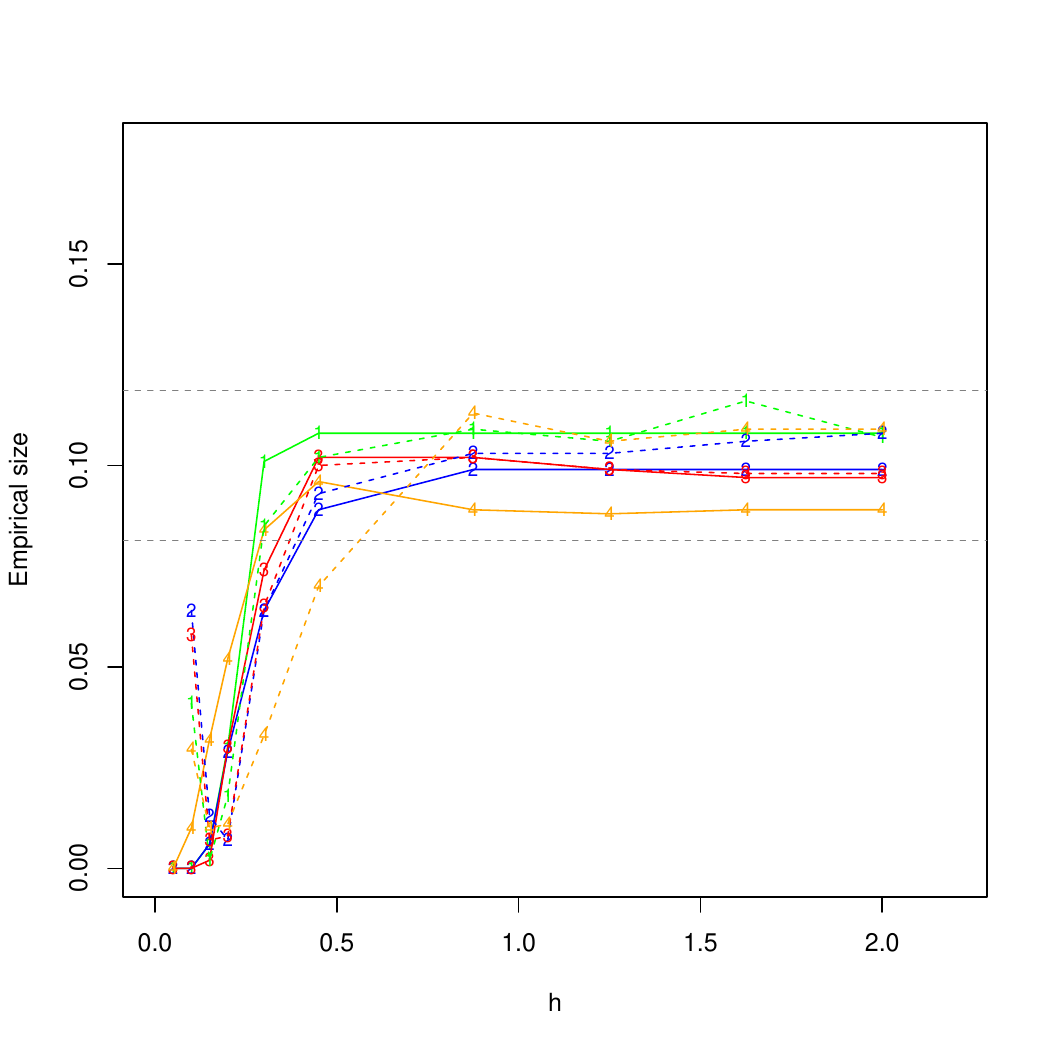}
	\caption{\small Empirical sizes for $\alpha=0.01$ (first row), $\alpha=0.05$ (second row) and $\alpha=0.10$ (third row) for the different scenarios, with $p=0$ (solid line) and $p=1$ (dashed line). From left to right, columns represent dimensions $q=1,2,3$ with sample size $n=250$. Green, blue, red and orange colors correspond to scenarios S1 to S4, respectively. \label{fig:size:2}}
\end{figure}

\vspace*{\fill}

\pagebreak

\vspace*{\fill}

\begin{figure}[H]
	\centering
	\includegraphics[width=0.32\textwidth]{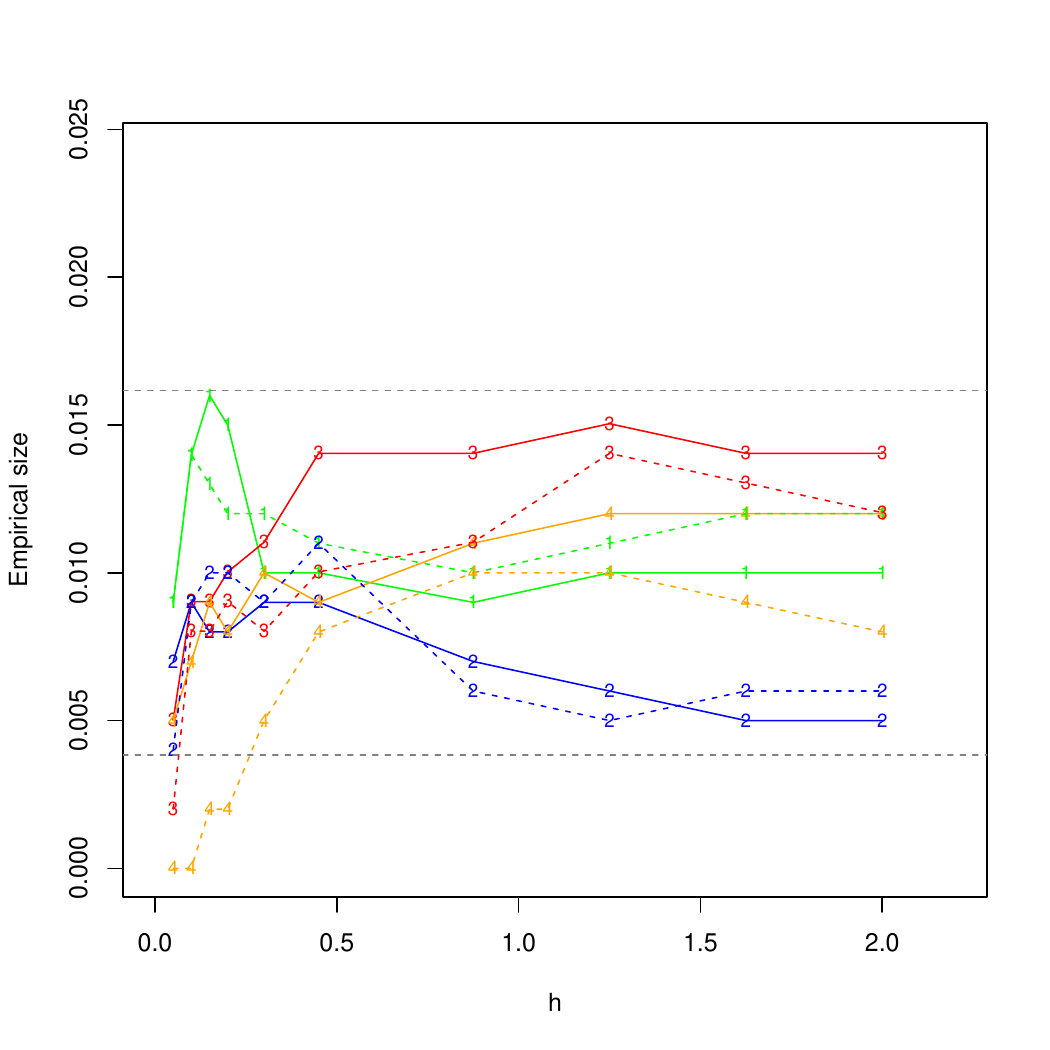}
	\includegraphics[width=0.32\textwidth]{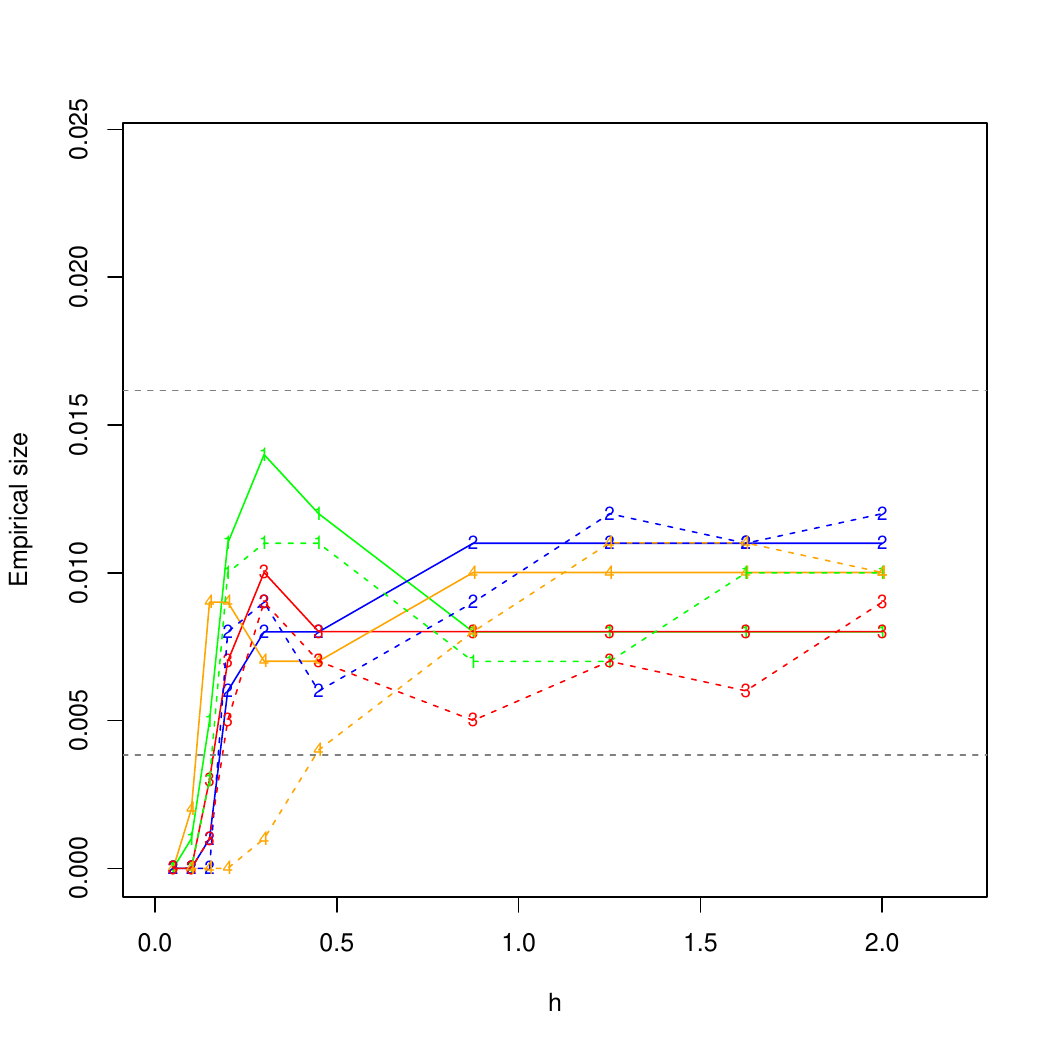}
	\includegraphics[width=0.32\textwidth]{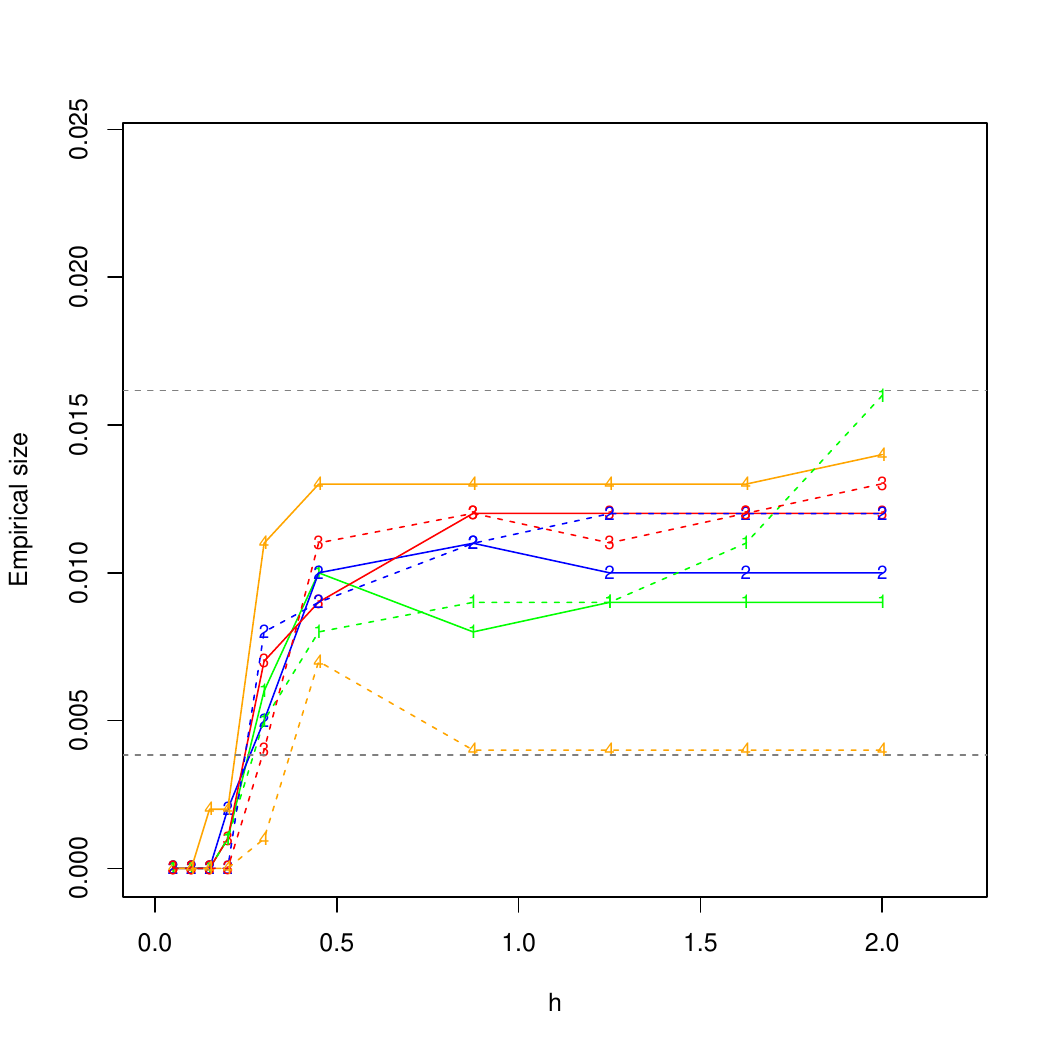}\\
	\includegraphics[width=0.32\textwidth]{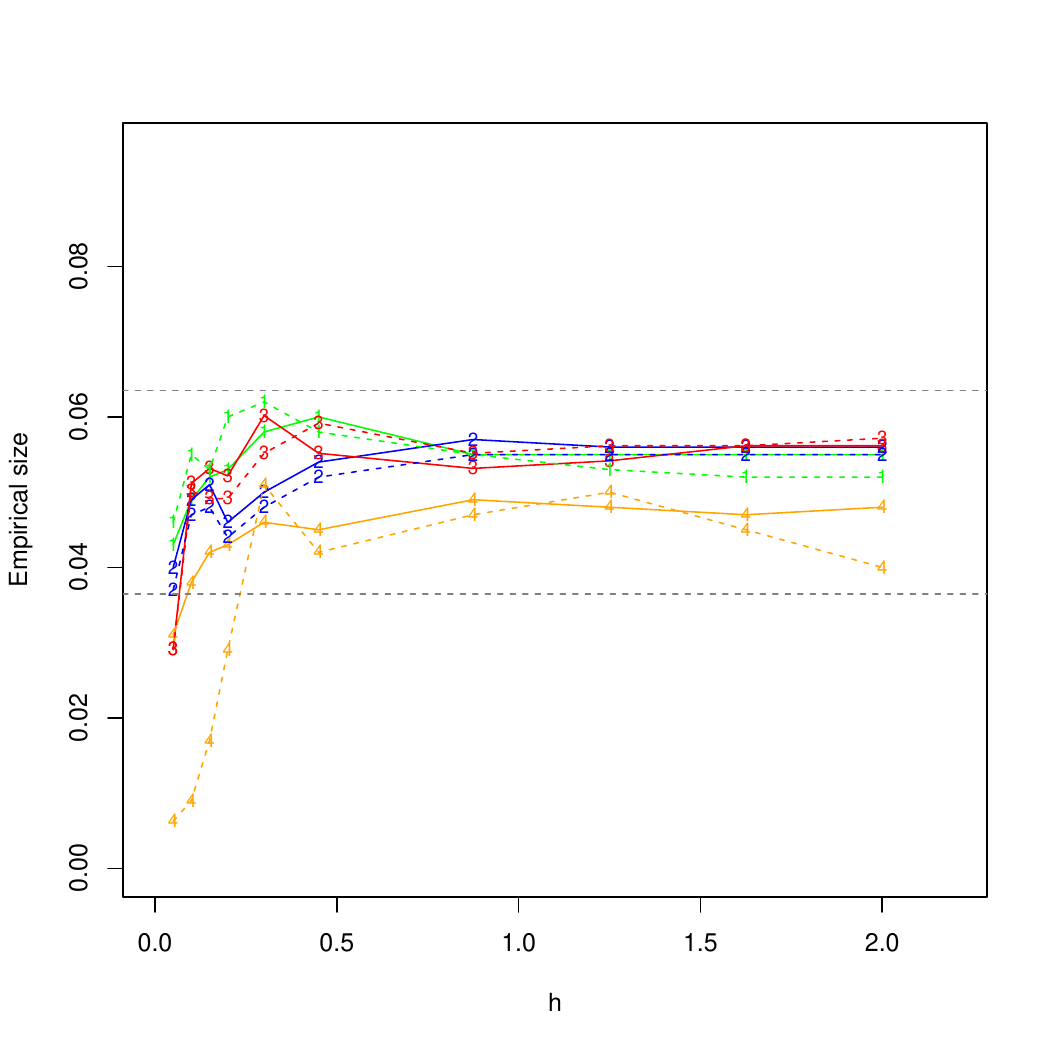}
	\includegraphics[width=0.32\textwidth]{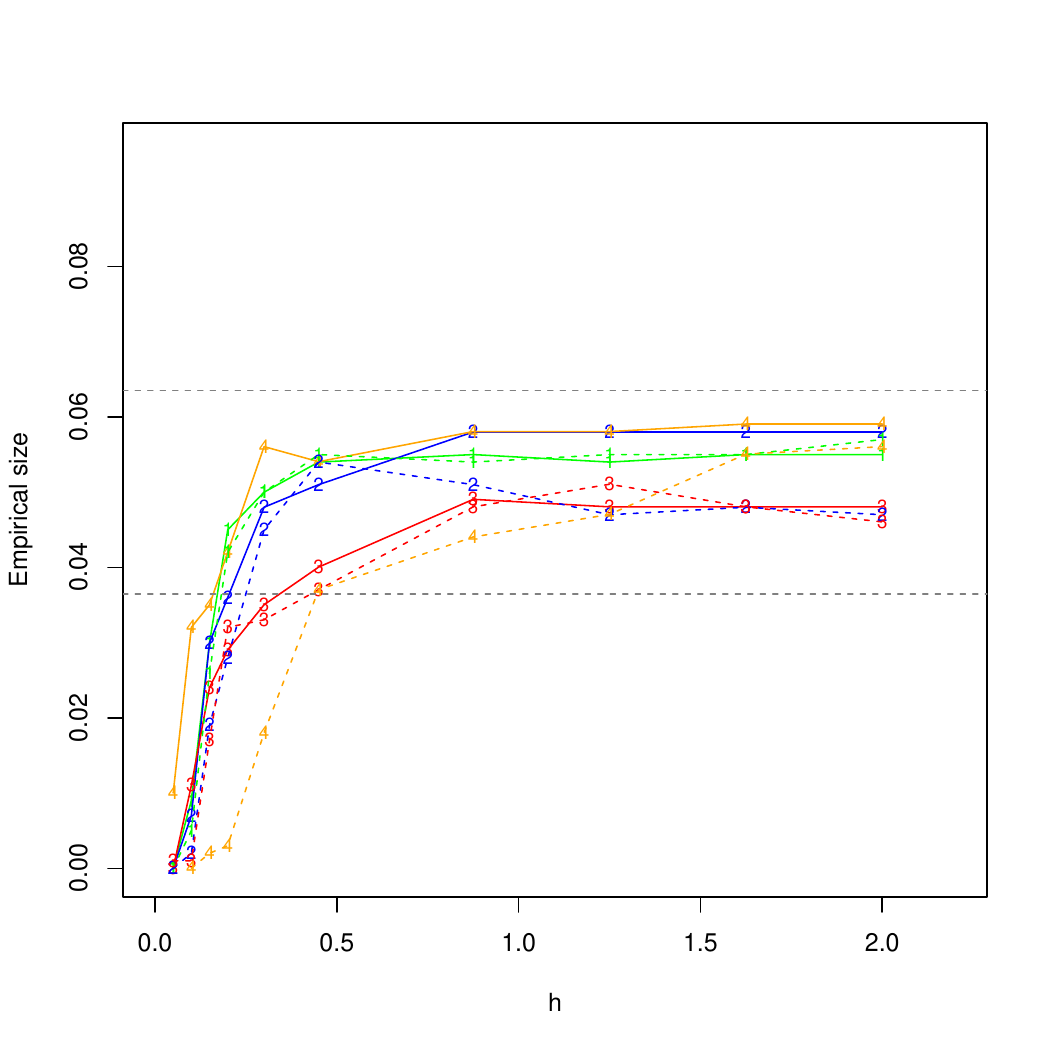}
	\includegraphics[width=0.32\textwidth]{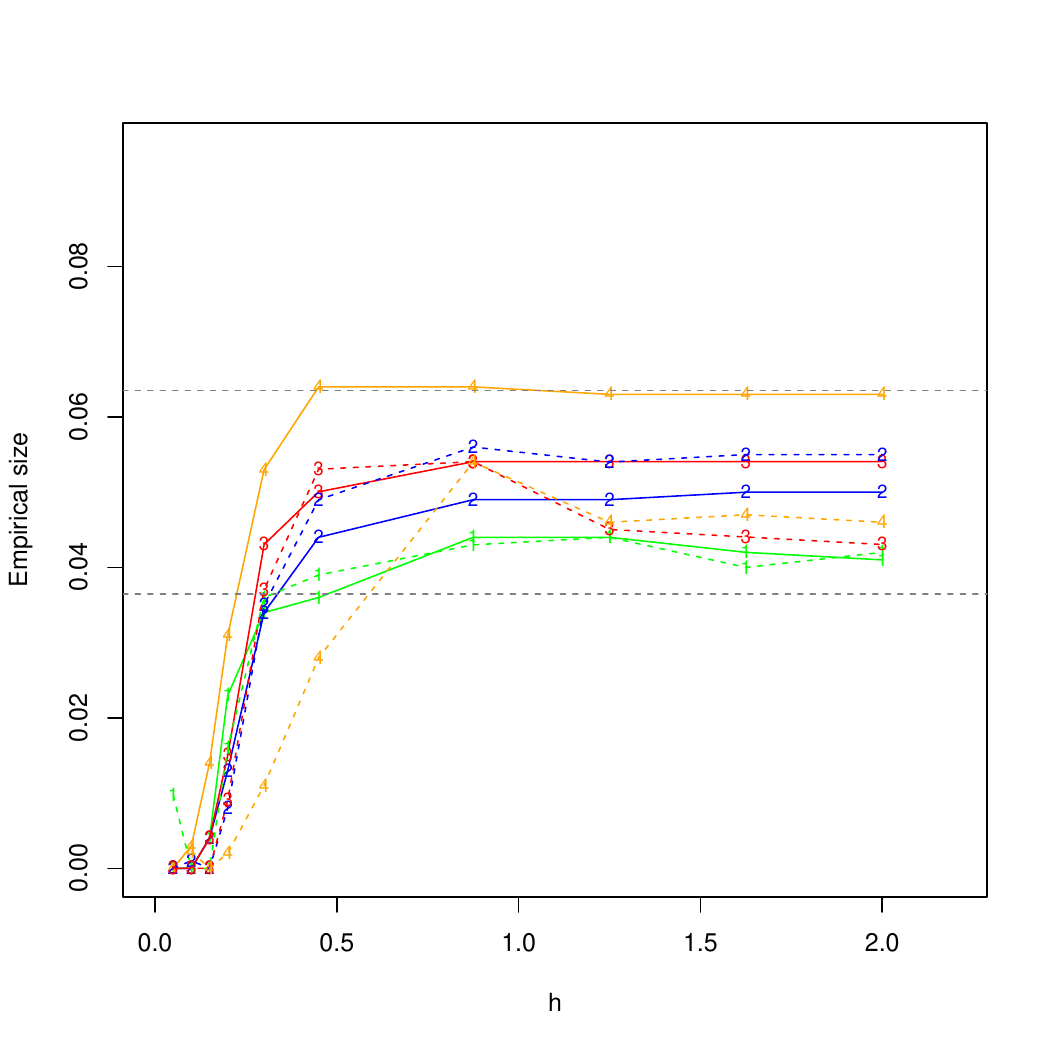}\\
	\includegraphics[width=0.32\textwidth]{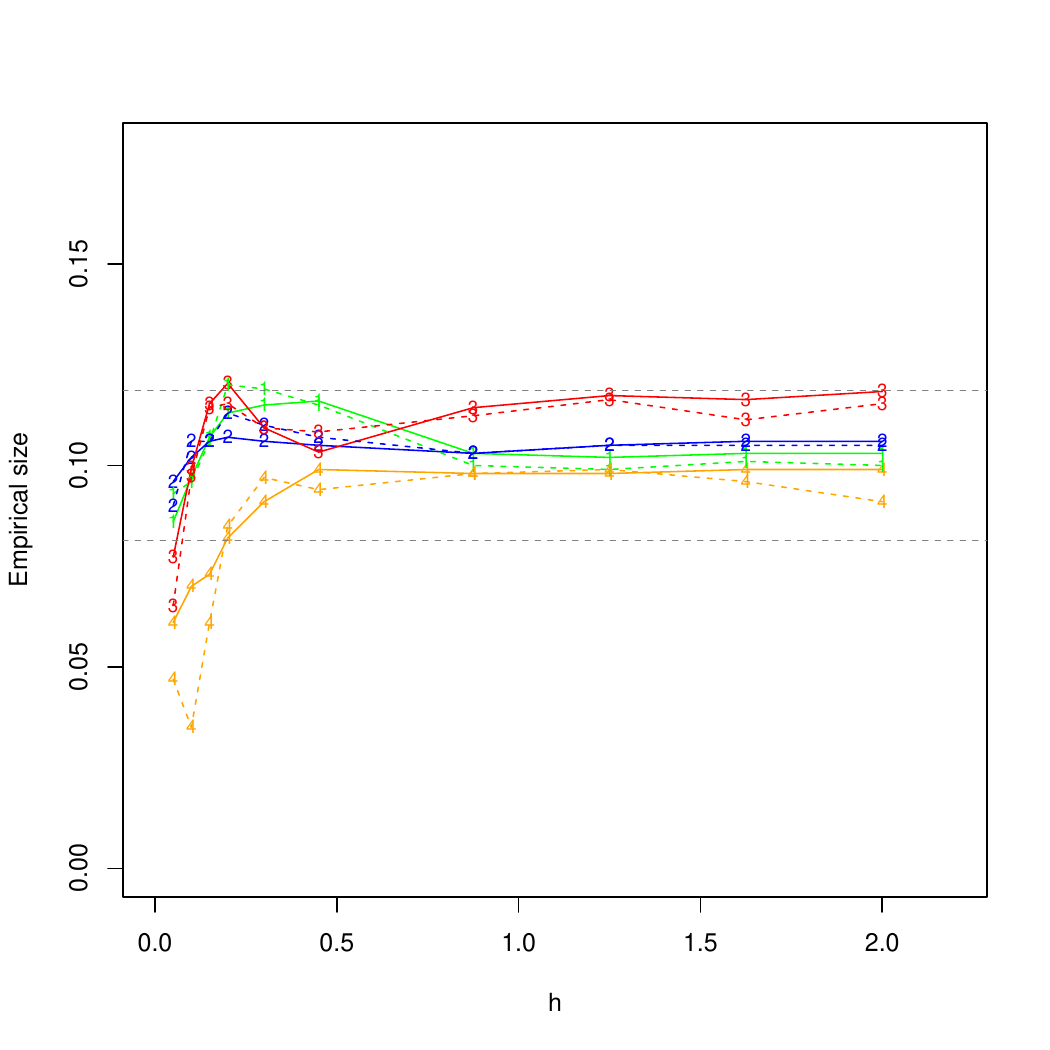}
	\includegraphics[width=0.32\textwidth]{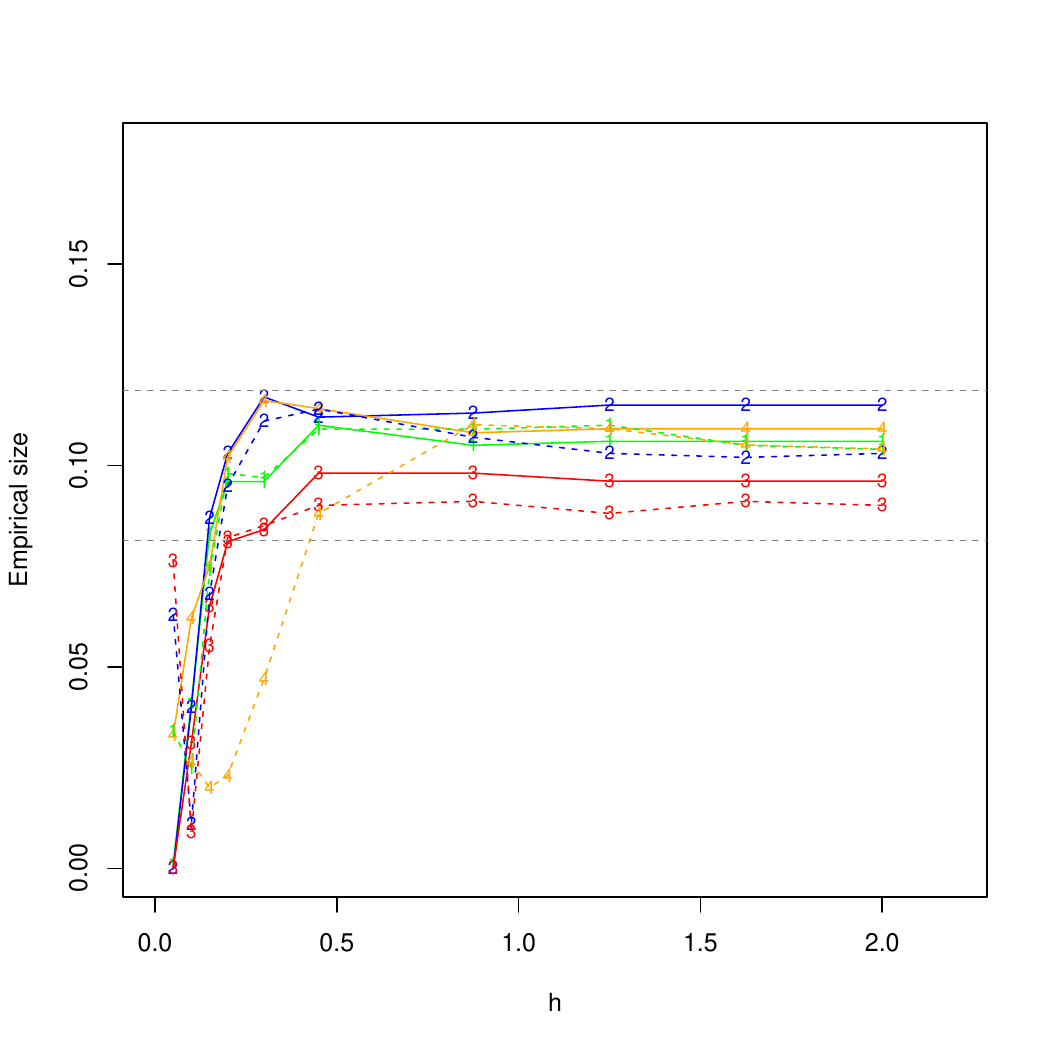}
	\includegraphics[width=0.32\textwidth]{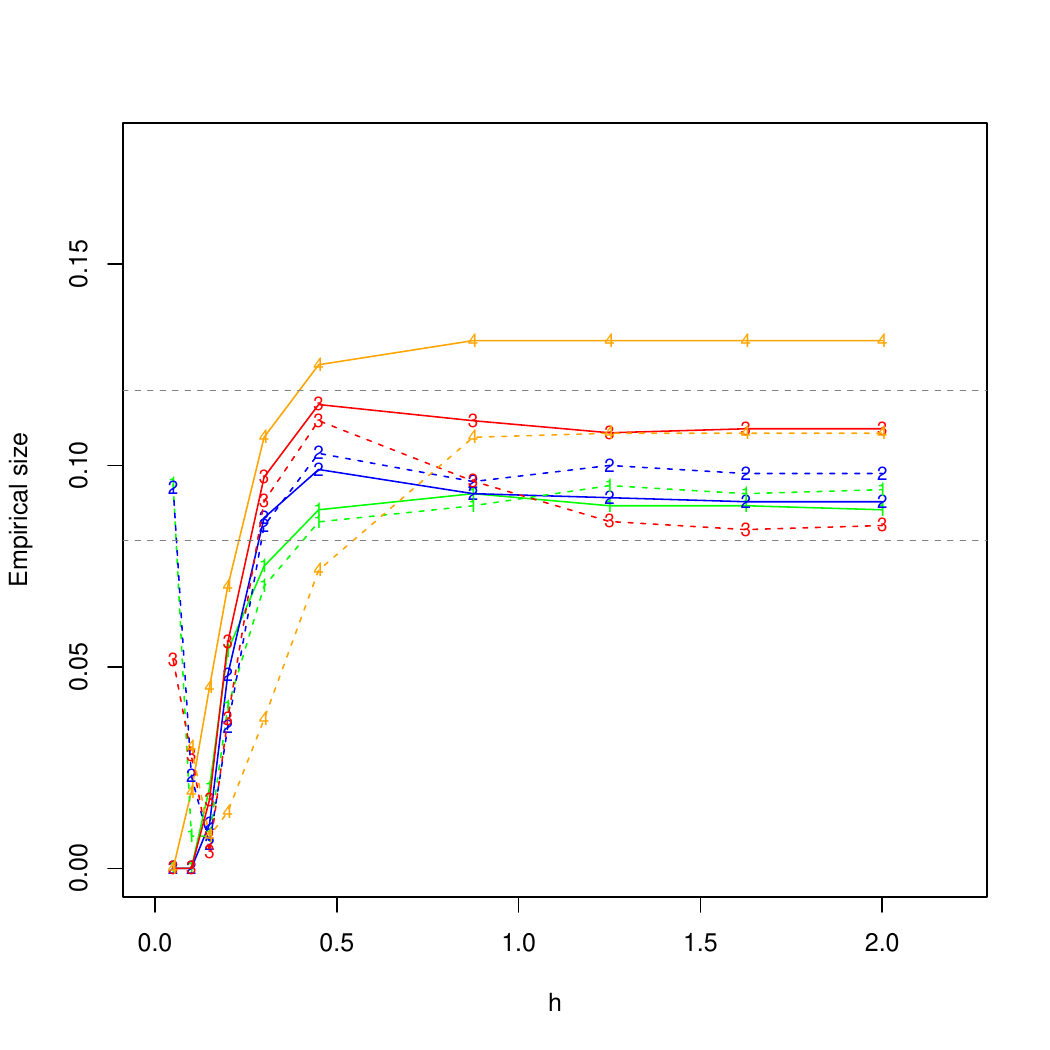}
	\caption{\small Empirical sizes for $\alpha=0.01$ (first row), $\alpha=0.05$ (second row) and $\alpha=0.10$ (third row) for the different scenarios, with $p=0$ (solid line) and $p=1$ (dashed line). From left to right, columns represent dimensions $q=1,2,3$ with sample size $n=500$. Green, blue, red and orange colors correspond to scenarios S1 to S4, respectively. \label{fig:size:3}}
\end{figure}

\vspace*{\fill}

\pagebreak

\vspace*{\fill}

\begin{figure}[H]
	\centering
	\includegraphics[width=0.32\textwidth]{img/fig12.pdf}
	\includegraphics[width=0.32\textwidth]{img/fig13.pdf}
	\includegraphics[width=0.32\textwidth]{img/fig14.pdf}\\
	\includegraphics[width=0.32\textwidth]{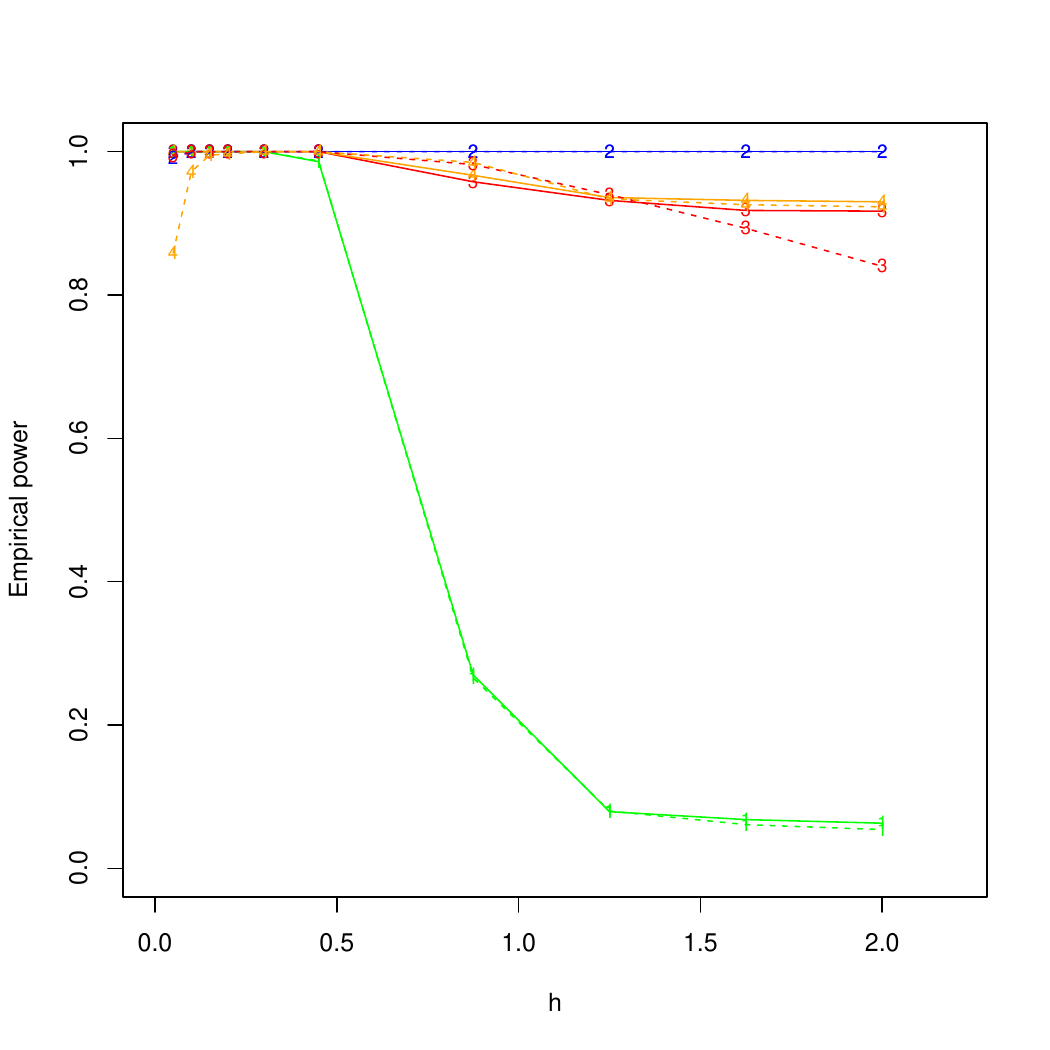}
	\includegraphics[width=0.32\textwidth]{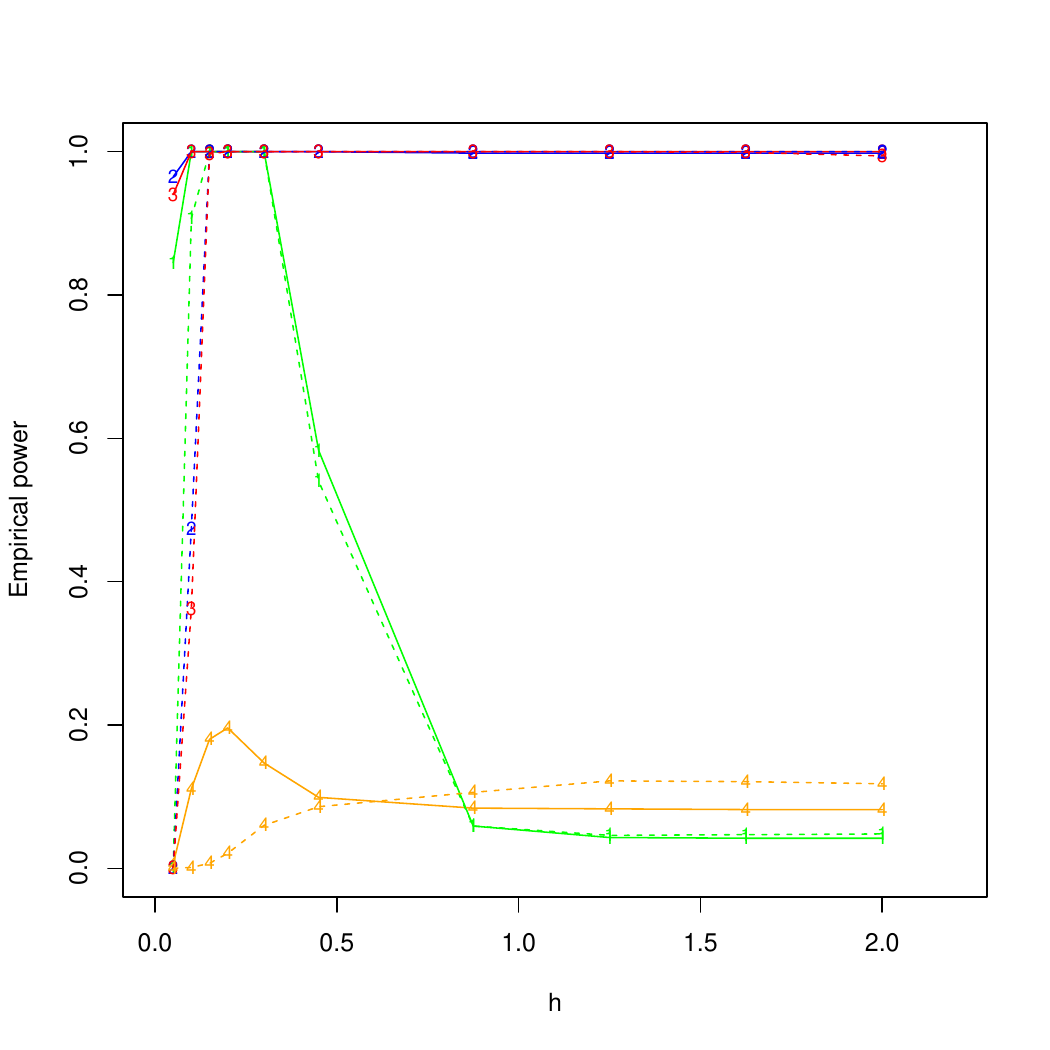}
	\includegraphics[width=0.32\textwidth]{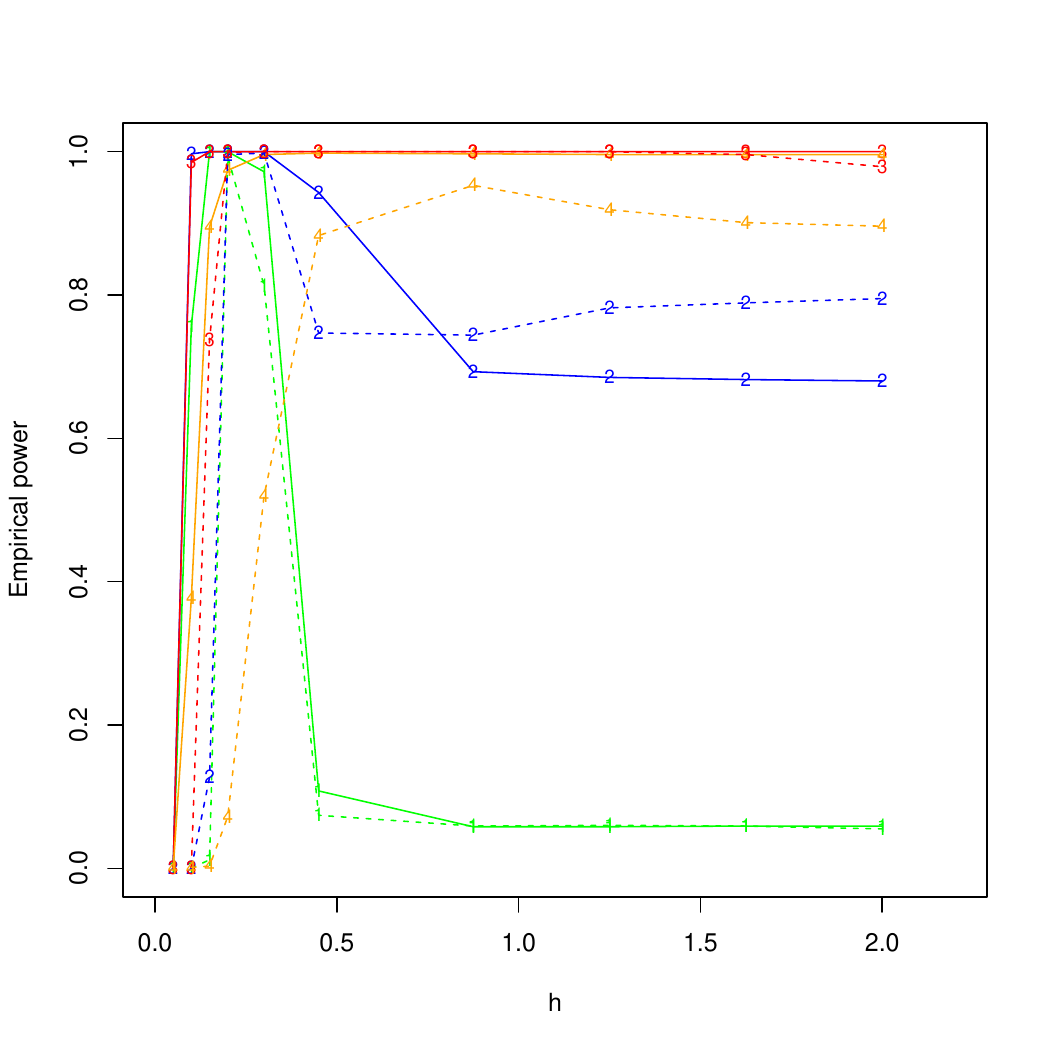}\\
	\includegraphics[width=0.32\textwidth]{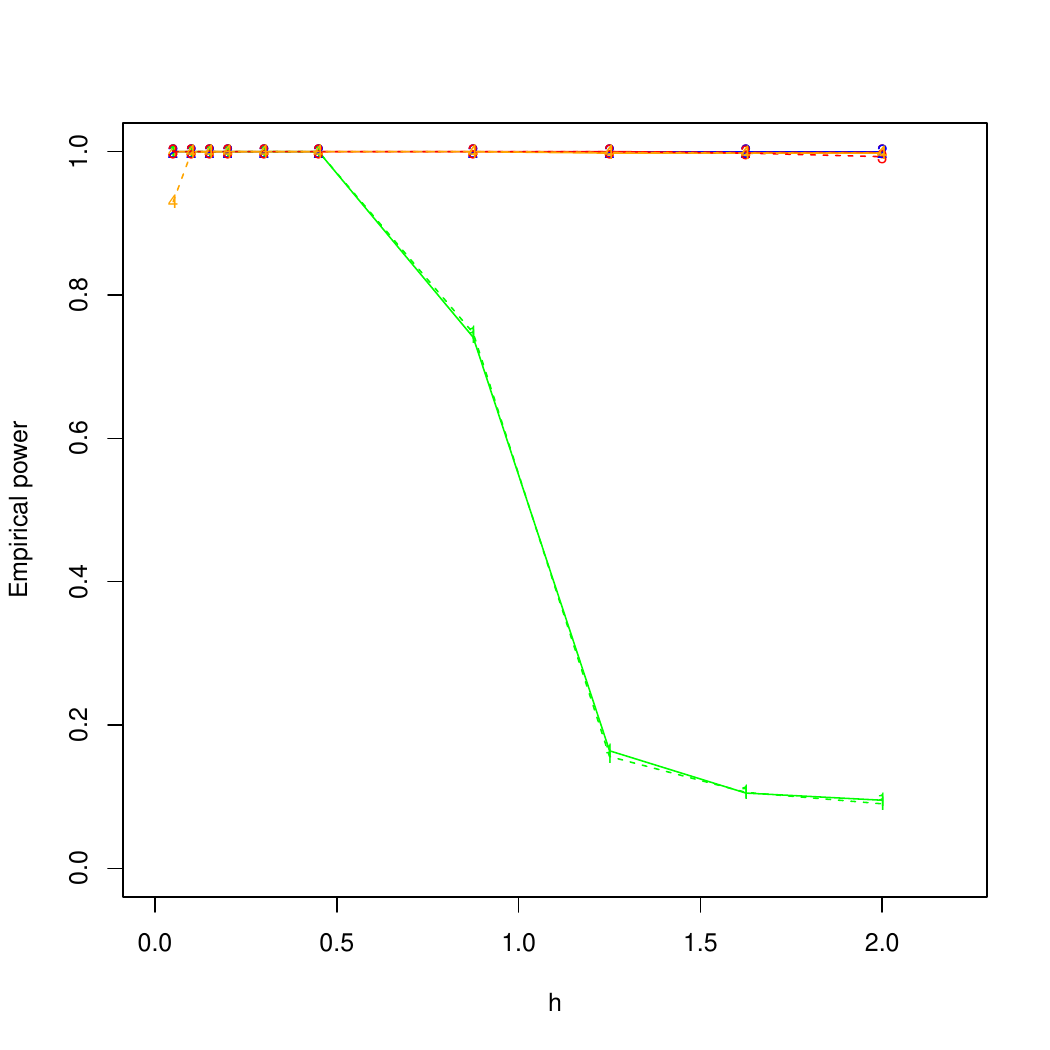}
	\includegraphics[width=0.32\textwidth]{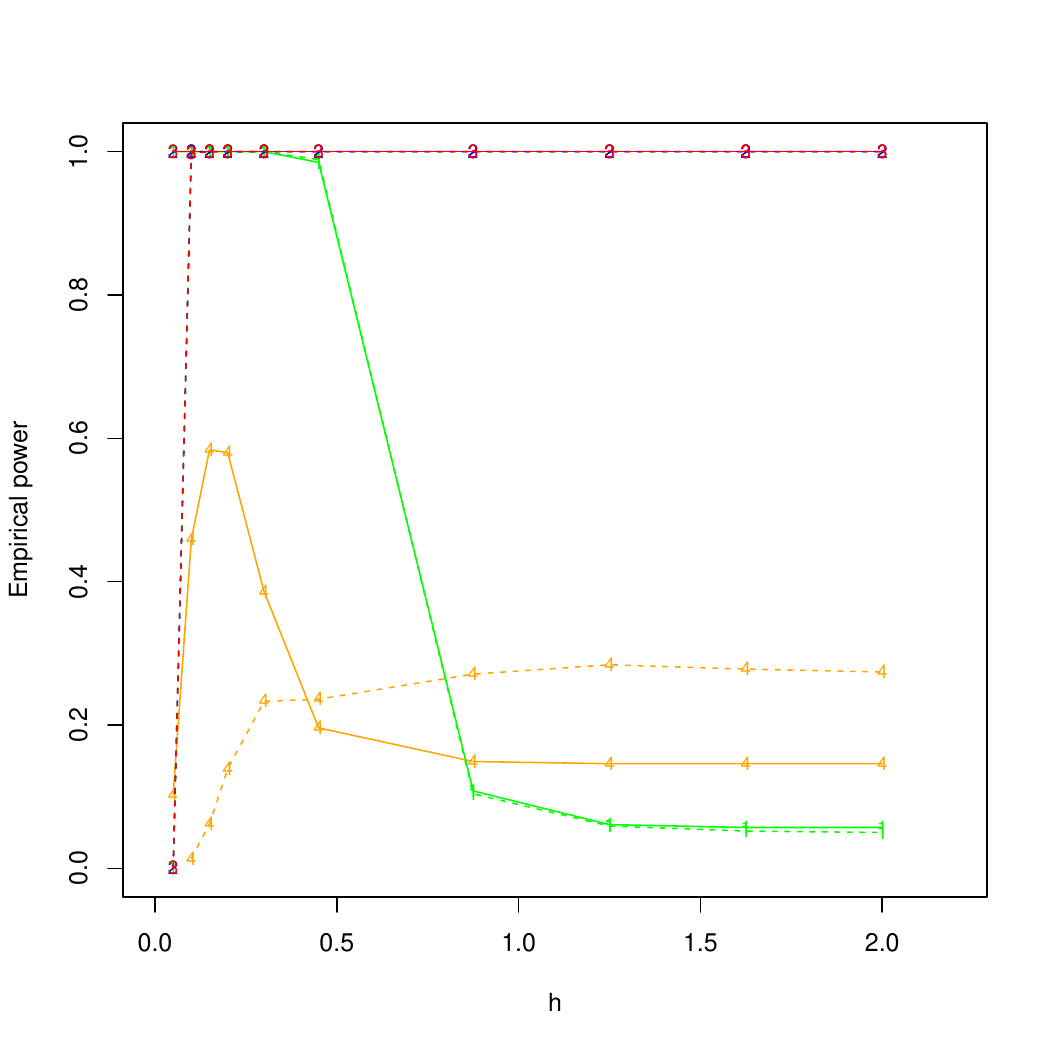}
	\includegraphics[width=0.32\textwidth]{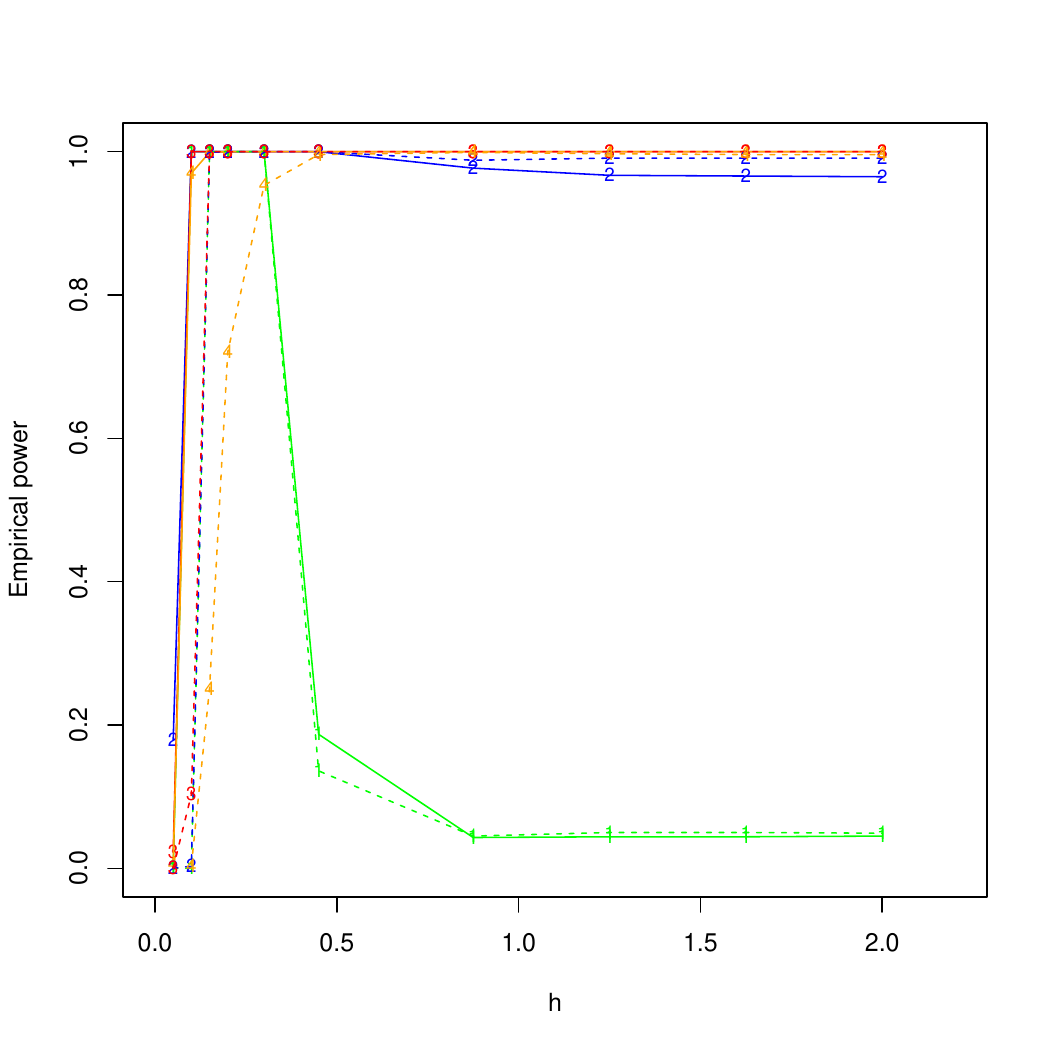}
	\caption{\small Empirical powers for the different scenarios, with $p=0$ (solid line) and $p=1$ (dashed line). From top to bottom, rows represent sample sizes $n=100,250,500$ and from left to right, columns represent dimensions $q=1,2,3$. Green, blue, red and orange colors correspond to scenarios S1 to S4, respectively. \label{fig:pow:2}}
\end{figure}

\vspace*{\fill}

\pagebreak

\section{Further information on the text mining application}
\label{appendix:data}

The acquisition and preprocessing of the dataset was done as follows. The titles, summaries and number of comments in each news appeared in the news aggregator \textit{Slashdot} (\url{wwww.slashdot.org}) in 2013 were downloaded from the website archive, resulting in a collection of $n=8121$ documents. After that, the next steps were performed with the help of the text mining \texttt{R} library \texttt{tm} \citep{Meyer2008}: 1) merge titles and summaries in the same document, omitting user submission details; 2) deletion of HTML codes; 3) conversion to lowercase; 4) deletion of stop words (defined in \texttt{tm} and \texttt{MySQL}), punctuation, white spaces and numbers; 5) stemming of words. The distribution of the \textit{document frequency} (number of documents containing a particular word) is highly right skewed and more than $50\%$ of the processed words only appeared in a single document, while in contrast a few words are repeated in many documents. To overcome this problem, a pruning was done such that only the words with document frequency between quantiles $95\%$ and $99.95\%$ were considered (words appearing within 58 and 1096 documents). After this process, the documents were represented in a document term matrix (normalized by rows with the Euclidean norm) formed by the $D=1508$~words.\\

Figure \ref{fig:text} shows the significance trace of the goodness-of-fit test with the local constant estimator for the constrained linear model. The minimum $p$-value is $0.120$. Table \ref{tab:model} gives the $77$ coefficients of the fitted model, each one linked to a stem in the dictionary of $D$ stemmed words.

\begin{figure}[H]
	\centering
	\includegraphics[scale=0.5]{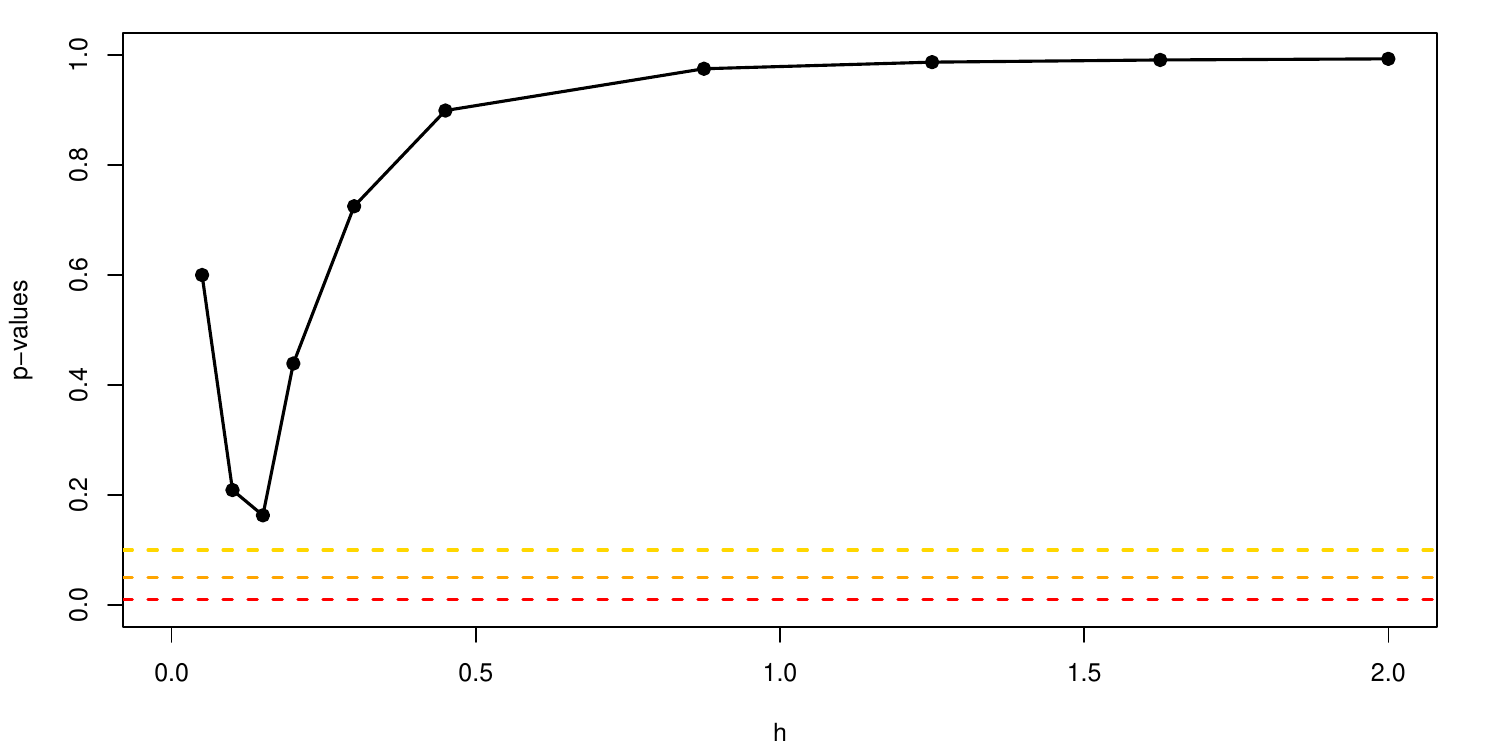}
	\caption{\small Significance trace of the local constant goodness-of-fit test for the constrained linear model. Dashed lines represents the significance levels $0.10$, $0.05$ and $0.01$. \label{fig:text}}
\end{figure}

\begin{table}[H]
	\centering	
	\scriptsize
	\setlength\tabcolsep{1.5pt}
	\renewcommand{\arraystretch}{1}
	\begin{tabular}{ccccccccccccc}\toprule\toprule
		(int) & conclud & gun & kill & refus & averag & lose & obama & declin & climat & snowden & stop & wrong \\
		$4.97$ & $2.56$ & $2.13$ & $1.86$ & $1.77$ & $1.74$ & $1.72$ & $1.68$ & $1.63$ & $1.53$ & $1.44$ & $1.43$ & $1.35$ \\\midrule
		war & polit & senat & tesla & violat & concern & slashdot & ban & reason & health & pay & window & american \\
		$1.34$ & $1.31$ & $1.27$ & $1.26$ & $1.25$ & $1.22$ & $1.22$ & $1.19$ & $1.15$ & $1.14$ & $1.14$ & $1.12$ & $1.10$ \\\midrule
		told & worker & man & comment & state & think & movi & ask & job & drive & know & problem & employe \\
		$1.10$ & $1.09$ & $1.09$ & $1.04$ & $1.00$ & $0.97$ & $0.96$ & $0.95$ & $0.94$ & $0.91$ & $0.87$ & $0.87$ & $0.87$ \\\midrule
		nsa & charg & feder & money & sale & need & microsoft & project & network & cell & imag & avail & video \\
		$0.84$ & $0.80$ & $0.80$ & $0.80$ & $0.78$ & $0.76$ & $0.52$ & $-0.46$ & $-0.51$ & $-0.69$ & $-0.70$ & $-0.73$ & $-0.78$ \\\midrule
		process & data & materi & nasa & launch & electron & robot & satellit & detect & planet & help & cloud & hack \\
		$-0.81$ & $-0.82$ & $-0.88$ & $-0.92$ & $-0.92$ & $-0.94$ & $-0.95$ & $-0.96$ & $-1.04$ & $-1.06$ & $-1.06$ & $-1.08$ & $-1.10$ \\\midrule
		open & lab & mobil & techniqu & vulner & mission & team & supercomput & abstract & simul & demo & guid &  \\
		$-1.15$ & $-1.15$ & $-1.16$ & $-1.17$ & $-1.21$ & $-1.23$ & $-1.50$ & $-1.89$ & $-1.97$ & $-1.99$ & $-2.01$ & $-2.02$ &  \\
		\bottomrule\bottomrule
	\end{tabular}
	\caption{\small Fitted constrained linear model on the Slashdot dataset, with $R^2=0.25$. The significances of each coefficient are lower than $0.002$.\label{tab:model}}
\end{table}

\fi

\end{document}